\documentclass[fleqn,usenatbib]{mnras}

\usepackage{newtxtext,newtxmath}

\usepackage[T1]{fontenc}

\DeclareRobustCommand{\VAN}[3]{#2}
\let\VANthebibliography\thebibliography
\def\thebibliography{\DeclareRobustCommand{\VAN}[3]{##3}\VANthebibliography}

\usepackage{graphicx}	
\usepackage{amsmath}	
\usepackage{subfloat}
\usepackage{booktabs}

\title[]{Correlation between optical flux and polarization variations in Flat Spectrum Radio Quasars on diverse timescales}
\author[A. Pandey et al.]{
Ashwani Pandey$^{1}$\thanks{E-mail: ashwanitapan@gmail.com},
Bhoomika Rajput$^{1}$
and C. S. Stalin$^{1}$
\\
$^{1}$Indian Institute of Astrophysics, Block II, Koramangala, Bangalore 560034, India
}

\date{Accepted XXX. Received YYY; in original form ZZZ}

\pubyear{2021}

\begin{document}
\label{firstpage}
\pagerange{\pageref{firstpage}--\pageref{lastpage}}
\maketitle

\begin{abstract}
Study of the polarization behaviour in blazars is a powerful tool to discern the role of magnetic field in the variable emission process in their relativistic jets. We  present here results of our systematic investigation on the correlation between optical flux and polarization variations for eight flat spectrum radio quasars on various timescales using data from the Steward Observatory that covers a period of $\sim$10 years. 
On long time scales ($\sim$several months), from a total of 79 observing cycles, in 34 observing cycles, we found a significant positive correlation between optical flux and optical polarization degree (PD), negative correlation in 3 cycles and no correlation in 42 cycles. On short time scales ($\sim$few days), in 47 out of a total of 55 epochs, we found a positive correlation between optical flux and PD, while, on the remaining 8 epochs, an anti-correlation was detected between the two quantities. Moreover, we noticed a significant positive correlation between optical and $\gamma-$ray fluxes on 14 epochs and a negative correlation between the two on one epoch. While the observed optical flux changes well fit in the shock-in-jet model, the observed changes in PD are not explainable by changes in the power-law spectral index of the relativistic electrons in the jet. Instead, the observed varied correlations between optical flux and PD could be due to multi-zone emission regions or the enhanced flux coinciding with the emergence of a new emission knot with its magnetic field either aligned or misaligned with the large scale magnetic field.
\end{abstract}



\section{Introduction}
Blazars are a special class of active galactic nuclei (AGN), whose radiation output is dominated by Doppler boosted non-thermal emission from their 
relativistic jets that are closely oriented to the observer's line of sight \citep{1995PASP..107..803U}. Doppler boosting amplifies the measured radiation resulting in the observed flux hundreds of times larger than that emitted by the source. The emission from blazars is highly variable in all bands of the electromagnetic (EM) spectrum i.e. from radio to very high energy $\gamma$-rays \citep[e.g.][]{1995ARA&A..33..163W,1997ARA&A..35..445U,2017ApJ...841..123P}. These flux variations are observed on a wide range of timescales from minutes to years \citep[e.g.][]{2004MNRAS.350..175S,2020ApJ...890...72P,2020A&A...634A..80R}. The broadband spectral energy distribution (SED) of blazars consists of a two hump structure \citep{1998MNRAS.299..433F,2016ApJS..224...26M}. The low energy hump is attributed to the synchrotron process and the high-energy hump is attributed to the inverse Compton (IC) process \citep{2010ApJ...716...30A}. The seed photons for IC scattering can be from within the jet (synchrotron-self Compton; SSC; \citealt{1981ApJ...243..700K,1985ApJ...298..114M,1989ApJ...340..181G,1996ApJ...461..657B}) or outside the jet such as accretion disk, broad-line region, and torus (external Compton; EC; \citealt{1987ApJ...322..650B,1994ApJ...421..153S}). In the flat spectrum radio quasar (FSRQ) category of blazars the high energy component of the SED is usually produced by the EC process, while in the BL Lacertae (BL Lac) category of blazars the high energy component of the SED is predominantly due to SSC process. However, some recent studies have also reported temporary SSC dominance on some FSRQs \citep[e.g.][]{2018MNRAS.479.2037P,2019A&A...630A..56P,2020ApJ...891...68C,2021ApJ...906....5A}.

In addition to flux variations, FSRQs are also characterized by a high degree ($>$3\%) of variable optical polarization. Optical polarization in quasars is known since their discovery about six decades ago \citep{1966ApJ...146..964K} and subsequently optical polarization were measured for many sources \citep[e.g.][]{1980ARA&A..18..321A,1990A&AS...83..183M,1998AJ....116.2119V,2008ApJ...672...40H,2016ApJ...833...77I}. The observed polarized emission in optical as well as at other longer wavelengths from these sources provide strong evidence for synchrotron radiation process as a source of the low energy component of their broadband SED \citep[e.g.][]{1978ApJ...220L..67S,1990A&AS...83..183M,1991ApJ...375...46I, 2001ApJ...562..208L,2002ApJ...577...85M,2011MNRAS.412..318M}. Observations of polarized emission, therefore, provide valuable information on the magnitude and direction of the magnetic field within relativistic jets. Changes in the observed polarization position angle could be related to changes in the direction of the magnetic field in the jet along the observer's line of sight. Moreover, as the flux variations in the optical and GeV $\gamma$-rays are connected (\citealt{2009ApJ...697L..81B,2012ApJ...749..191C,2014ApJ...783...83L, 2015MNRAS.450.2677C,2018MNRAS.479.2037P,2020ApJ...891...68C,2021ApJ...906....5A}, however, see also \cite{2019MNRAS.486.1781R,2020MNRAS.498.5128R} and references therein for no correlation between optical and $\gamma$-ray flux variations) flux and polarization variability observations at different wavelengths can provide important inputs on the connection between different emission regions in the jets of these sources. Also, observations with the {\it Fermi} Gamma-ray Space Telescope (hereinafter {\it Fermi}) have revealed a close association between the $\gamma$-ray flare and the rotation of the optical polarization position angle, which again can constrain the nature of the high energy $\gamma$-ray emission process \citep{2008Natur.452..966M,2010Natur.463..919A,2010ApJ...710L.126M}.

\begin{table}
\centering
\caption{\label{tab:src_list}Details of the sources studied.}
\begin{tabular} {lcccr} \hline
Source  &   RA ($\alpha_{\rm 2000}$)  &  Dec ($\delta_{\rm 2000}$)  & Redshift (z)  & \\ \hline
OJ 248 & $08^h30^m52^s$ & $+24^{\circ}10^{\prime}60^{\prime\prime}$ & 0.941 & \\
PKS 1222$+$216 & $12^h24^m54^s$ & $+21^{\circ}22^{\prime}46^{\prime\prime}$ & 0.432 & \\
3C 273 & $12^h29^m07^s$ & $+02^{\circ}03^{\prime}09^{\prime\prime}$ & 0.158 & \\
3C 279 & $12^h56^m11^s$ & $-05^{\circ}47^{\prime}22^{\prime\prime}$ & 0.538 & \\
PKS 1510$-$089 & $15^h12^m51^s$ & $-09^{\circ}05^{\prime}60^{\prime\prime}$ & 0.360 & \\
B2 1633$+$382 & $16^h35^m15^s$ & $+38^{\circ}08^{\prime}05^{\prime\prime}$ & 1.814 & \\
CTA 102 & $22^h32^m36^s$ & $+11^{\circ}43^{\prime}51^{\prime\prime}$ & 1.037 & \\
3C 454.3 & $22^h53^m58^s$ & $+16^{\circ}08^{\prime}54^{\prime\prime}$ & 0.859 & \\
\hline
\end{tabular}
\end{table}

\begin{subfigures}\label{fig:multi_lc}
\begin{figure*}
\centering
\includegraphics[width=18cm, height=10cm]{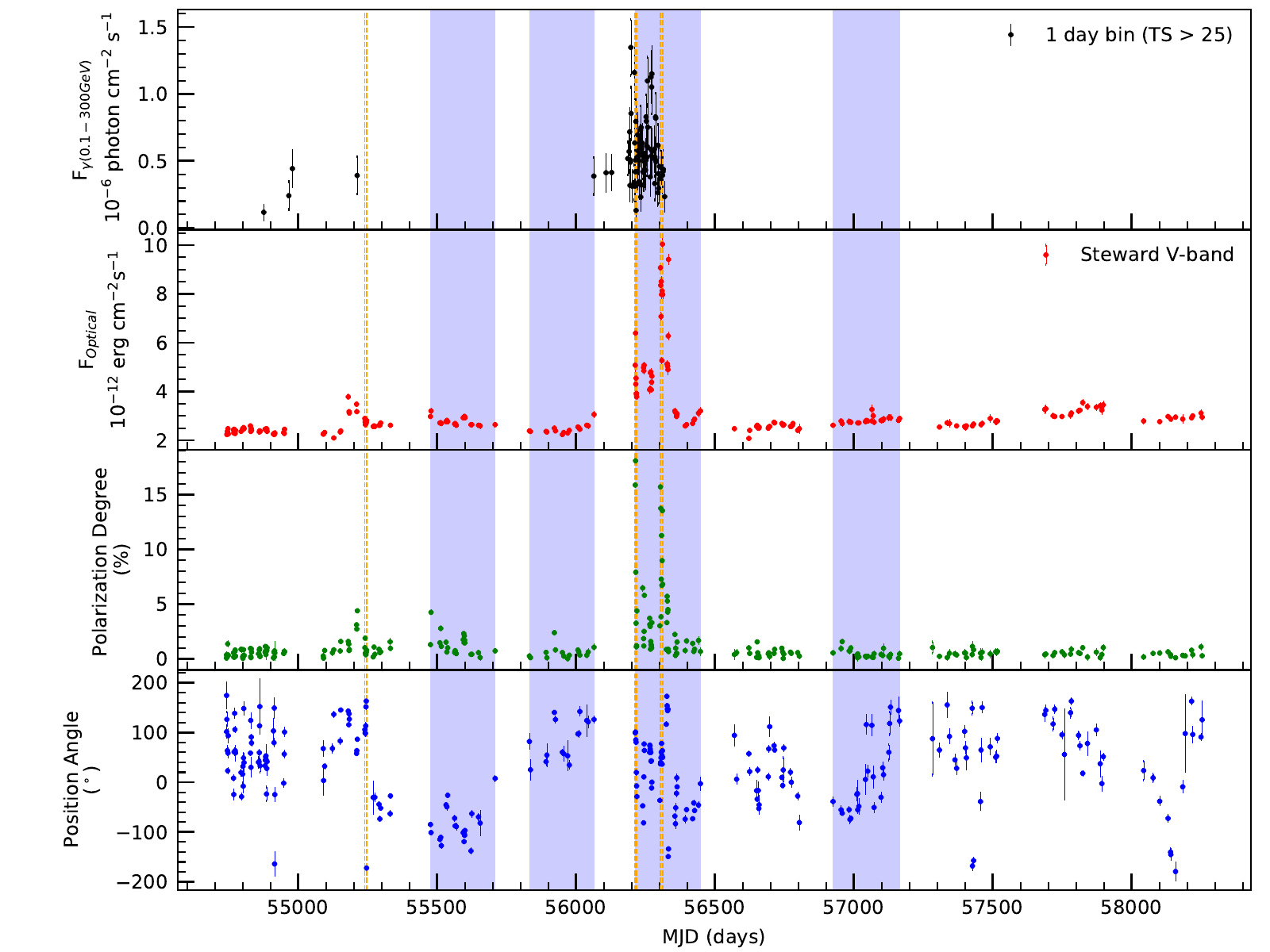}
\caption{\label{fig:multi_lc_oj248}Multi-wavelength light curves of OJ 248. From the top, the first panel shows the 1 day binned $\gamma$-ray light curve, the second panel shows the optical light curve in V-band, the third panel shows the variation of the degree of polarization, and the fourth panel shows the variation of PA (corrected for the $180^{\circ}$ ambiguity). The blue shaded regions represent the cycles (long-term time scales) where a correlation (either positive or negative) between
optical flux and PD was observed. The dashed orange vertical lines indicate the epochs (short-term time scales) where a correlation between optical flux and polarization was observed.}
\end{figure*} 

\begin{figure*}
\centering
\includegraphics[width=18cm, height=10cm]{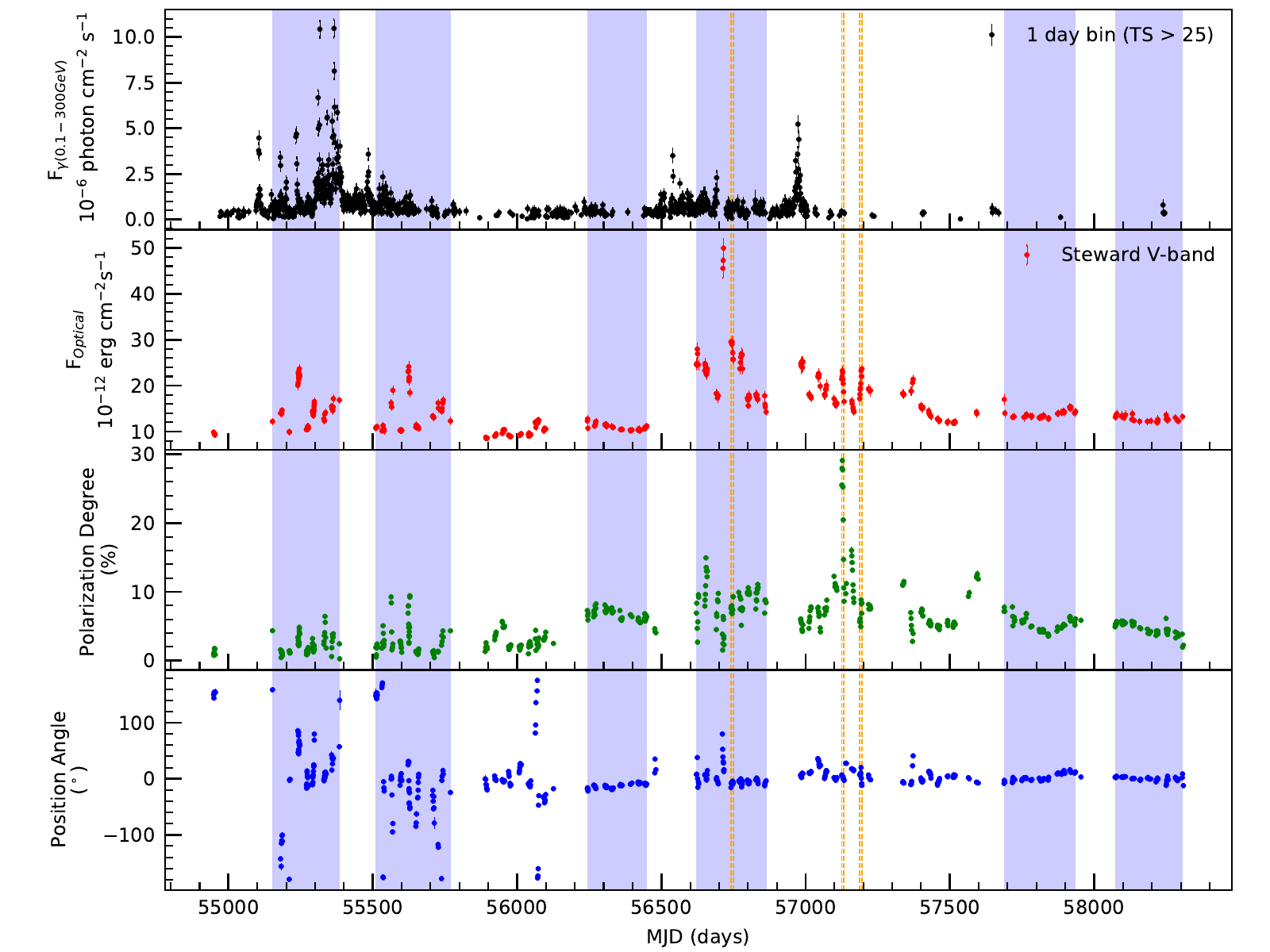}
\caption{\label{fig:multi_lc_1222}Multi-wavelength light curves of FSRQ PKS 1222$+$216. The details of the panels are the same as in Figure \ref{fig:multi_lc_oj248}.}
\end{figure*}

\begin{figure*}
\centering
\includegraphics[width=18cm, height=10cm]{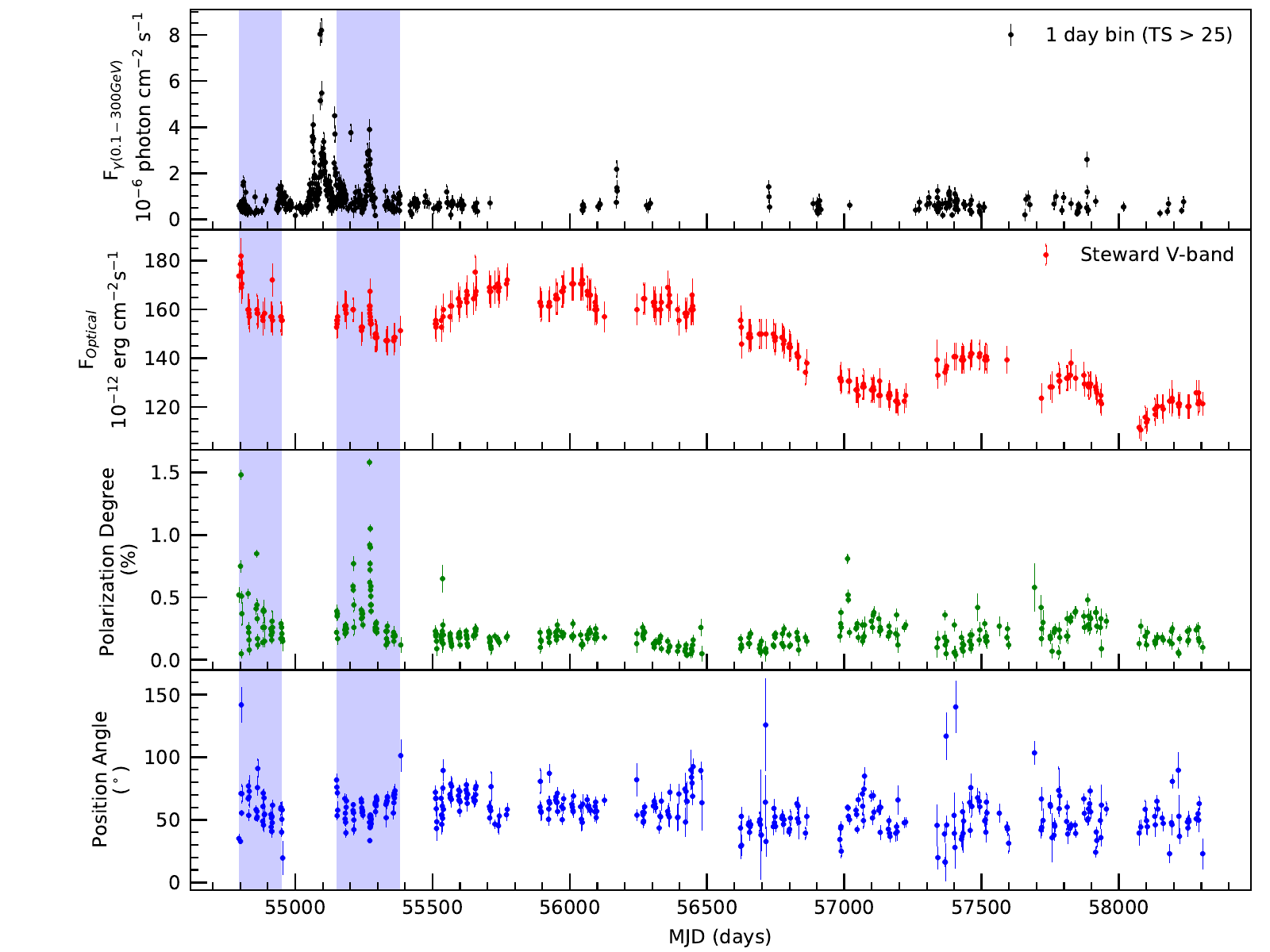}
\caption{\label{fig:multi_lc_3c273}Multi-wavelength light curves of the source 3C 273. The details of the panels are the same as in Figure \ref{fig:multi_lc_oj248}.}
\end{figure*}

\begin{figure*}
\centering
\includegraphics[width=18cm, height=10cm]{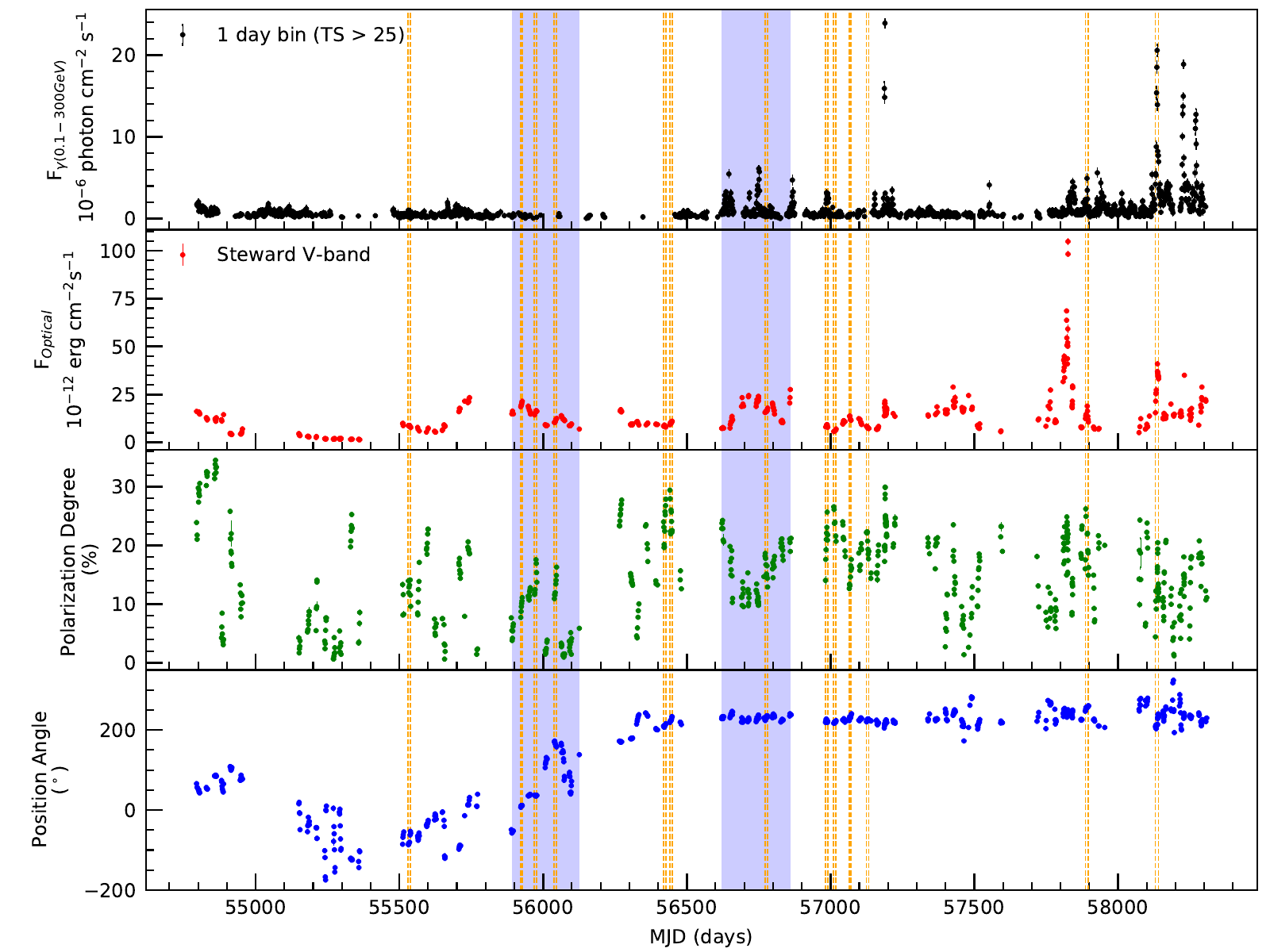}
\caption{\label{fig:multi_lc_3c379}Multi-wavelength light curves of FSRQ 3C 279. The details of the panels are the same as in Figure \ref{fig:multi_lc_oj248}.}
\end{figure*} 

\begin{figure*}
\centering
\includegraphics[width=18cm, height=10cm]{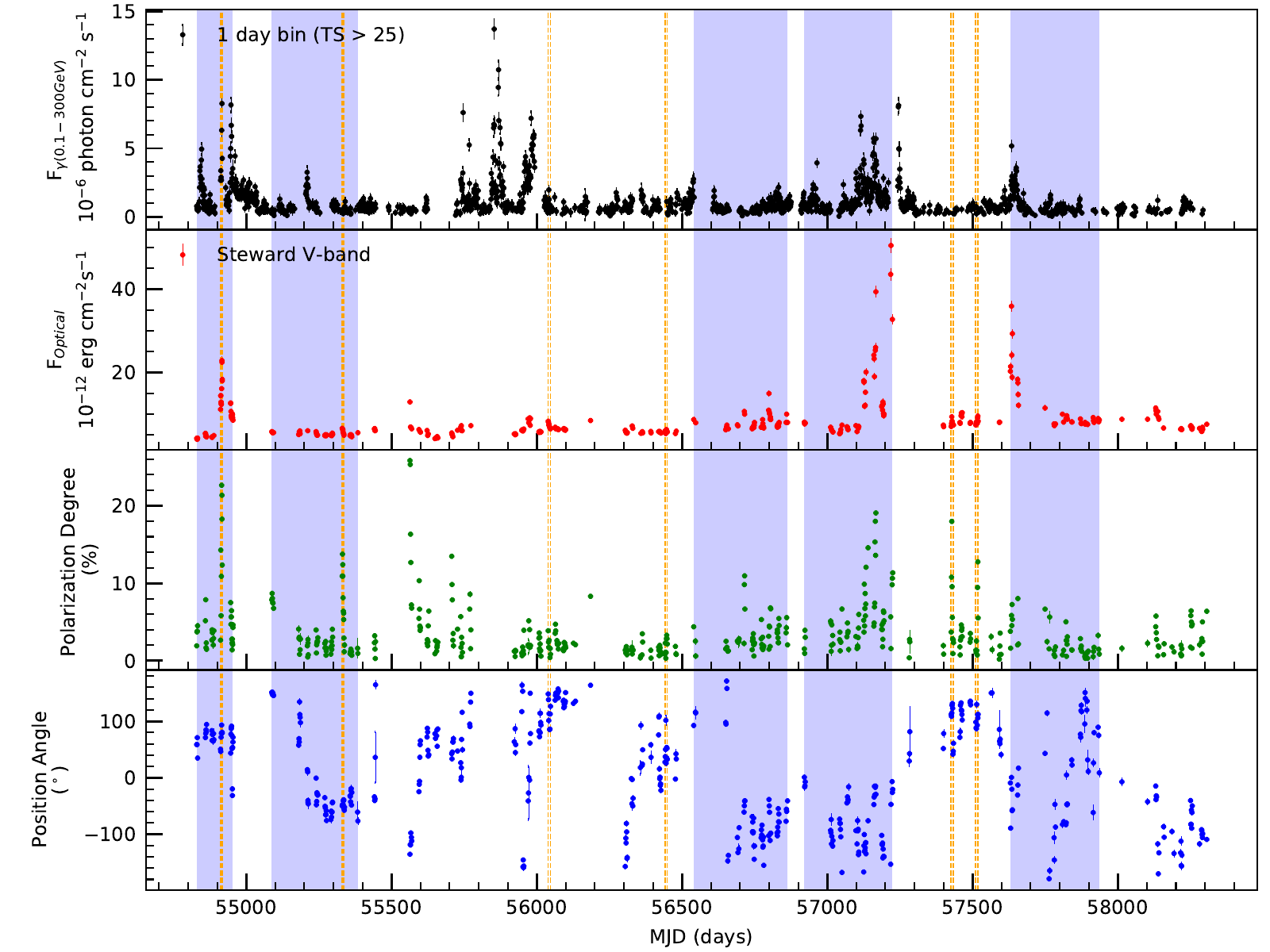}
\caption{\label{fig:multi_lc_1510}Multi-wavelength light curves of FSRQ PKS 1510$-$089. The details of the panels are the same as in Figure \ref{fig:multi_lc_oj248}.}
\end{figure*} 

\begin{figure*}
\centering
\includegraphics[width=18cm, height=10cm]{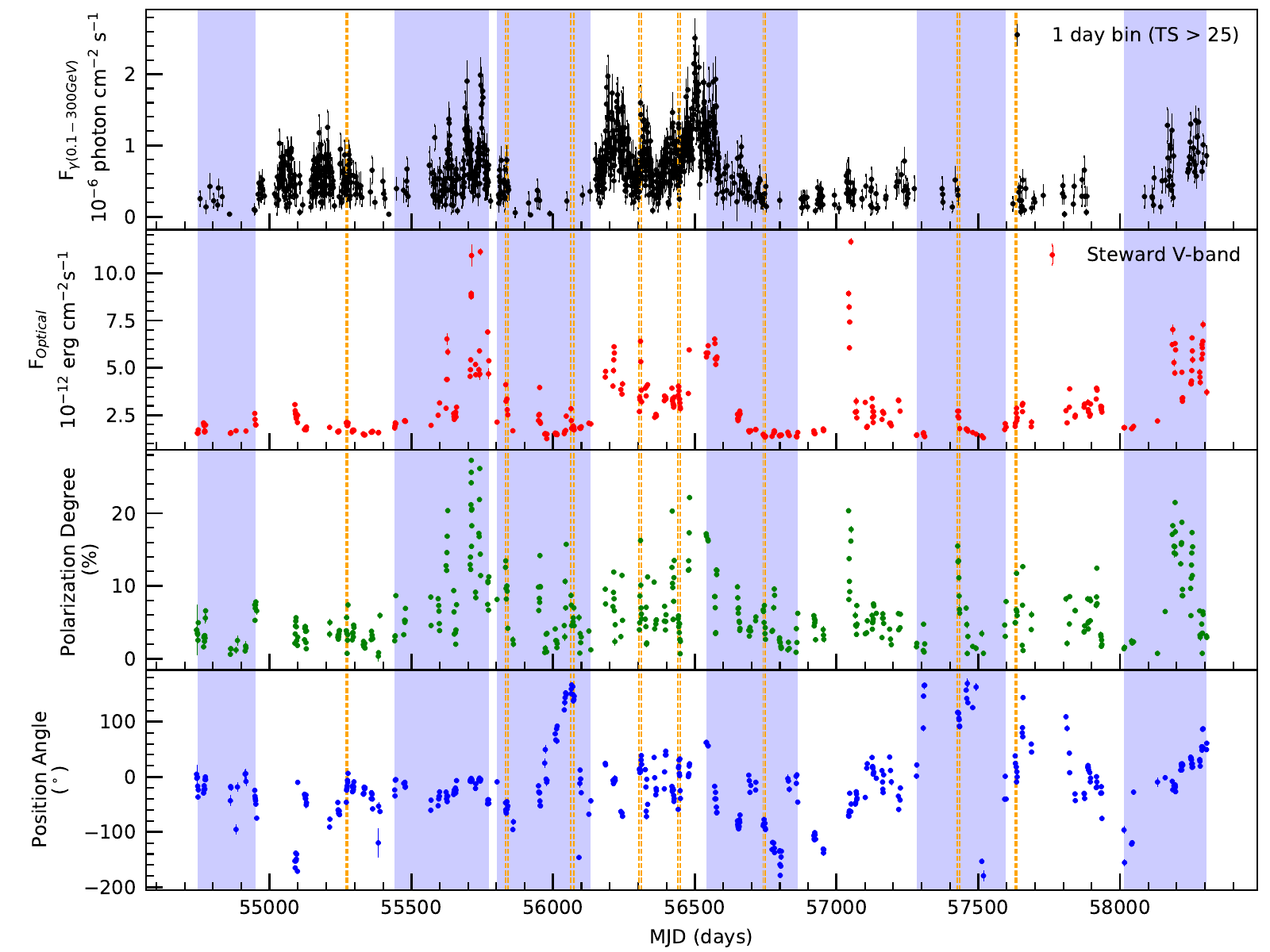}
\caption{\label{fig:multi_lc_b21633}Multi-wavelength light curves of FSRQ B2 1633$+$382. The details of the panels are the same as in Figure \ref{fig:multi_lc_oj248}.}
\end{figure*} 

\begin{figure*}
\centering
\includegraphics[width=18cm, height=10cm]{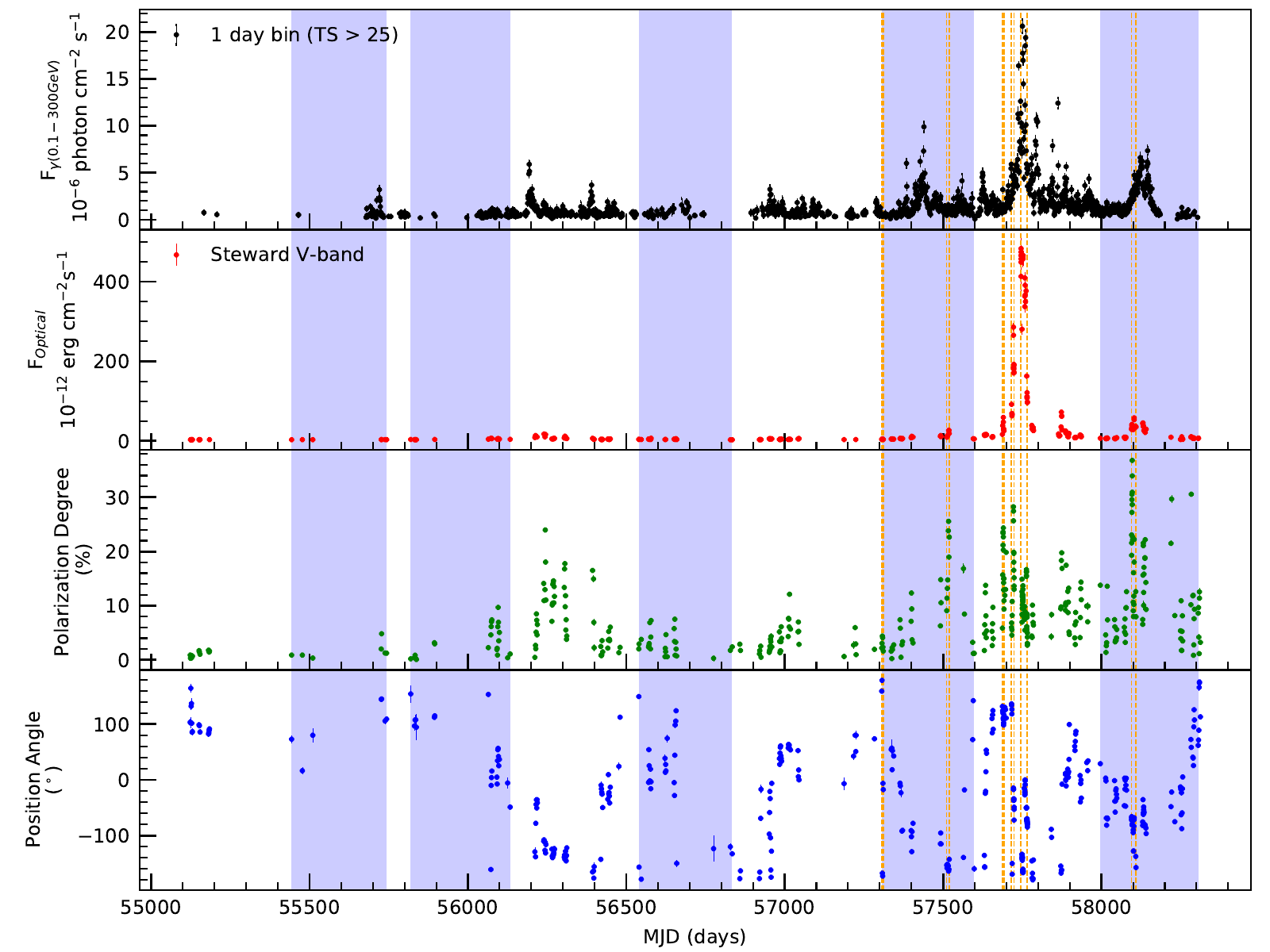}
\caption{\label{fig:multi_lc_cta102}Multi-wavelength light curves of the source CTA 102. The details of the panels are the same as in Figure \ref{fig:multi_lc_oj248}.}
\end{figure*} 

\begin{figure*}
\centering
\includegraphics[width=18cm, height=10cm]{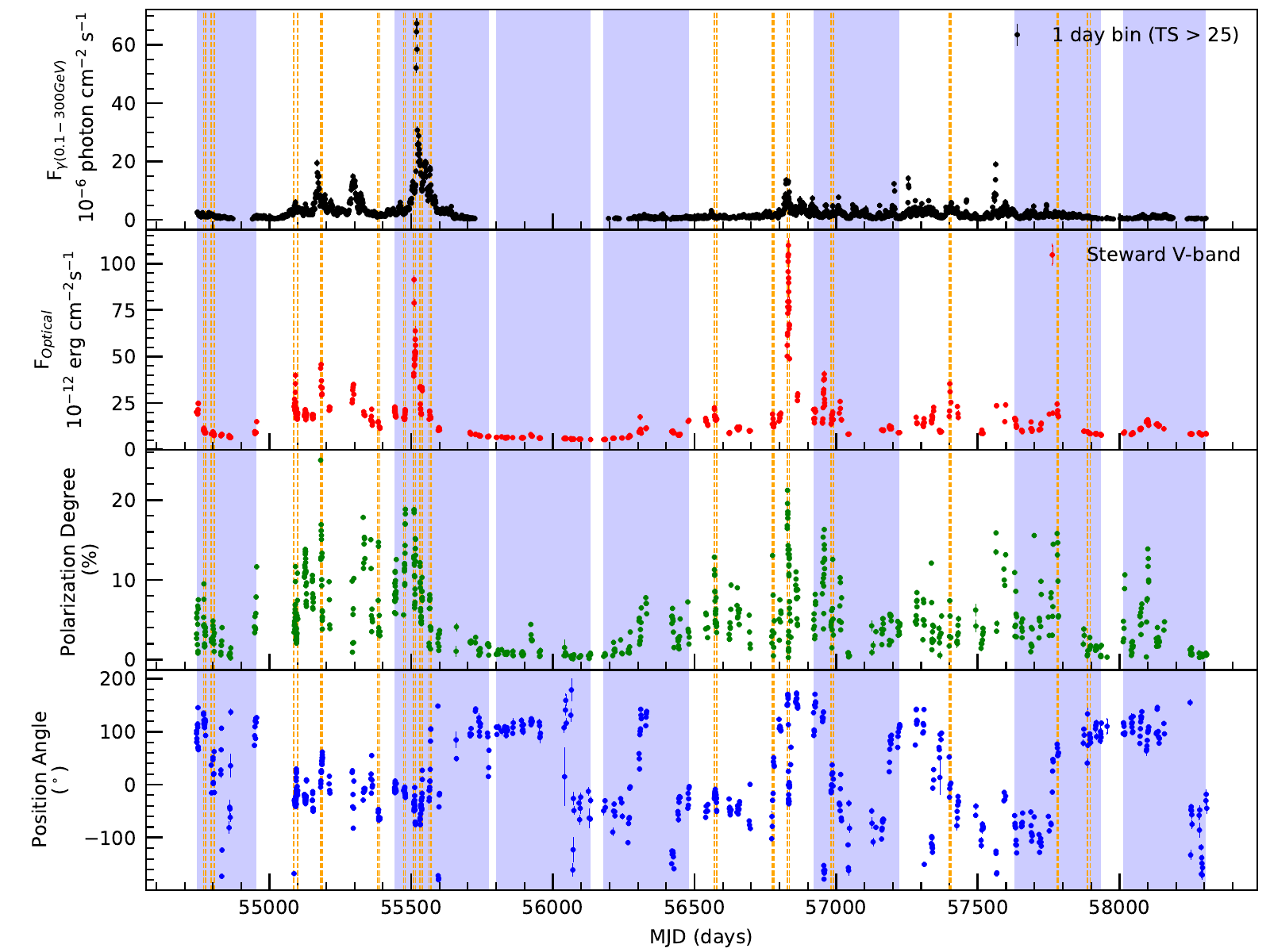}
\caption{\label{fig:multi_lc_3c454}Multi-wavelength light curves of FSRQ 3C 454.3. The details of the panels are the same as in Figure \ref{fig:multi_lc_oj248}.}
\end{figure*} 
\end{subfigures}

Investigations, available in the literature, on the relation between the optical flux and the polarization degree (PD) in blazars, point to varied correlations. For example, \cite{2002A&A...385...55H} found an average tendency of negative correlation between PD and flux on long-term timescales, while on shorter timescales they observed varied correlations for the source BL Lacertae using data from 1969 to 1991. For the same source using data collected from 2008 to 2010, \cite{2014ApJ...781L...4G} found that the V-band flux anti-correlates with PD on a timescale of $\sim$ 4 months in one observing season, while no trend between flux and PD was observed during the second observing season. \cite{2006ChJAS...6a.247J} found a positive correlation between PD and total flux in the sources 0420$-$014, 3C 279, 3C 345 and OJ 287 while a negative correlation was noticed in 3C 66A and BL Lac. On intranight timescales, in the source A0 0235+164, \cite{2008ApJ...672...40H} found a positive correlation between total flux and PD. For the TeV blazar 1ES 1959+650, \cite{2013ApJS..206...11S} detected a strong positive correlation between optical flux and PD using data accumulated over the period of 2009 August to November. \cite{2017ApJ...835..275R} monitored the blazar OJ 287 during its outburst in 2016 and found a significant anti-correlation between optical flux and PD on internight timescales. \cite{2019MNRAS.486.1781R} observed different behaviours between optical V-band flux and degree of polarization for the source 3C 454.3. 
Recently, on intra-night timescales too, \cite{2020MNRAS.492.1295P} found different behaviours between optical flux and PD that includes correlation as well as anti-correlation between the two quantities. 

Available polarimetric observations, therefore, indicate that the polarization behaviours shown by blazars are complex. Coordinated polarization observations along with flux monitoring at multiple wavelengths and on diverse timescales are needed to unravel the physical processes in the jets of blazars. The complexity of polarization results we know as of today is based on limited observations on a few blazars, and more sources need to be investigated to understand the relation between optical flux and polarization. Polarization observations do exist in the Steward Observatory archives for $\gamma$-ray detected blazars and our motivation here is to exploit this database to investigate the relation between optical flux and polarization for a large number of blazars. 

Towards this objective, we examined the data available in the Steward Observatory archive for the sources that have reasonable time resolution data for correlation analysis between flux and polarization. We arrived at a sample of nineteen sources having observations for at least nine years and are therefore suitable for the analysis of flux and polarization variations on diverse timescales. Our final sample includes eight FSRQs and eleven BL Lac objects. We present here the results of our analysis of the correlation between optical flux and polarization variations in eight FSRQs on short-term timescales as well as on long-term timescales using data for a period of $\sim$ 10 years from 2008 to 2018. The details of these eight FSRQs are given in Table \ref{tab:src_list}. The results of the correlation study of eleven BL Lac objects will be presented separately. 

Our paper is organized as follows. In Section \ref{sec:data} we discuss the multiwavelength data used in this work. Section \ref{sec:multi_lc} describes the multiwavelength light curves of the source. In Section \ref{sec:analysis}, we discuss the analysis steps followed in this work. The results are given in Section \ref{sec:res} followed by the discussion in Section \ref{sec:dis}. A summary of our work is given in Section \ref{sec:summary}.
\begin{table*}
\centering
\caption{\label{tab:LTV_stats}Statistics of the observed flux and polarization properties of all the sources on long-term timescales (or during each observing cycle). Here, $R$ is the Spearman rank correlation coefficient and $P$ is the null hypothesis probability.}
\resizebox{0.7\textwidth} {!}{  
\begin{tabular} {lccccccccccc} \hline
Blazar  &  Cycle & Time-period  &  \multicolumn{3}{c}{F$_{opt}$(10$^{-12}$ erg cm$^{-2}$ s$^{-1}$)}     & \multicolumn{3}{c}{PD (\%)} & \multicolumn{3}{c}{F$_{opt}$ v/s PD} \\
	&        & 	MJD	&       Min & Max & Average          & Min  & Max  & Average  &         R & P   & Status   \\ \hline
OJ 248  & C1 & 54743$-$54951 & 2.24 & 2.59 & 2.39 $\pm$ 0.01 & 0.06 & 1.38 & 0.53 $\pm$ 0.03  &	0.13 &  0.42 & - \\ 
	& C2 & 55090$-$55332 & 2.10 & 3.78 & 2.72 $\pm$ 0.01 & 0.07 & 4.40 & 1.10 $\pm$ 0.02  &	0.48 &  0.01 & - \\ 
	& C3 & 55476$-$55709 & 2.59 & 3.21 & 2.79 $\pm$ 0.01 & 0.14 & 4.26 & 1.33 $\pm$ 0.02  &	0.82 &  <0.01 & PC \\ 
	& C4 & 55832$-$56065 & 2.24 & 3.06 & 2.46 $\pm$ 0.01 & 0.03 & 2.40 & 0.59 $\pm$ 0.04  &	0.56 &  0.04 & PC \\ 
	& C5 & 56213$-$56448 & 2.59 & 10.04 & 5.03 $\pm$ 0.02 & 0.32 & 18.09 & 4.20 $\pm$ 0.02  &	0.66 &  <0.01 & PC \\ 
	& C6 & 56570$-$56804 & 2.08 & 2.74 & 2.55 $\pm$ 0.01 & 0.05 & 1.56 & 0.51 $\pm$ 0.03  &	0.09 &  0.67 & - \\ 
	& C7 & 56925$-$57165 & 2.62 & 3.27 & 2.83 $\pm$ 0.01 & 0.07 & 1.59 & 0.47 $\pm$ 0.03  &	-0.51 &  0.02 & NC \\ 
	& C8 & 57308$-$57517 & 2.52 & 2.90 & 2.67 $\pm$ 0.02 & 0.13 & 1.15 & 0.53 $\pm$ 0.06  &	0.13 &  0.66 & - \\ 
	& C9 & 57687$-$57898 & 2.98 & 3.55 & 3.23 $\pm$ 0.02 & 0.12 & 1.04 & 0.55 $\pm$ 0.04  &	0.02 &  0.94 & - \\ 
	& C10 & 58043$-$58254 & 2.77 & 3.12 & 2.92 $\pm$ 0.02 & 0.11 & 1.11 & 0.46 $\pm$ 0.04  &	0.39 &  0.26 & - \\ 

PKS 1222$+$216  & C1 & 54948$-$54953 & 9.25 & 9.87 & 9.55 $\pm$ 0.16 & 0.75 & 1.76 & 1.29 $\pm$ 0.03  &	-0.67 &  0.22 & - \\ 
		& C2 & 55152$-$55384 & 9.87 & 23.67 & 15.33 $\pm$ 0.07 & 0.25 & 6.43 & 2.29 $\pm$ 0.01  &	0.73 &  <0.01 & PC \\ 
		& C3 & 55510$-$55769 & 10.14 & 24.11 & 14.14 $\pm$ 0.08 & 0.43 & 9.42 & 3.10 $\pm$ 0.01  &	0.57 &  <0.01 & PC \\ 
		& C4 & 55891$-$56099 & 8.52 & 12.54 & 9.90 $\pm$ 0.05 & 0.99 & 5.71 & 2.82 $\pm$ 0.01  &	0.40 &  <0.01 & - \\ 
		& C5 & 56243$-$56449 & 10.14 & 12.77 & 11.00 $\pm$ 0.06 & 4.05 & 8.25 & 6.56 $\pm$ 0.01  &	0.73 &  <0.01 & PC \\ 
		& C6 & 56621$-$56863 & 14.26 & 49.92 & 23.47 $\pm$ 0.12 & 1.53 & 14.94 & 8.24 $\pm$ 0.01  &	-0.58 &  <0.01 & NC \\ 
		& C7 & 56983$-$57224 & 14.26 & 25.25 & 19.68 $\pm$ 0.10 & 4.18 & 29.10 & 9.56 $\pm$ 0.01  &	-0.42 &  <0.01 & - \\ 
		& C8 & 57336$-$57593 & 11.86 & 21.39 & 14.46 $\pm$ 0.09 & 2.79 & 12.61 & 6.72 $\pm$ 0.02  &	0.43 &  <0.01 & - \\ 
		& C9 & 57688$-$57935 & 12.77 & 16.99 & 13.94 $\pm$ 0.09 & 3.55 & 7.80 & 5.38 $\pm$ 0.02  &	0.52 &  <0.01 & PC \\ 
		& C10 & 58073$-$58306 & 11.97 & 13.88 & 12.82 $\pm$ 0.08 & 1.94 & 6.13 & 4.56 $\pm$ 0.02  &	0.54 &  <0.01 & PC \\ 
		
3C 273	& C1 & 54794$-$54952 & 155.50 & 181.90 & 162.15 $\pm$ 1.07 & 0.05 & 1.48 & 0.32 $\pm$ 0.01  &	0.59 &  <0.01 & PC \\ 
	& C2 & 55150$-$55383 & 147.20 & 167.40 & 153.54 $\pm$ 0.75 & 0.12 & 1.58 & 0.38 $\pm$ 0.00  &	0.66 &  <0.01 & PC \\ 
	& C3 & 55510$-$55772 & 152.70 & 175.30 & 163.64 $\pm$ 1.03 & 0.09 & 0.65 & 0.19 $\pm$ 0.00  &	-0.22 &  0.20 & - \\ 
	& C4 & 55891$-$56126 & 157.00 & 172.10 & 165.09 $\pm$ 1.04 & 0.10 & 1.31 & 0.22 $\pm$ 0.01  &	-0.18 &  0.28 & - \\ 
	& C5 & 56244$-$56449 & 155.50 & 169.00 & 161.29 $\pm$ 1.19 & 0.04 & 0.26 & 0.13 $\pm$ 0.01  &	0.25 &  0.22 & - \\ 
	& C6 & 56621$-$56863 & 134.20 & 155.50 & 147.02 $\pm$ 0.99 & 0.06 & 0.25 & 0.14 $\pm$ 0.01  &	-0.19 &  0.33 & - \\ 
	& C7 & 56983$-$57223 & 121.30 & 131.80 & 126.70 $\pm$ 0.87 & 0.12 & 0.81 & 0.28 $\pm$ 0.01  &	0.05 &  0.81 & - \\ 
	& C8 & 57337$-$57592 & 133.00 & 141.90 & 139.43 $\pm$ 1.10 & 0.04 & 0.42 & 0.17 $\pm$ 0.01  &	0.07 &  0.77 & - \\ 
	& C9 & 57717$-$57936 & 121.30 & 138.00 & 129.30 $\pm$ 0.95 & 0.06 & 0.58 & 0.28 $\pm$ 0.01  &	0.16 &  0.43 & - \\ 
	& C10 & 58073$-$58305 & 110.60 & 125.80 & 119.62 $\pm$ 0.91 & 0.05 & 0.27 & 0.17 $\pm$ 0.01  &	0.01 &  0.97 & - \\

3C 279  & C1 & 54794$-$54954 & 4.06 & 16.15 & 10.02 $\pm$ 0.02 & 3.07 & 34.50 & 20.71 $\pm$ 0.05  &	0.44 &  0.02 & - \\ 
	& C2 & 55150$-$55362 & 1.37 & 4.57 & 2.22 $\pm$ 0.01 & 0.63 & 25.25 & 7.43 $\pm$ 0.03  &	-0.19 &  0.26 & - \\ 
	& C3 & 55511$-$55744 & 5.30 & 23.35 & 11.62 $\pm$ 0.03 & 0.63 & 22.76 & 12.08 $\pm$ 0.01  &	0.45 &  0.01 & - \\ 
	& C4 & 55891$-$56126 & 6.92 & 21.49 & 13.06 $\pm$ 0.02 & 0.96 & 17.55 & 7.40 $\pm$ 0.01  &	0.52 &  <0.01 & PC \\ 
	& C5 & 56268$-$56449 & 7.95 & 17.07 & 10.53 $\pm$ 0.02 & 4.24 & 29.40 & 19.37 $\pm$ 0.01  &	0.18 &  0.29 & - \\ 
	& C6 & 56621$-$56860 & 7.25 & 27.56 & 17.12 $\pm$ 0.02 & 9.56 & 24.25 & 14.97 $\pm$ 0.01  &	-0.72 &  <0.01 & NC \\ 
	& C7 & 56983$-$57224 & 5.65 & 21.49 & 11.95 $\pm$ 0.02 & 12.64 & 29.92 & 20.72 $\pm$ 0.01  &	0.05 &  0.70 & - \\ 
	& C8 & 57339$-$57593 & 5.55 & 28.86 & 15.37 $\pm$ 0.04 & 1.38 & 23.50 & 13.26 $\pm$ 0.02  &	-0.27 &  0.09 & - \\ 
	& C9 & 57722$-$57934 & 6.86 & 104.80 & 26.11 $\pm$ 0.06 & 5.84 & 26.20 & 16.12 $\pm$ 0.01  &	0.26 &  0.07 & - \\ 
	& C10 & 58073$-$58306 & 5.06 & 40.95 & 18.89 $\pm$ 0.04 & 1.19 & 24.30 & 13.03 $\pm$ 0.03  &	-0.02 &  0.89 & - \\

PKS 1510$-$089  & C1 & 54829$-$54954 & 3.94 & 22.86 & 9.89 $\pm$ 0.04 & 1.40 & 22.63 & 5.89 $\pm$ 0.03  &	0.60 &  <0.01 & PC \\ 
		& C2 & 55088$-$55384 & 4.56 & 6.59 & 5.29 $\pm$ 0.02 & 0.48 & 13.75 & 3.51 $\pm$ 0.03  &	0.58 &  <0.01 & PC \\ 
		& C3 & 55441$-$55773 & 4.12 & 12.91 & 5.94 $\pm$ 0.03 & 0.29 & 25.82 & 5.30 $\pm$ 0.02  &	0.18 &  0.30 & - \\ 
		& C4 & 55922$-$56099 & 5.05 & 9.10 & 6.56 $\pm$ 0.03 & 0.41 & 5.15 & 2.04 $\pm$ 0.02  &	0.39 &  0.01 & - \\ 
		& C5 & 56185$-$56480 & 5.33 & 8.45 & 5.84 $\pm$ 0.03 & 0.32 & 8.31 & 1.59 $\pm$ 0.02  &	0.40 &  0.02 & - \\ 
		& C6 & 56540$-$56863 & 6.24 & 14.96 & 7.89 $\pm$ 0.03 & 0.61 & 10.95 & 3.19 $\pm$ 0.02  &	0.55 &  <0.01 & PC \\ 
		& C7 & 56921$-$57224 & 5.29 & 50.47 & 13.24 $\pm$ 0.07 & 0.95 & 19.06 & 5.53 $\pm$ 0.02  &	0.57 &  <0.01 & PC \\ 
		& C8 & 57399$-$57593 & 6.97 & 10.35 & 8.26 $\pm$ 0.05 & 0.20 & 17.96 & 3.60 $\pm$ 0.04  &	0.50 &  <0.01 & - \\ 
		& C9 & 57630$-$57936 & 7.23 & 35.90 & 11.68 $\pm$ 0.06 & 0.31 & 8.02 & 2.35 $\pm$ 0.03  &	0.75 &  <0.01 & PC \\ 
		& C10 & 58014$-$58306 & 5.85 & 11.46 & 7.68 $\pm$ 0.05 & 0.63 & 6.44 & 2.76 $\pm$ 0.03  &	0.33 &  0.10 & - \\

B2 1633$+$382   & C1 & 54745$-$54951 & 1.55 & 2.60 & 1.82 $\pm$ 0.00 & 0.63 & 7.82 & 3.70 $\pm$ 0.13  &	0.52 &  0.02 & PC \\ 
		& C2 & 55089$-$55384 & 1.47 & 3.06 & 1.88 $\pm$ 0.00 & 0.36 & 7.41 & 3.26 $\pm$ 0.02  &	0.49 &  <0.01 & - \\ 
		& C3 & 55441$-$55773 & 1.83 & 11.12 & 4.41 $\pm$ 0.02 & 2.01 & 27.26 & 11.24 $\pm$ 0.01  &	0.82 &  <0.01 & PC \\ 
		& C4 & 55802$-$56133 & 1.26 & 4.11 & 2.07 $\pm$ 0.00 & 0.79 & 15.73 & 5.93 $\pm$ 0.02  &	0.76 &  <0.01 & PC \\ 
		& C5 & 56184$-$56480 & 2.39 & 6.40 & 3.78 $\pm$ 0.01 & 0.70 & 22.15 & 7.60 $\pm$ 0.02  &	0.45 &  <0.01 & - \\ 
		& C6 & 56540$-$56863 & 1.35 & 6.52 & 2.65 $\pm$ 0.01 & 0.90 & 17.18 & 6.02 $\pm$ 0.02  &	0.68 &  <0.01 & PC \\ 
		& C7 & 56920$-$57224 & 1.54 & 11.65 & 3.16 $\pm$ 0.01 & 1.95 & 20.34 & 6.20 $\pm$ 0.03  &	0.45 &  <0.01 & - \\ 
		& C8 & 57282$-$57598 & 1.31 & 2.72 & 1.79 $\pm$ 0.01 & 0.74 & 15.49 & 5.26 $\pm$ 0.04  &	0.85 &  <0.01 & PC \\ 
		& C9 & 57630$-$57936 & 1.90 & 3.93 & 2.82 $\pm$ 0.01 & 1.15 & 12.67 & 5.82 $\pm$ 0.03  &	0.24 &  0.16 & - \\ 
		& C10 & 58014$-$58306 & 1.80 & 7.28 & 4.59 $\pm$ 0.01 & 0.75 & 21.47 & 9.37 $\pm$ 0.03  &	0.53 &  <0.01 & PC \\ 
		
CTA 102	& C2 & 55124$-$55184 & 3.13 & 3.53 & 3.26 $\pm$ 0.03 & 0.31 & 1.77 & 1.04 $\pm$ 0.03  &	0.45 &  0.19 & - \\ 
		& C3 & 55443$-$55743 & 3.19 & 3.53 & 3.39 $\pm$ 0.04 & 0.34 & 4.83 & 1.64 $\pm$ 0.05  &	0.91 &  <0.01 & PC \\ 
		& C4 & 55819$-$56133 & 3.31 & 6.80 & 4.84 $\pm$ 0.03 & 0.08 & 9.69 & 3.33 $\pm$ 0.02  &	0.87 &  <0.01 & PC \\ 
		& C5 & 56211$-$56449 & 3.60 & 17.55 & 7.31 $\pm$ 0.04 & 0.47 & 23.97 & 8.00 $\pm$ 0.02  &	0.50 &  <0.01 & - \\ 
		& C6 & 56539$-$56834 & 3.05 & 6.92 & 4.33 $\pm$ 0.03 & 0.30 & 7.51 & 2.94 $\pm$ 0.03  &	0.55 &  0.01 & PC \\ 
		& C7 & 56920$-$57224 & 3.31 & 6.31 & 4.49 $\pm$ 0.02 & 0.49 & 12.11 & 3.71 $\pm$ 0.02  &	0.25 &  0.20 & - \\ 
		& C8 & 57306$-$57598 & 3.98 & 26.81 & 8.96 $\pm$ 0.06 & 0.24 & 25.56 & 7.53 $\pm$ 0.03  &	0.79 &  <0.01 & PC \\ 
		& C9 & 57630$-$57936 & 8.17 & 483.30 & 126.98 $\pm$ 0.82 & 1.73 & 28.25 & 10.80 $\pm$ 0.01  &	0.27 &  <0.01 & - \\ 
		& C10 & 57996$-$58306 & 3.95 & 58.11 & 17.79 $\pm$ 0.09 & 0.85 & 36.80 & 12.26 $\pm$ 0.02  &	0.68 &  <0.01 & PC \\ 

3C 454.3	& C1 & 54743$-$54954 & 6.44 & 24.93 & 11.35 $\pm$ 0.05 & 0.23 & 11.64 & 3.56 $\pm$ 0.01  &	0.62 &  <0.01 & PC \\ 
		& C2 & 55086$-$55389 & 11.39 & 45.78 & 21.75 $\pm$ 0.06 & 0.95 & 24.97 & 7.44 $\pm$ 0.01  &	0.36 &  <0.01 & - \\ 
		& C3 & 55441$-$55774 & 6.80 & 91.34 & 25.57 $\pm$ 0.09 & 0.55 & 18.83 & 7.60 $\pm$ 0.01  &	0.54 &  <0.01 & PC \\ 
		& C4 & 55800$-$56133 & 5.40 & 8.10 & 6.26 $\pm$ 0.03 & 0.13 & 4.43 & 0.93 $\pm$ 0.02  &	0.75 &  <0.01 & PC \\ 
		& C5 & 56179$-$56480 & 5.40 & 17.56 & 8.86 $\pm$ 0.05 & 0.43 & 7.77 & 3.18 $\pm$ 0.01  &	0.76 &  <0.01 & PC \\ 
		& C6 & 56539$-$56864 & 8.64 & 109.80 & 36.29 $\pm$ 0.15 & 0.29 & 21.21 & 7.20 $\pm$ 0.01  &	0.19 &  0.01 & - \\ 
		& C7 & 56920$-$57224 & 8.25 & 40.61 & 17.97 $\pm$ 0.07 & 0.36 & 16.30 & 5.71 $\pm$ 0.01  &	0.74 &  <0.01 & PC \\ 
		& C8 & 57282$-$57598 & 8.64 & 35.37 & 16.51 $\pm$ 0.09 & 1.25 & 15.88 & 4.99 $\pm$ 0.02  &	-0.07 &  0.63 & - \\ 
		& C9 & 57630$-$57936 & 7.74 & 24.47 & 12.26 $\pm$ 0.05 & 0.33 & 15.80 & 4.53 $\pm$ 0.02  &	0.79 &  <0.01 & PC \\ 
		& C10 & 58014$-$58306 & 7.96 & 16.02 & 10.41 $\pm$ 0.05 & 0.30 & 13.87 & 3.44 $\pm$ 0.02  &	0.72 &  <0.01 & PC \\  
\hline
\end{tabular}}
\end{table*}
\section{Multiwavelength Observations and Data Reduction}\label{sec:data}
In this work, we used the publicly available optical photometric and polarimetric data as well as $\gamma$-ray data of all the eight FSRQs that span a period of $\sim$ 10 years from 2008 to 2018. 
\subsection{Optical Observations}
Steward Observatory, at the University of Arizona, monitored a list of $\gamma$-ray emitting blazars as part of the blazar monitoring program supporting Fermi observations. We downloaded the optical ($V$ and $R$ bands) photometric, and polarimetric data of eight FSRQs from the public archive of Steward Observatory\footnote{\url{http://james.as.arizona.edu/~psmith/Fermi}}. Details of the observations and data reductions can be found in \cite{2009arXiv0912.3621S} and \cite{2016Galax...4...27S}. We first dereddened the magnitudes of the sources by subtracting the values of Galactic extinction\footnote{taken from \url{https://ned.ipac.caltech.edu/}}, $A_{\lambda}$, from them. We then converted the dereddened magnitudes into fluxes using the coefficients of \cite{1998A&A...333..231B}. 

\subsection{$\gamma-$ray Observations}  
In addition to the optical data, we also used the $\gamma-$ray data of the sources in the 100 MeV to 300 GeV energy range observed by the Large Area Telescope (LAT) onboard {\it Fermi} \citep{2009ApJ...697.1071A}. We generated one day binned light curves of all the eight FSRQS following the procedures given in \cite{2019MNRAS.486.1781R}. In the one day binned light curve, the source is considered to be detected if the test statistics (TS) $>$ 25, which corresponds to detection at around 5 $\sigma$ level \citep{1996ApJ...461..396M}. 

\begin{table*}
\centering
\caption{\label{tab:STV_stats}Statistics of the observed flux and polarization properties on short timescales. Here '-' indicates that source was not detected with {\it Fermi} during the epoch.}
\resizebox{0.9\textwidth} {!}{  
\begin{tabular} {lcccccccccccccc} \hline
Blazar 	& Epoch & Time-period    & \multicolumn{3}{c}{F$_{\gamma}$ (10$^{-6}$ ph cm$^{-2}$ s$^{-1}$)}  &  \multicolumn{3}{c}{F$_{opt}$ (10$^{-12}$ erg cm$^{-2}$ s$^{-1}$)}     & \multicolumn{3}{c}{PD (\%)} & \multicolumn{3}{c}{PA (degrees)} \\
       	&       &  MJD              & Min & Max & Average               & Min  & Max  & Average              & Min  & Max  & Average  & Min & Max  & Average        \\ \hline
OJ 248	& A & 55240$-$55246 & -    & -    & -               & 2.64 & 2.90 & 2.75 $\pm$ 0.02 & 0.39 & 1.90 & 0.81 $\pm$ 0.03 & -172.20 & 163.00 & 79.73 $\pm$ 1.42 \\ 
	& B & 56213$-$56218 & 0.13 & 0.80 & 0.46 $\pm$ 0.06 & 3.78 & 6.40 & 4.56 $\pm$ 0.04 & 1.11 & 18.09 & 7.91 $\pm$ 0.05 & -7.40 & 100.90 & 63.02 $\pm$ 0.79 \\ 
	& C & 56304$-$56312 & 0.39 & 0.45 & 0.42 $\pm$ 0.09 & 5.27 & 10.00 & 8.04 $\pm$ 0.05 & 3.85 & 15.71 & 9.77 $\pm$ 0.02 & 36.40 & 77.90 & 52.04 $\pm$ 0.09 \\ 
		
PKS 1222$+$216 & A & 56741$-$56749 & 0.08 & 0.84 & 0.55 $\pm$ 0.06 & 25.70 & 29.50 & 28.20 $\pm$ 0.37 & 6.82 & 9.29  & 7.85 $\pm$ 0.01 & -15.60 & -4.20 & -8.19 $\pm$ 0.05 \\ 
& B & 57125$-$57133 & -    & -    & -               & 16.50 & 23.20 & 20.79 $\pm$ 0.27 & 8.63 & 29.10 & 21.10 $\pm$ 0.03 & -2.90 & 6.90 & 2.36 $\pm$ 0.05 \\ 
& C & 57187$-$57195 & -    & -    & -               & 17.20 & 23.70 & 20.58 $\pm$ 0.25 & 4.94 & 8.84  & 6.70 $\pm$ 0.03 & -11.80 & 20.10 & 4.59 $\pm$ 0.12 \\ 
                
3C 279 	& A & 55531$-$55539 & 0.06 & 1.06 & 0.61 $\pm$ 0.11 & 7.66 & 8.80 & 8.40 $\pm$ 0.05 & 9.61 & 14.11 & 12.46 $\pm$ 0.04 & -86.80 & -53.20 & -72.46 $\pm$ 0.10 \\ 
& B & 55922$-$55927 & 0.57 & 0.57 & 0.57 $\pm$ 0.19 & 18.50 & 21.50 & 19.90 $\pm$ 0.07 & 7.71 & 11.09 & 9.61 $\pm$ 0.02 & 6.90 & 11.40 & 10.13 $\pm$ 0.07 \\ 
& C & 55970$-$55978 & 0.18 & 0.18 & 0.18 $\pm$ 0.13 & 14.30 & 16.60 & 15.85 $\pm$ 0.06 & 11.63 & 17.55 & 14.22 $\pm$ 0.02 & 35.10 & 37.30 & 36.06 $\pm$ 0.05 \\ 
& D & 56039$-$56047 & - & - & - & 10.40 & 12.50 & 11.46 $\pm$ 0.04 & 10.91 & 16.29 & 13.07 $\pm$ 0.03 & 157.90 & 172.30 & 164.23 $\pm$ 0.08 \\ 
& E & 56419$-$56427 & - & - & - & 7.95 & 8.96 & 8.47 $\pm$ 0.03 & 19.61 & 27.86 & 23.45 $\pm$ 0.03 & 207.00 & 215.80 & 210.48 $\pm$ 0.04 \\ 
& F & 56441$-$56449 & - & - & - & 9.30 & 11.00 & 10.21 $\pm$ 0.04 & 21.95 & 29.40 & 24.69 $\pm$ 0.03 & 218.80 & 233.40 & 225.40 $\pm$ 0.03 \\ 
& G & 56772$-$56781 & 0.25 & 0.95 & 0.52 $\pm$ 0.08 & 15.60 & 17.90 & 16.39 $\pm$ 0.06 & 12.91 & 18.63 & 16.10 $\pm$ 0.02 & 223.60 & 235.40 & 230.29 $\pm$ 0.04 \\ 
& H & 56983$-$56990 & 1.13 & 3.19 & 2.51 $\pm$ 0.11 & 7.88 & 9.21 & 8.59 $\pm$ 0.05 & 17.60 & 25.66 & 21.42 $\pm$ 0.08 & 218.70 & 227.20 & 223.22 $\pm$ 0.12 \\ 
& I & 57011$-$57019 & 0.33 & 0.56 & 0.43 $\pm$ 0.09 & 5.65 & 6.99 & 6.20 $\pm$ 0.03 & 20.68 & 26.57 & 23.77 $\pm$ 0.04 & 216.60 & 223.20 & 219.11 $\pm$ 0.05 \\ 
& J & 57066$-$57071 & 0.35 & 0.57 & 0.46 $\pm$ 0.12 & 11.50 & 13.80 & 12.50 $\pm$ 0.09 & 13.14 & 16.54 & 14.69 $\pm$ 0.04 & 229.40 & 236.50 & 232.68 $\pm$ 0.07 \\ 
& K & 57125$-$57133 & - & - & - & 7.18 & 8.40 & 7.77 $\pm$ 0.03 & 17.70 & 22.29 & 20.45 $\pm$ 0.04 & 220.30 & 227.60 & 225.06 $\pm$ 0.06 \\ 
& L & 57887$-$57897 & 0.73 & 4.94 & 2.02 $\pm$ 0.10 & 10.60 & 18.90 & 14.00 $\pm$ 0.06 & 14.90 & 26.20 & 20.67 $\pm$ 0.04 & 248.00 & 260.00 & 256.14 $\pm$ 0.06 \\ 
& M & 58130$-$58141 & 4.22 & 20.60 & 10.25 $\pm$ 0.15 & 15.40 & 40.90 & 31.05 $\pm$ 0.12 & 4.41 & 20.46 & 13.15 $\pm$ 0.01 & 203.00 & 240.10 & 223.97 $\pm$ 0.04 \\ 

PKS 1510$-$089  & A & 54911$-$54917 & 2.67 & 8.26 & 4.35 $\pm$ 0.13 & 11.10 & 22.90 & 15.95 $\pm$ 0.10 & 2.65 & 22.63 & 13.53 $\pm$ 0.04 & 46.40 & 93.30 & 72.47 $\pm$ 0.11 \\ 
& B & 55330$-$55336 & 0.24 & 0.58 & 0.41 $\pm$ 0.08 & 4.91 & 6.59 & 5.75 $\pm$ 0.05 & 2.93 & 13.75 & 8.54 $\pm$ 0.03 & -54.30 & -39.10 & -46.17 $\pm$ 0.15 \\ 
& C & 56039$-$56047 & 0.34 & 1.97 & 0.83 $\pm$ 0.06 & 6.47 & 8.30 & 7.22 $\pm$ 0.07 & 0.41 & 3.87 & 1.89 $\pm$ 0.05 & 85.30 & 148.30 & 113.96 $\pm$ 1.16 \\ 
& D & 56441$-$56449 & 0.30 & 1.13 & 0.72 $\pm$ 0.17 & 5.33 & 6.30 & 5.77 $\pm$ 0.05 & 0.32 & 3.29 & 1.86 $\pm$ 0.04 & 25.70 & 101.70 & 46.28 $\pm$ 1.35 \\ 
& E & 57426$-$57434 & 0.25 & 0.27 & 0.26 $\pm$ 0.06 & 7.03 & 9.35 & 7.82 $\pm$ 0.08 & 0.92 & 17.96 & 6.26 $\pm$ 0.05 & 41.60 & 131.50 & 96.38 $\pm$ 0.58 \\ 
& F & 57511$-$57519 & 0.23 & 0.75 & 0.49 $\pm$ 0.16 & 7.36 & 9.53 & 8.21 $\pm$ 0.10 & 0.74 & 12.75 & 4.34 $\pm$ 0.06 & 86.80 & 129.90 & 103.30 $\pm$ 1.21 \\ 

B2 1633$+$382   & A & 55270$-$55276 & 0.36 & 0.53 & 0.42 $\pm$ 0.07 & 1.95 & 2.12 & 2.04 $\pm$ 0.01 & 0.75 & 7.41 & 3.86 $\pm$ 0.04 & -46.20 & 6.30 & -16.91 $\pm$ 0.53 \\ 
& B & 55833$-$55840 & 0.36 & 0.80 & 0.51 $\pm$ 0.09 & 2.52 & 4.11 & 3.20 $\pm$ 0.02 & 4.20 & 13.46 & 9.14 $\pm$ 0.04 & -66.10 & -45.10 & -57.58 $\pm$ 0.15 \\ 
& C & 56063$-$56074 & - & - & - & 1.73 & 2.85 & 1.99 $\pm$ 0.01 & 4.56 & 8.71 & 5.96 $\pm$ 0.06 & 137.80 & 166.80 & 151.81 $\pm$ 0.32 \\ 
& D & 56304$-$56312 & 0.35 & 1.60 & 0.87 $\pm$ 0.06 & 2.69 & 6.40 & 4.03 $\pm$ 0.03 & 4.00 & 16.25 & 7.83 $\pm$ 0.04 & 7.70 & 38.60 & 18.65 $\pm$ 0.21 \\ 
& E & 56441$-$56449 & 0.25 & 1.30 & 0.75 $\pm$ 0.06 & 2.85 & 4.04 & 3.45 $\pm$ 0.01 & 0.70 & 5.90 & 3.88 $\pm$ 0.04 & -59.10 & 32.50 & -1.81 $\pm$ 0.53 \\ 
& F & 56741$-$56749 & 0.17 & 0.32 & 0.25 $\pm$ 0.07 & 1.35 & 1.49 & 1.42 $\pm$ 0.01 & 2.68 & 7.33 & 5.15 $\pm$ 0.05 & -95.60 & -77.40 & -87.09 $\pm$ 0.32 \\ 
& G & 57427$-$57434 & 0.30 & 0.30 & 0.30 $\pm$ 0.12 & 1.80 & 2.72 & 2.41 $\pm$ 0.02 & 6.28 & 15.49 & 10.73 $\pm$ 0.11 & 91.10 & 117.10 & 106.07 $\pm$ 0.38 \\ 
& H & 57630$-$57637 & - & - & - & 1.91 & 2.87 & 2.35 $\pm$ 0.02 & 4.96 & 11.74 & 6.81 $\pm$ 0.06 & -9.20 & 38.00 & 13.38 $\pm$ 0.26 \\ 

CTA 102	& A & 57306$-$57311 & 0.38 & 0.46 & 0.42 $\pm$ 0.11 & 3.98 & 4.97 & 4.38 $\pm$ 0.05 & 1.63 & 4.40 & 2.96 $\pm$ 0.04 & -173.00 & 178.70 & -4.43 $\pm$ 0.45 \\ 
& B & 57511$-$57519 & 0.47 & 1.58 & 0.91 $\pm$ 0.07 & 9.73 & 26.80 & 16.96 $\pm$ 0.19 & 9.08 & 25.56 & 17.43 $\pm$ 0.05 & -164.10 & -142.90 & -154.95 $\pm$ 0.11 \\ 
& C & 57687$-$57692 & 1.05 & 3.21 & 1.58 $\pm$ 0.12 & 16.60 & 59.20 & 34.40 $\pm$ 0.37 & 5.87 & 24.35 & 16.75 $\pm$ 0.02 & 98.40 & 131.90 & 115.18 $\pm$ 0.06 \\ 
& D & 57716$-$57724 & 2.20 & 4.74 & 3.59 $\pm$ 0.12 & 63.10 & 286.00 & 162.85 $\pm$ 1.36 & 4.57 & 28.25 & 14.95 $\pm$ 0.01 & -169.60 & 136.50 & -10.59 $\pm$ 0.04 \\ 
& E & 57746$-$57766 & 3.57 & 20.60 & 10.57 $\pm$ 0.11 & 96.40 & 483.00 & 340.76 $\pm$ 2.97 & 2.72 & 16.67 & 9.64 $\pm$ 0.01 & -166.90 & 0.00 & -94.70 $\pm$ 0.06 \\ 
& F & 58095$-$58109 & 2.02 & 4.74 & 3.15 $\pm$ 0.10 & 28.90 & 58.10 & 37.01 $\pm$ 0.30 & 7.96 & 36.80 & 20.59 $\pm$ 0.02 & -157.60 & -66.20 & -87.26 $\pm$ 0.03 \\ 

3C 454.3	& A & 54767$-$54774 & 1.05 & 1.70 & 1.31 $\pm$ 0.08 & 8.89 & 11.50 & 9.84 $\pm$ 0.07 & 1.62 & 9.50 & 3.96 $\pm$ 0.01 & 92.80 & 135.30 & 118.00 $\pm$ 0.15 \\ 
& B & 54794$-$54805 & 1.18 & 1.91 & 1.50 $\pm$ 0.06 & 7.53 & 9.56 & 8.35 $\pm$ 0.13 & 1.04 & 4.86 & 2.97 $\pm$ 0.03 & -15.50 & 62.50 & 21.41 $\pm$ 0.37 \\ 
& C & 55086$-$55099 & 2.46 & 6.05 & 3.80 $\pm$ 0.10 & 16.30 & 39.90 & 21.75 $\pm$ 0.10 & 2.07 & 11.66 & 4.23 $\pm$ 0.01 & -167.90 & 29.70 & -20.76 $\pm$ 0.06 \\ 
& D & 55180$-$55185 & 5.71 & 7.79 & 6.73 $\pm$ 0.18 & 29.40 & 45.80 & 35.06 $\pm$ 0.25 & 9.27 & 24.97 & 14.50 $\pm$ 0.01 & -4.80 & 44.90 & 24.40 $\pm$ 0.03 \\ 
& E & 55383$-$55389 & 1.12 & 1.83 & 1.50 $\pm$ 0.10 & 11.40 & 15.00 & 13.06 $\pm$ 0.11 & 3.01 & 14.71 & 6.57 $\pm$ 0.03 & -66.90 & -47.10 & -58.81 $\pm$ 0.18 \\ 
& F & 55473$-$55479 & 1.82 & 2.59 & 2.27 $\pm$ 0.12 & 16.20 & 21.50 & 18.23 $\pm$ 0.13 & 5.62 & 18.83 & 13.04 $\pm$ 0.02 & -24.20 & -4.60 & -13.85 $\pm$ 0.05 \\ 
& G & 55509$-$55515 & 9.20 & 12.00 & 10.53 $\pm$ 0.20 & 39.50 & 91.30 & 54.19 $\pm$ 0.34 & 6.49 & 18.78 & 12.38 $\pm$ 0.01 & -75.50 & -5.30 & -47.37 $\pm$ 0.03 \\ 
& H & 55531$-$55539 & 9.87 & 19.20 & 14.02 $\pm$ 0.22 & 18.90 & 33.80 & 26.40 $\pm$ 0.23 & 4.51 & 12.15 & 7.53 $\pm$ 0.01 & -75.70 & 27.20 & -40.22 $\pm$ 0.07 \\ 
& I & 55563$-$55569 & 11.10 & 17.80 & 14.13 $\pm$ 0.31 & 16.30 & 20.70 & 18.30 $\pm$ 0.21 & 1.35 & 8.14 & 4.83 $\pm$ 0.02 & -30.40 & 105.40 & 23.08 $\pm$ 0.22 \\ 
& J & 56570$-$56578 & 0.53 & 1.10 & 0.83 $\pm$ 0.07 & 15.70 & 22.50 & 17.48 $\pm$ 0.11 & 3.41 & 12.84 & 7.20 $\pm$ 0.02 & -50.50 & -8.70 & -23.02 $\pm$ 0.10 \\ 
& K & 56774$-$56781 & 0.99 & 1.52 & 1.35 $\pm$ 0.08 & 12.10 & 19.10 & 14.54 $\pm$ 0.17 & 0.50 & 13.05 & 5.00 $\pm$ 0.04 & -78.70 & 50.80 & 5.84 $\pm$ 1.30 \\ 
& L & 56827$-$56835 & 4.93 & 13.20 & 8.47 $\pm$ 0.16 & 48.80 & 110.00 & 79.32 $\pm$ 0.44 & 0.29 & 21.21 & 10.69 $\pm$ 0.01 & -36.20 & 170.50 & 53.48 $\pm$ 0.13 \\ 
& M & 56982$-$56990 & 1.38 & 4.71 & 2.80 $\pm$ 0.16 & 13.70 & 20.20 & 16.92 $\pm$ 0.15 & 1.50 & 12.56 & 5.15 $\pm$ 0.02 & -15.20 & 37.20 & 6.21 $\pm$ 0.17 \\ 
& N & 57399$-$57405 & 2.95 & 4.66 & 3.91 $\pm$ 0.15 & 17.90 & 35.40 & 26.10 $\pm$ 0.33 & 1.72 & 7.40 & 4.07 $\pm$ 0.04 & -23.60 & 52.20 & 4.70 $\pm$ 0.41 \\ 
& O & 57780$-$57785 & 1.26 & 2.42 & 1.69 $\pm$ 0.12 & 17.70 & 24.50 & 19.75 $\pm$ 0.25 & 5.36 & 15.80 & 10.71 $\pm$ 0.06 & 55.00 & 76.50 & 62.68 $\pm$ 0.22 \\ 
& P & 57887$-$57897 & 0.65 & 1.25 & 0.90 $\pm$ 0.07 & 8.49 & 9.65 & 8.92 $\pm$ 0.11 & 0.66 & 2.79 & 1.52 $\pm$ 0.07 & 74.20 & 133.70 & 95.33 $\pm$ 2.00 \\ 
\hline
\end{tabular}}
\end{table*}

\begin{subfigures}\label{fig:pol_flx_cor}
\begin{figure*}
\centering
\includegraphics[width=15cm, height=13cm]{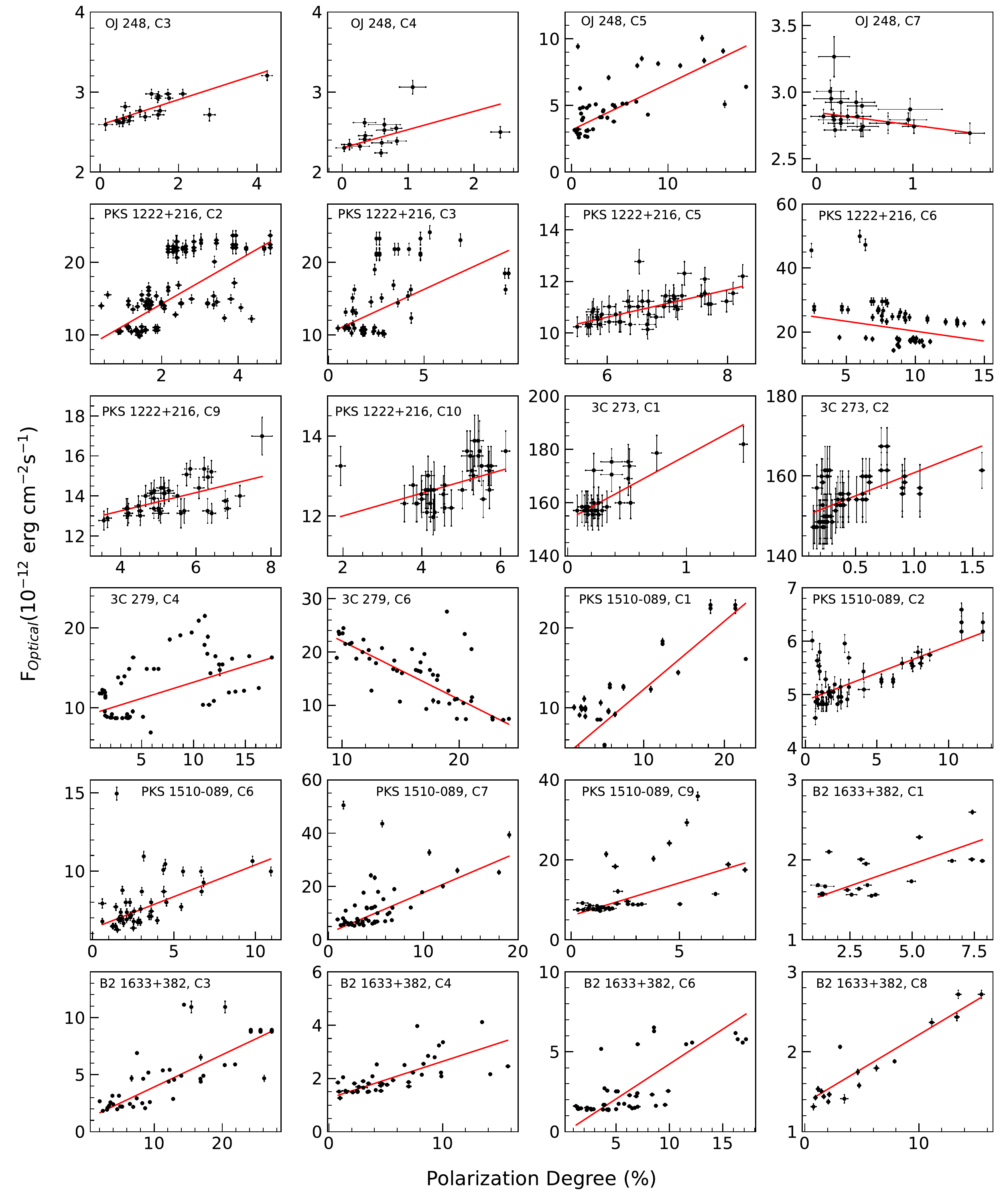}
\caption{\label{fig:pol_flx_cor1}Degree of polarization vs optical V-band flux plots on long-term timescales for OJ 248, PKS 1222$+$216, 3C 273, 3C 279, PKS 1510-089, and B2 1633$+$382. The name of the source and the observing cycle are given in each panel. The solid red line represents the straight line fit to the data.}
\end{figure*}

\begin{figure*}
\centering
\includegraphics[width=12cm, height=10cm]{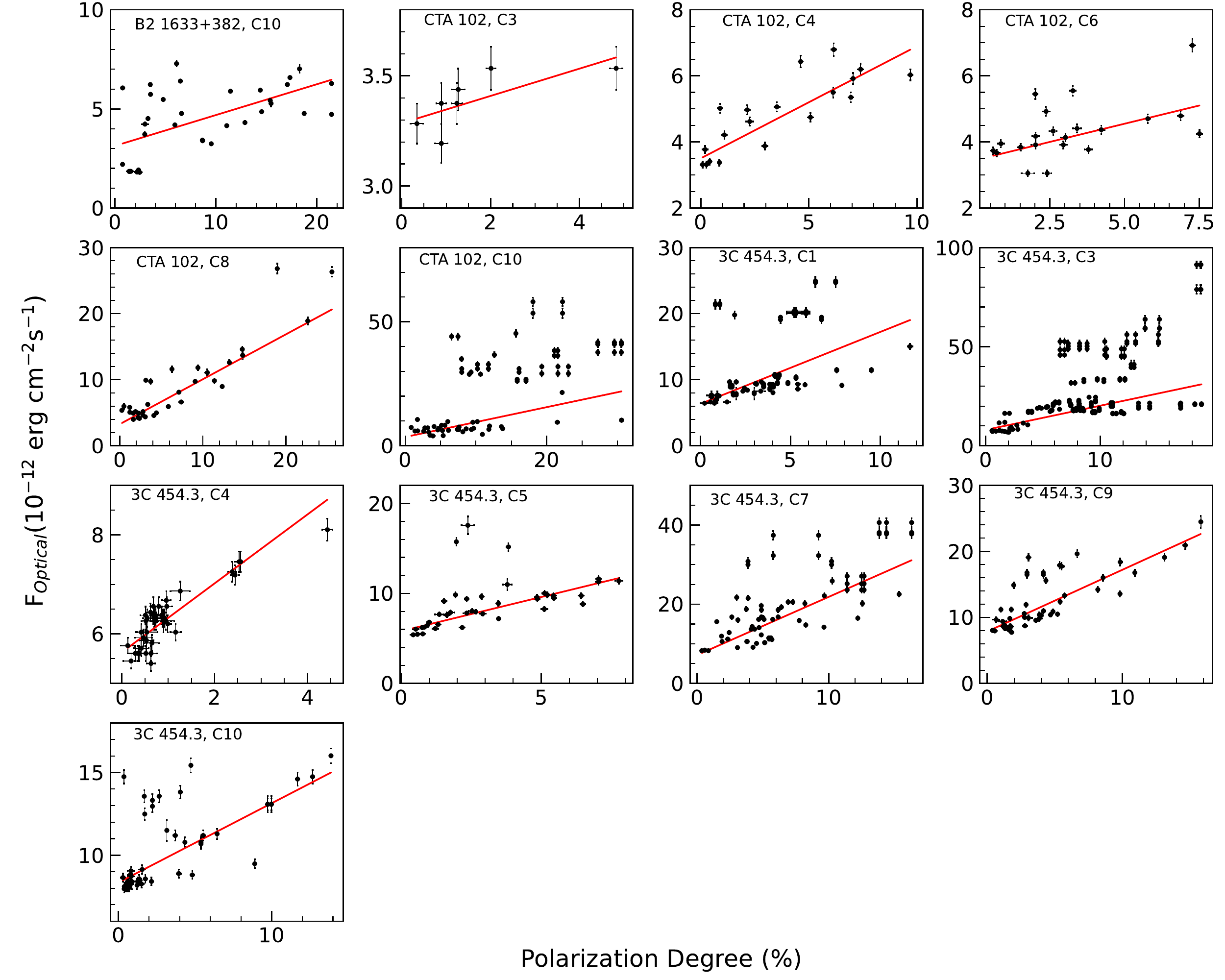}
\caption{\label{fig:pol_flx_cor2}Same as in Figure \ref{fig:pol_flx_cor1} but for B2 1633$+$382, CTA 102 and 3C 453.3.}
\end{figure*}
\end{subfigures}

\section{Multiwavelength Lightcurves}\label{sec:multi_lc}
The $\gamma$-ray and the Steward Observatory’s optical $V-$band light curves as well as optical PD and electric vector position angle (PA) for the time interval 2008$-$2018 (MJD 54743$-$58306) are plotted from top to bottom panels, respectively, for each of the FSRQ in Figures \ref{fig:multi_lc_oj248} to \ref{fig:multi_lc_3c454}, respectively. We adopted the standard procedure to overcome the $180^{\circ}$ ambiguity in the PA measurements \citep{2010Natur.463..919A,2011PASJ...63..639I,2015MNRAS.453.1669B}, wherein changes in the PA between consecutive measurements should be minimum i.e. $-90^{\circ}$ $\leq$ $\Delta\theta$ $\leq$ $90^{\circ}$. Here, $\Delta\theta$ is defined as $|\theta_{n}-\theta_{n-1}|$-$\sqrt{\sigma(\theta_{n})^2+\sigma(\theta_{n-1})^2}$, where $\theta_{n-1}$ and $\theta_{n}$ are the $(n-1)^{th}$ and n$^{th}$ measurement of PA and $\sigma(\theta_{n-1})$, $\sigma(\theta_{n})$ are the corresponding errors of the PA measurements. For $\Delta\theta$ $>$ $90^{\circ}$, $\theta_{n}$ is shifted to $\theta_{n}$-$180^{\circ}$ and for $\Delta\theta$ $<$ $-90^{\circ}$, $\theta_{n}$ is shifted to $\theta_{n}$+$180^{\circ}$. 

\begin{table}
\centering
\caption{\label{tab:epoch_cor}Results of correlation analysis for all the epochs of eight FSRQs. Here, PC represents a positive correlation between optical flux and polarization, while NC indicates that optical flux and polarization are anti-correlated. Here `$-$' indicates that no correlation coefficient could be estimated as the number of quasi-simultaneous data points in the $\gamma-$ray and optical bands are less than 5. }
\resizebox{0.5\textwidth} {!}{ 
\begin{tabular} {lccccccccccccc} \hline
Blazar		&  Epoch &   \multicolumn{2}{c}{F$_\gamma$ v/s F$_{opt}$} & \multicolumn{3}{c}{F$_{opt}$ v/s PD} \\
		&        &     R     & P                                 & R        & P     & Status  \\ \hline  
OJ 248	& A & - & - & 0.77 & 0.04 & PC \\ 
	& B & - & - & 0.85 & 0.02 & PC \\ 
	& C & - & - & 0.78 & 0.01 & PC \\ 
		
PKS 1222$+$216	& A & -0.56 & 0.32 & -0.73 & 0.04 & NC \\ 
& B & - & - & 0.97 & $<$0.01 & PC \\ 
& C & - & - & 0.73 & 0.03 & PC \\ 
		      
3C 279	& A & - & - & 0.97 & $<$ 0.01 & PC \\ 
& B & - & - & 1.00 & $<$ 0.01 & PC \\ 
& C & - & - & 0.90 & 0.01 & PC \\ 
& D & - & - & 0.99 & $<$ 0.01 & PC \\ 
& E & - & - & -0.98 & $<$ 0.01 & NC \\ 
& F & - & - & -0.82 & 0.01 & NC \\ 
& G & - & - & -0.91 & $<$ 0.01 & NC \\ 
& H & -0.03 & 0.96 & 0.81 & 0.05 & PC \\ 
& I & - & - & -1.00 & $<$ 0.01 & NC \\ 
& J & - & - & -0.90 & 0.04 & NC \\ 
& K & - & - & 0.73 & 0.03 & PC \\ 
& L & 0.83 & 0.01 & 0.74 & 0.04 & PC \\ 
& M & 0.74 & $<$ 0.01 & 0.67 & $<$ 0.01 & PC \\ 
		
PKS 1510$-$089  & A & 0.97 & $<$ 0.01 & 0.78 & $<$ 0.01 & PC \\ 
& B & - & - & 0.86 & $<$ 0.01 & PC \\ 
& C & 0.21 & 0.61 & 0.90 & $<$ 0.01 & PC \\ 
& D & - & - & 0.81 & 0.01 & PC \\ 
& E & - & - & 0.76 & 0.03 & PC \\ 
& F & - & - & 0.81 & 0.01 & PC \\ 
 
B2 1633$+$382   & A & 0.39 & 0.51 & 0.93 & $<$ 0.01  & PC\\ 
& B & - & - & 1.00 & $<$ 0.01  & PC\\ 
& C & - & - & 0.88 & $<$ 0.01  & PC\\ 
& D & 0.71 & 0.07 & 0.82 & 0.02 & PC \\ 
& E & 0.76 & 0.02 & 0.85 & $<$ 0.01  & PC\\ 
& F & - & - & 0.94 & $<$ 0.01  & PC\\ 
& G & - & - & 0.97 & $<$ 0.01  & PC\\ 
& H & - & - & 0.94 & $<$ 0.01  & PC\\ 

CTA 102 & A & - & - & 0.94 & $<$ 0.01  & PC\\ 
& B & 0.71 & 0.07 & 0.89 & 0.01 & PC \\ 
& C & 0.50 & 0.20 & 0.81 & $<$ 0.01  & PC\\ 
& D & 0.57 & 0.04 & 0.72 & $<$ 0.01  & PC\\ 
& E & 0.58 & $<$ 0.01 & 0.75 & $<$ 0.01  & PC\\ 
& F & 0.56 & 0.04 & 0.53 & $<$ 0.01  & PC\\ 

3C 454.3 & A & 0.91 & $<$ 0.01 & 0.60 & $<$ 0.01  & PC\\ 
& B & 0.49 & 0.12 & 0.82 & $<$ 0.01  & PC\\ 
& C & 0.74 & $<$ 0.01 & 0.66 & $<$ 0.01  & PC\\ 
& D & 0.71 & 0.03 & 0.83 & $<$ 0.01  & PC\\ 
& E & 0.63 & 0.25 & 0.84 & 0.02 & PC \\ 
& F & -0.29 & 0.36 & 0.68 & $<$ 0.01  & PC\\ 
& G & 0.61 & 0.01 & 0.62 & $<$ 0.01  & PC\\ 
& H & 0.09 & 0.81 & 0.92 & $<$ 0.01  & PC\\ 
& I & 0.79 & 0.06 & 0.90 & $<$ 0.01  & PC\\ 
& J & -0.04 & 0.88 & 0.81 & $<$ 0.01  & PC\\ 
& K & -0.41 & 0.42 & 0.86 & 0.01 & PC \\ 
& L & 0.69 & $<$ 0.01 & -0.52 & $<$ 0.01  & NC\\ 
& M & 0.77 & 0.01 & 0.63 & 0.02 & PC \\ 
& N & 0.90 & 0.04 & 0.90 & 0.04 & PC \\ 
& O & -0.83 & 0.04 & 0.94 & $<$ 0.01  & PC\\ 
& P & 0.72 & 0.10 & -0.81 & 0.05 & NC \\ 
\hline 
\end{tabular}} 
\end{table}

\begin{subfigures}
\begin{figure*}
\includegraphics[width=16cm, height=8cm]{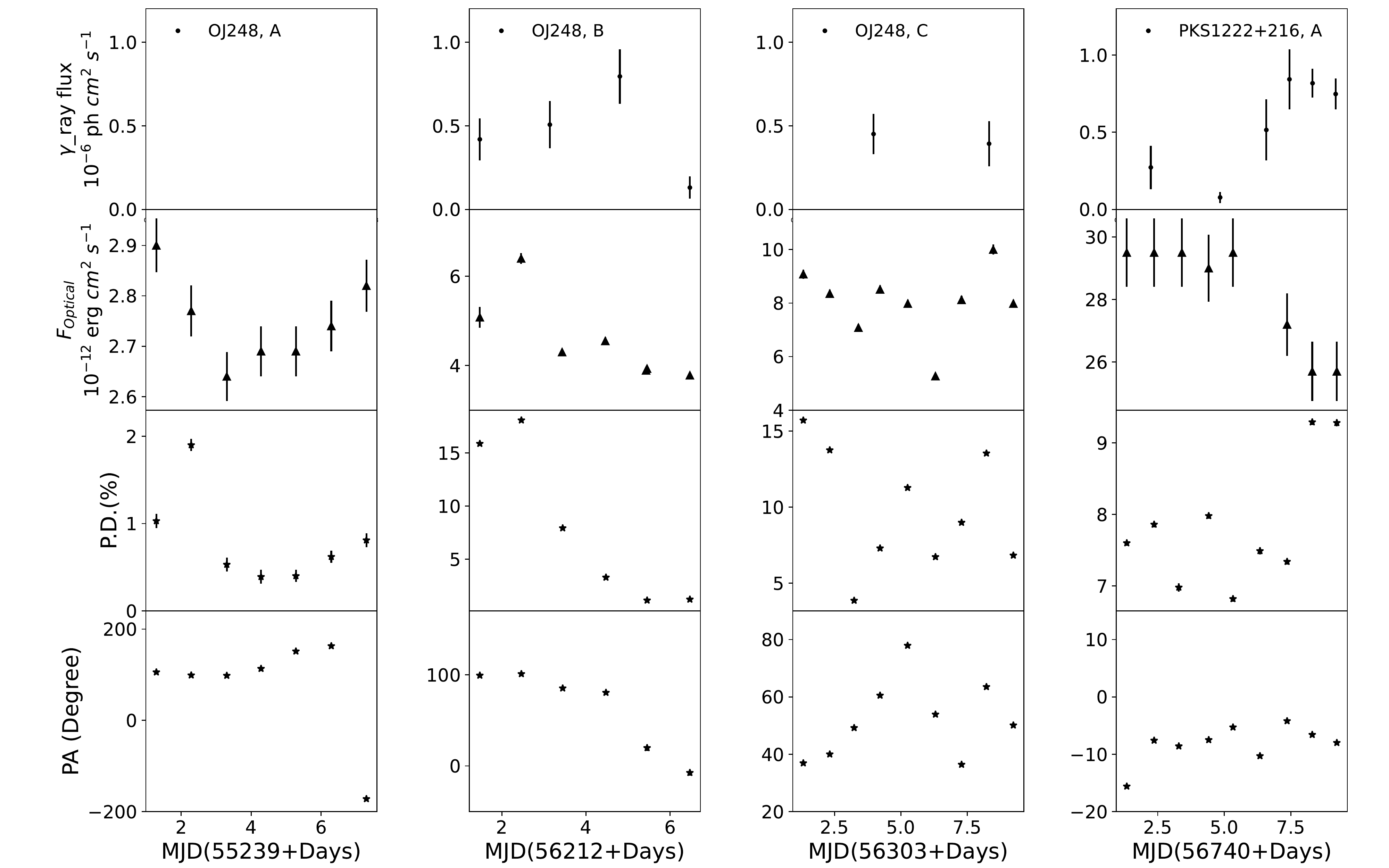}
\caption{\label{fig:epoch_lc1}Multi-wavelength light curves along with polarization measurements. The name of the source and its epoch are given in each panel. In each lightcurve, from the top, the first panel shows the one day binned $\gamma$-ray light curve, the second panel shows the light curve in the optical V-band, the third and fourth panels show the variation in the degree of optical polarization and position angle respectively.}
\end{figure*}

\begin{figure*}
\includegraphics[width=16cm, height=8cm]{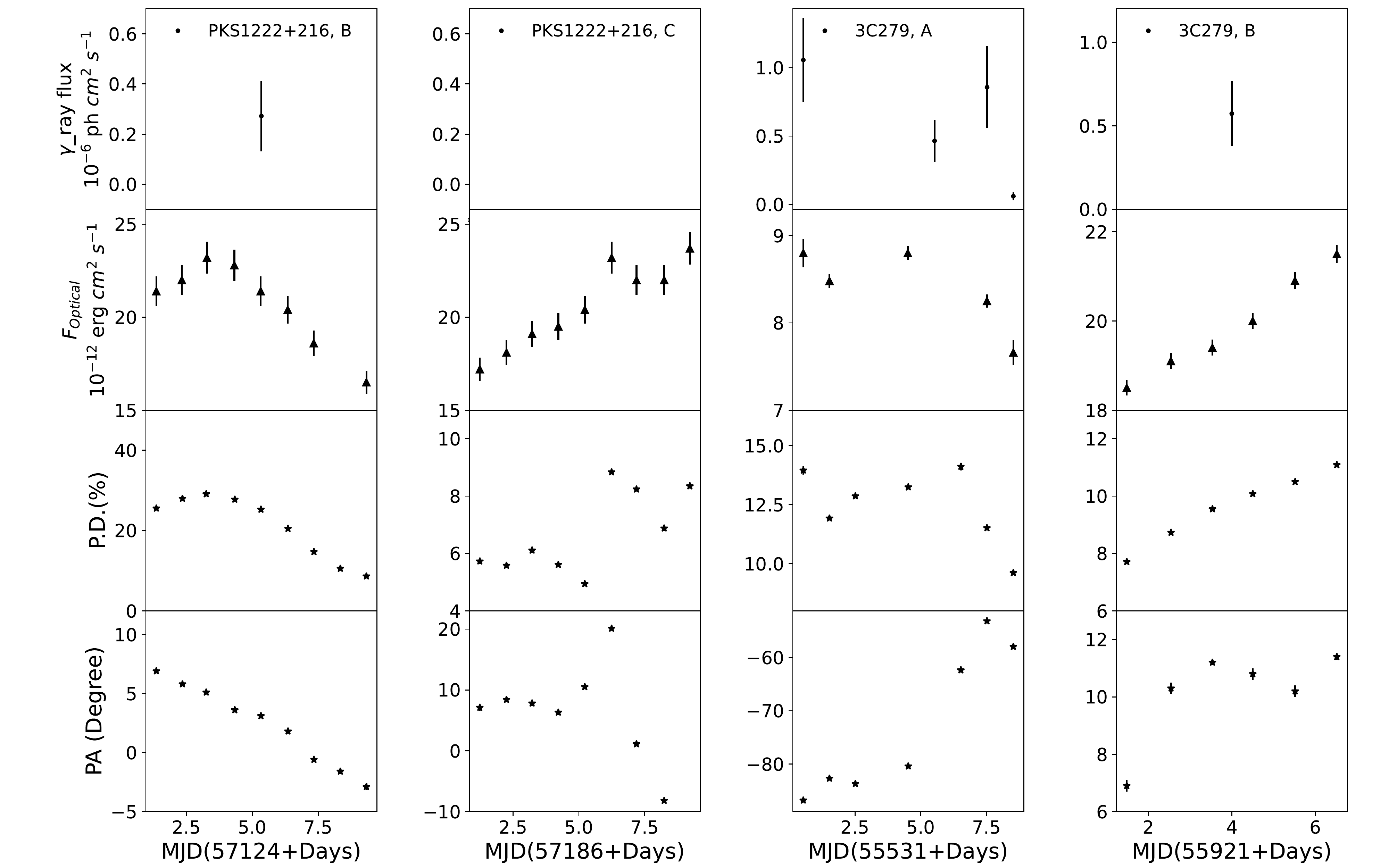}
\caption{\label{fig:epoch_lc2}Multi-wavelength light curves along with polarization measurements. The panels have the same meaning as in Fig. \ref{fig:epoch_lc1}.}
\end{figure*}

\begin{figure*}
\includegraphics[width=16cm, height=8cm]{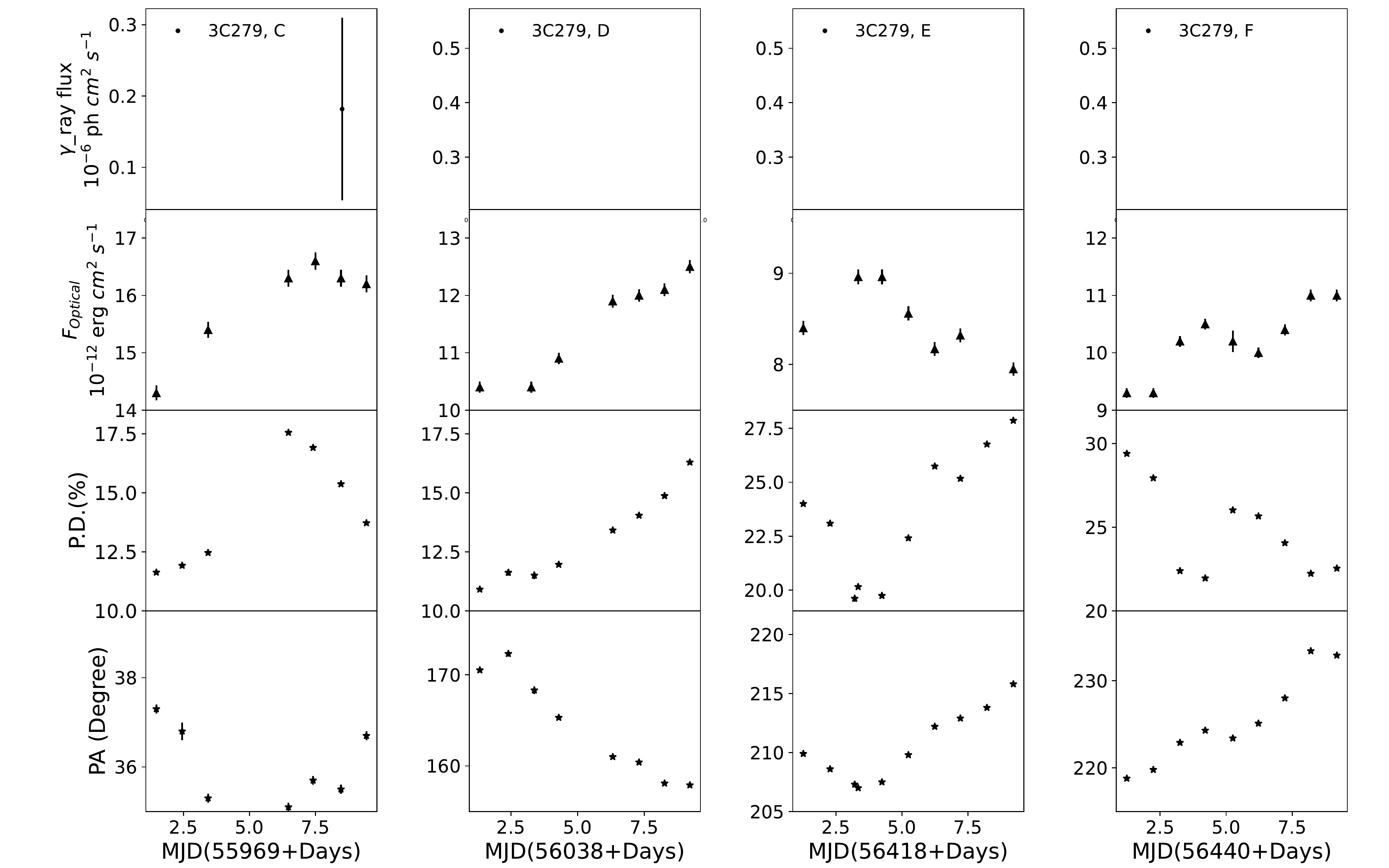}
\caption{\label{fig:epoch_lc3}Multi-wavelength light curves along with polarization measurements. The panels have the same meaning as in Fig. \ref{fig:epoch_lc1}.}
\end{figure*}

\begin{figure*}
\includegraphics[width=16cm, height=8cm]{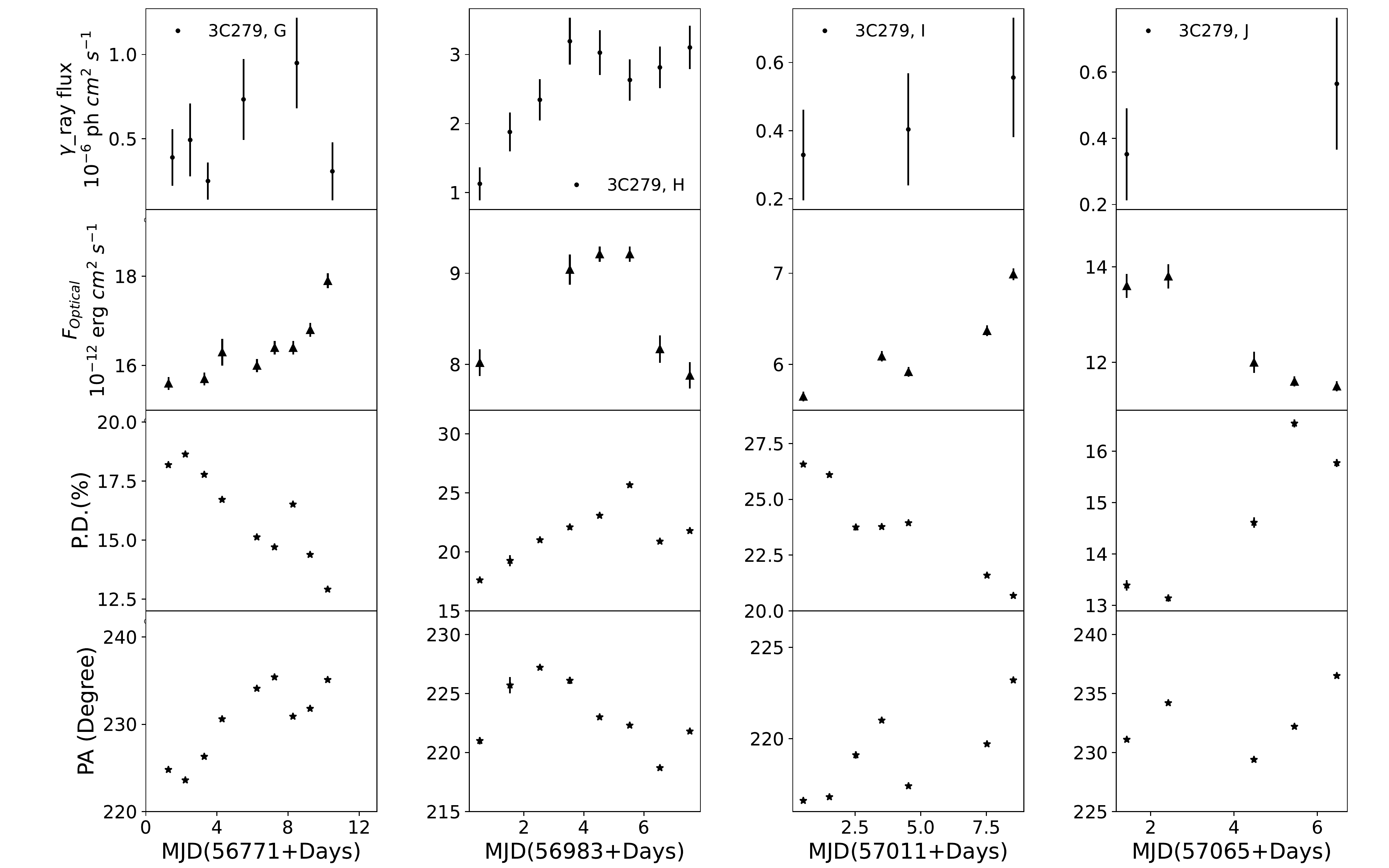}
\caption{\label{fig:epoch_lc4}Multi-wavelength light curves along with polarization measurements. The panels have the same meaning as in Fig. \ref{fig:epoch_lc1}.}
\end{figure*}

\begin{figure*}
\includegraphics[width=16cm, height=8cm]{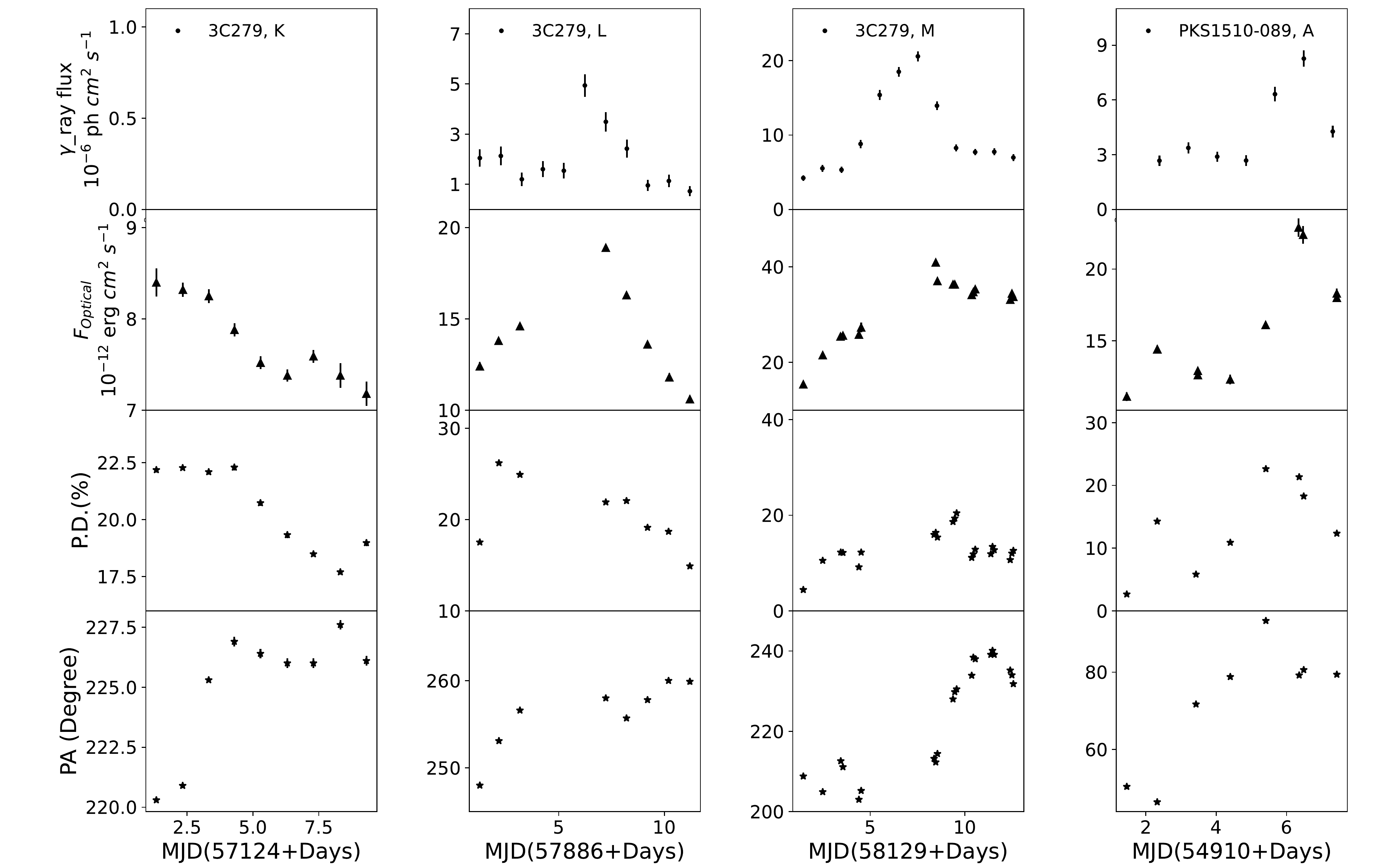}
\caption{\label{fig:epoch_lc5}Multi-wavelength light curves along with polarization measurements. The panels have the same meaning as in Fig. \ref{fig:epoch_lc1}.}
\end{figure*}
\begin{figure*}
\includegraphics[width=16cm, height=8cm]{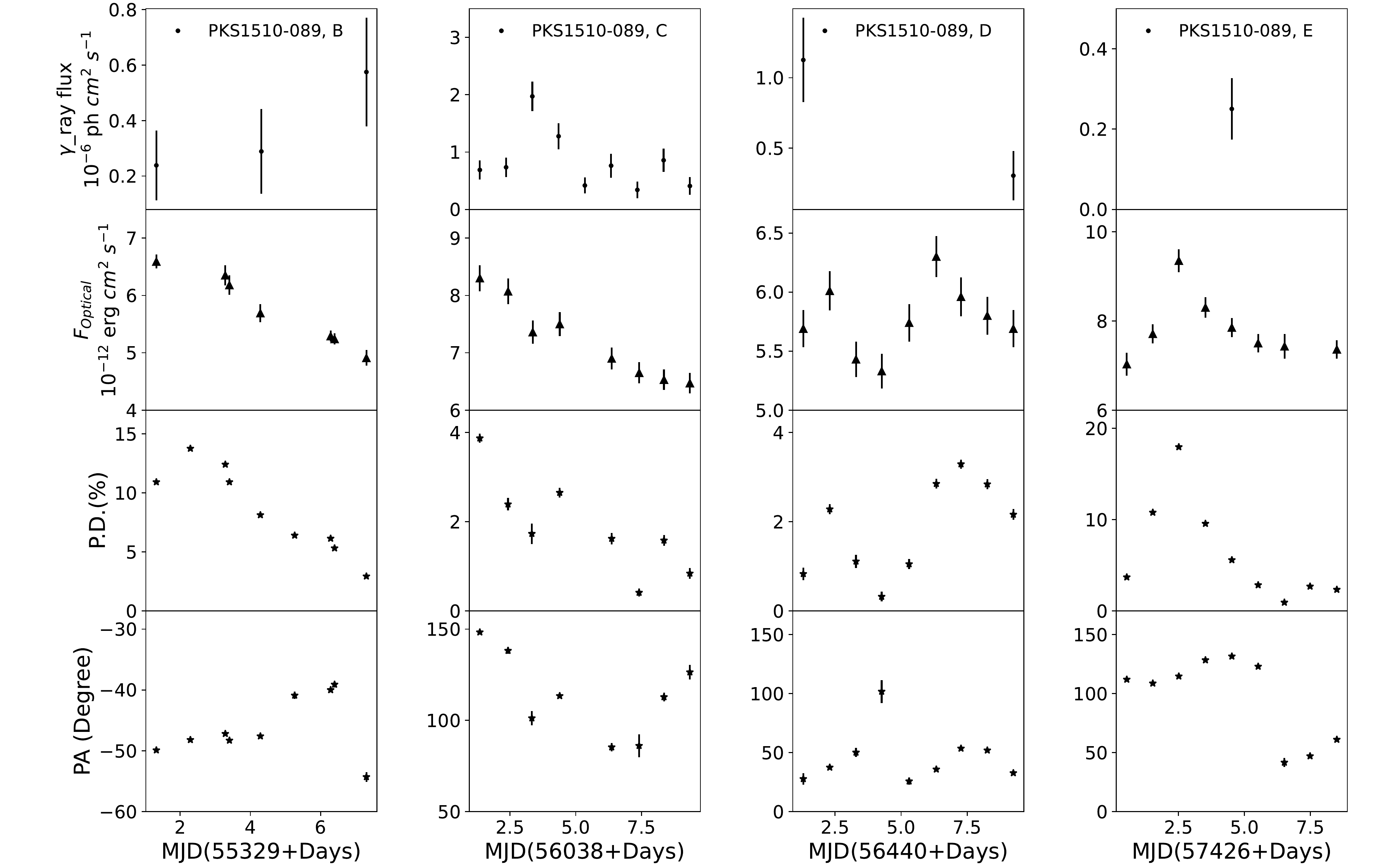}
\caption{\label{fig:epoch_lc6}Multi-wavelength light curves along with polarization measurements. The panels have the same meaning as in Fig. \ref{fig:epoch_lc1}.}
\end{figure*}
\begin{figure*}
\includegraphics[width=16cm, height=8cm]{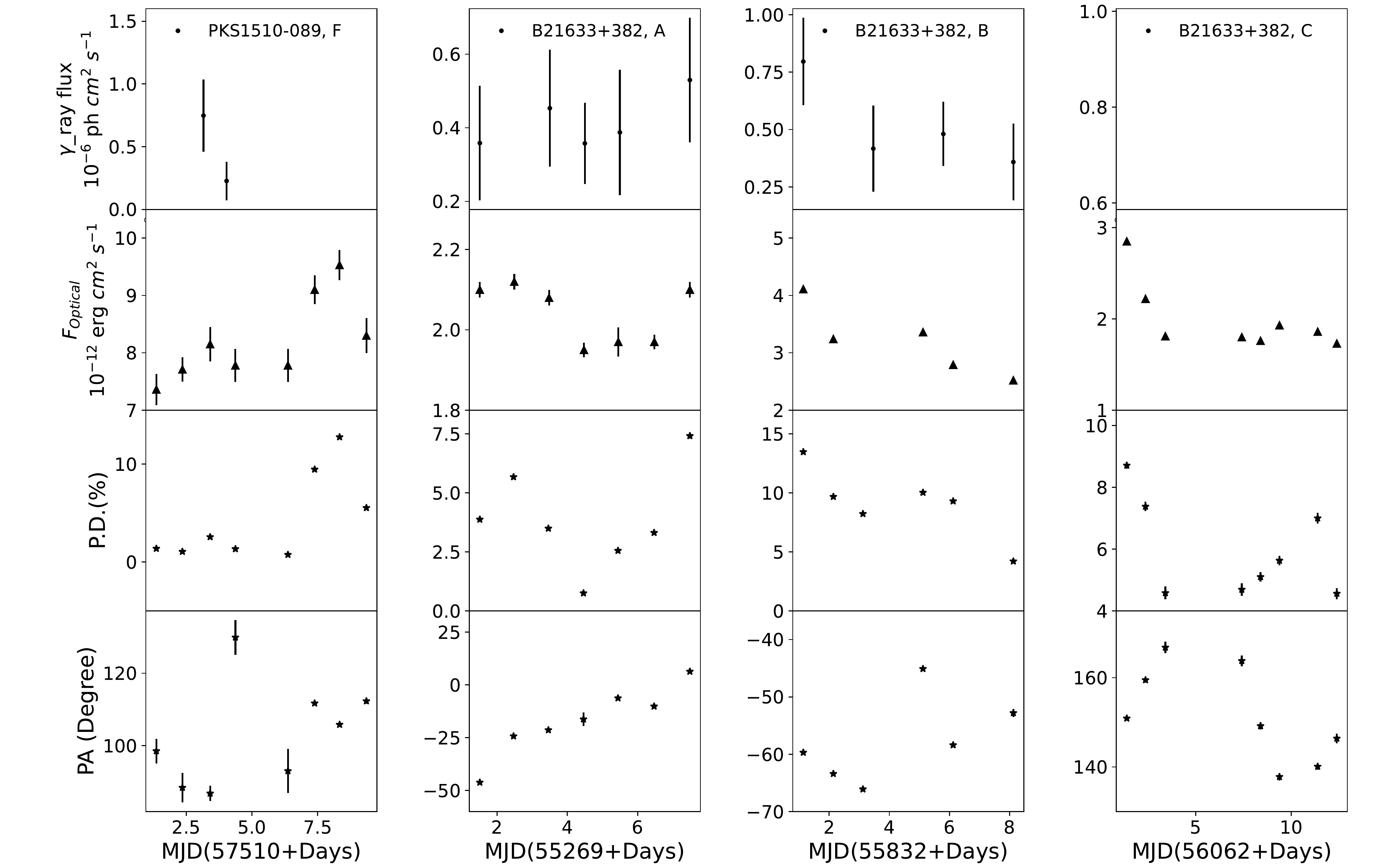}
\caption{\label{fig:epoch_lc7}Multi-wavelength light curves along with polarization measurements. The panels have the same meaning as in Fig. \ref{fig:epoch_lc1}.}
\end{figure*}
\begin{figure*}
\includegraphics[width=16cm, height=8cm]{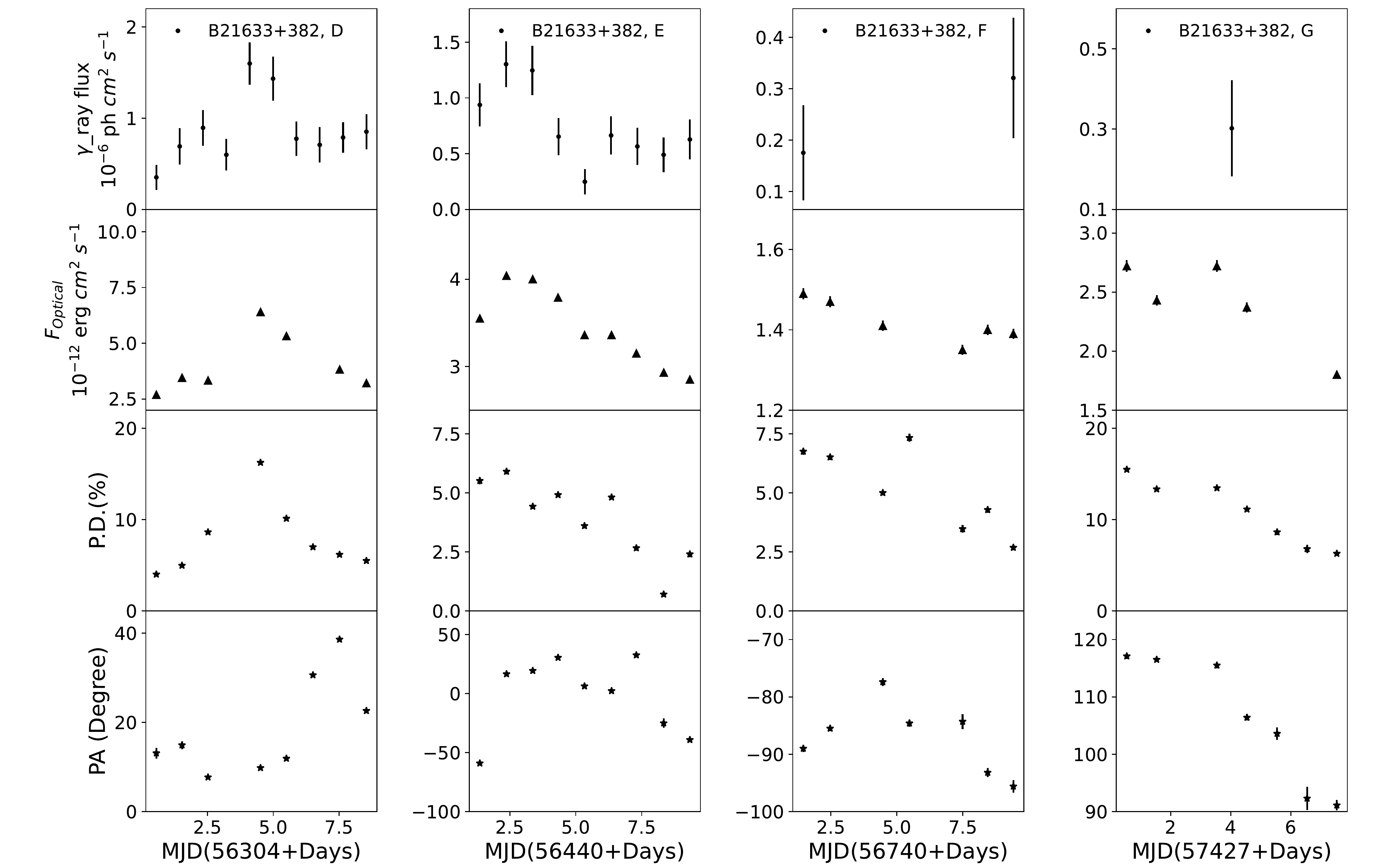}
\caption{\label{fig:epoch_lc8}Multi-wavelength light curves along with polarization measurements. The panels have the same meaning as in Fig. \ref{fig:epoch_lc1}.}
\end{figure*}
\begin{figure*}
\includegraphics[width=16cm, height=8cm]{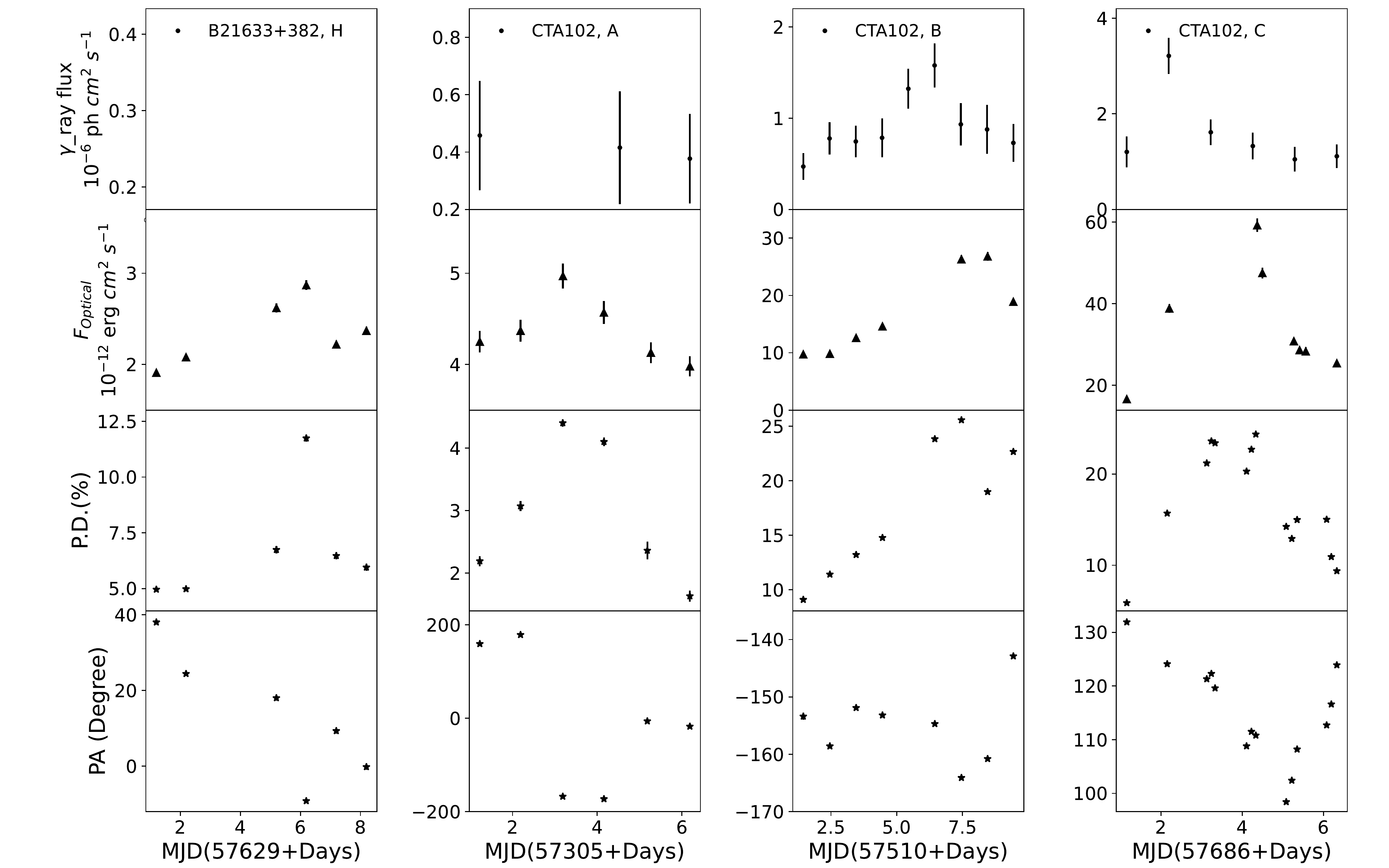}
\caption{\label{fig:epoch_lc9}Multi-wavelength light curves along with polarization measurements. The panels have the same meaning as in Fig. \ref{fig:epoch_lc1}.}
\end{figure*}
\begin{figure*}
\includegraphics[width=16cm, height=8cm]{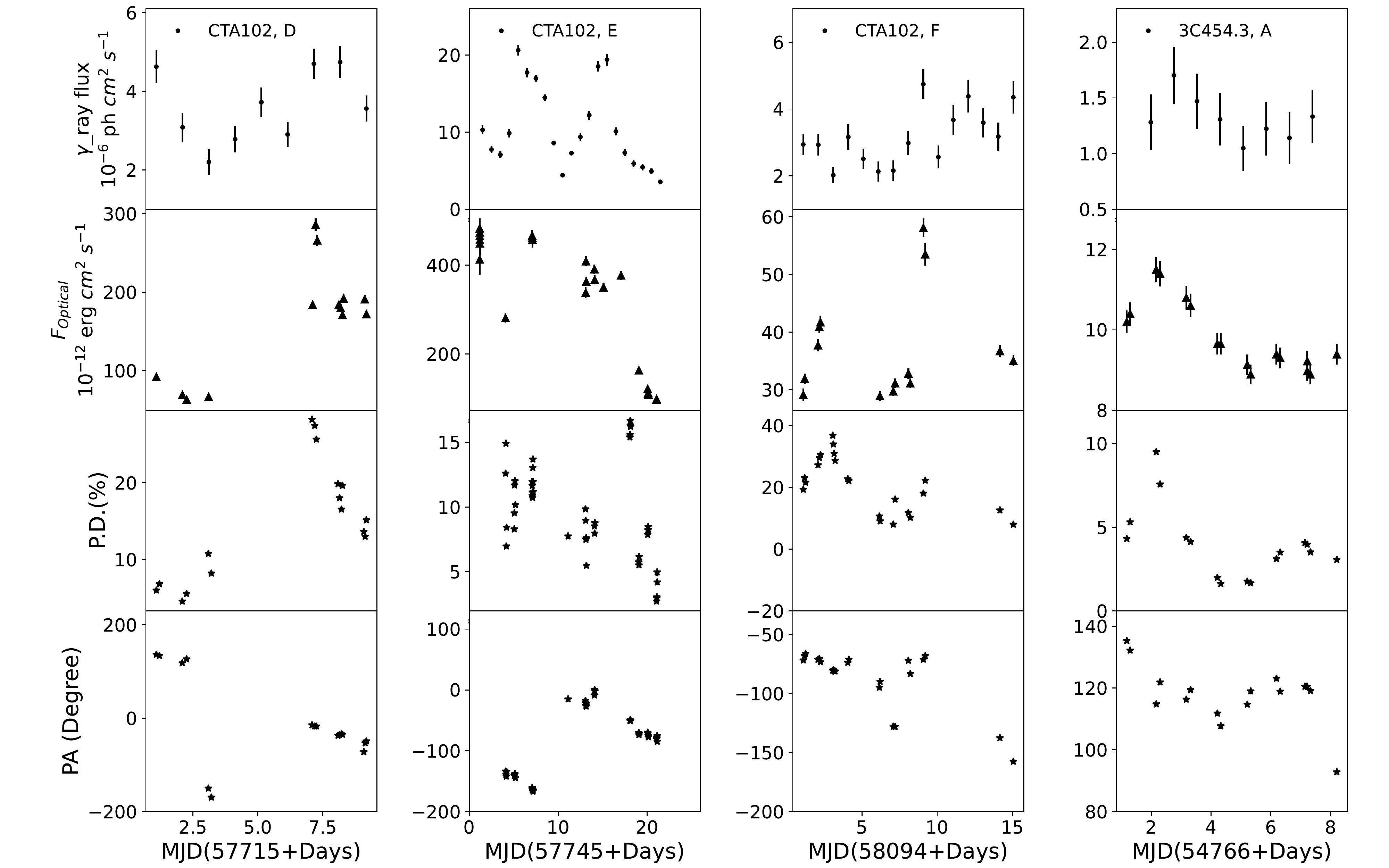}
\caption{\label{fig:epoch_lc10}Multi-wavelength light curves along with polarization measurements. The panels have the same meaning as in Fig. \ref{fig:epoch_lc1}.}
\end{figure*}
\begin{figure*}
\includegraphics[width=16cm, height=8cm]{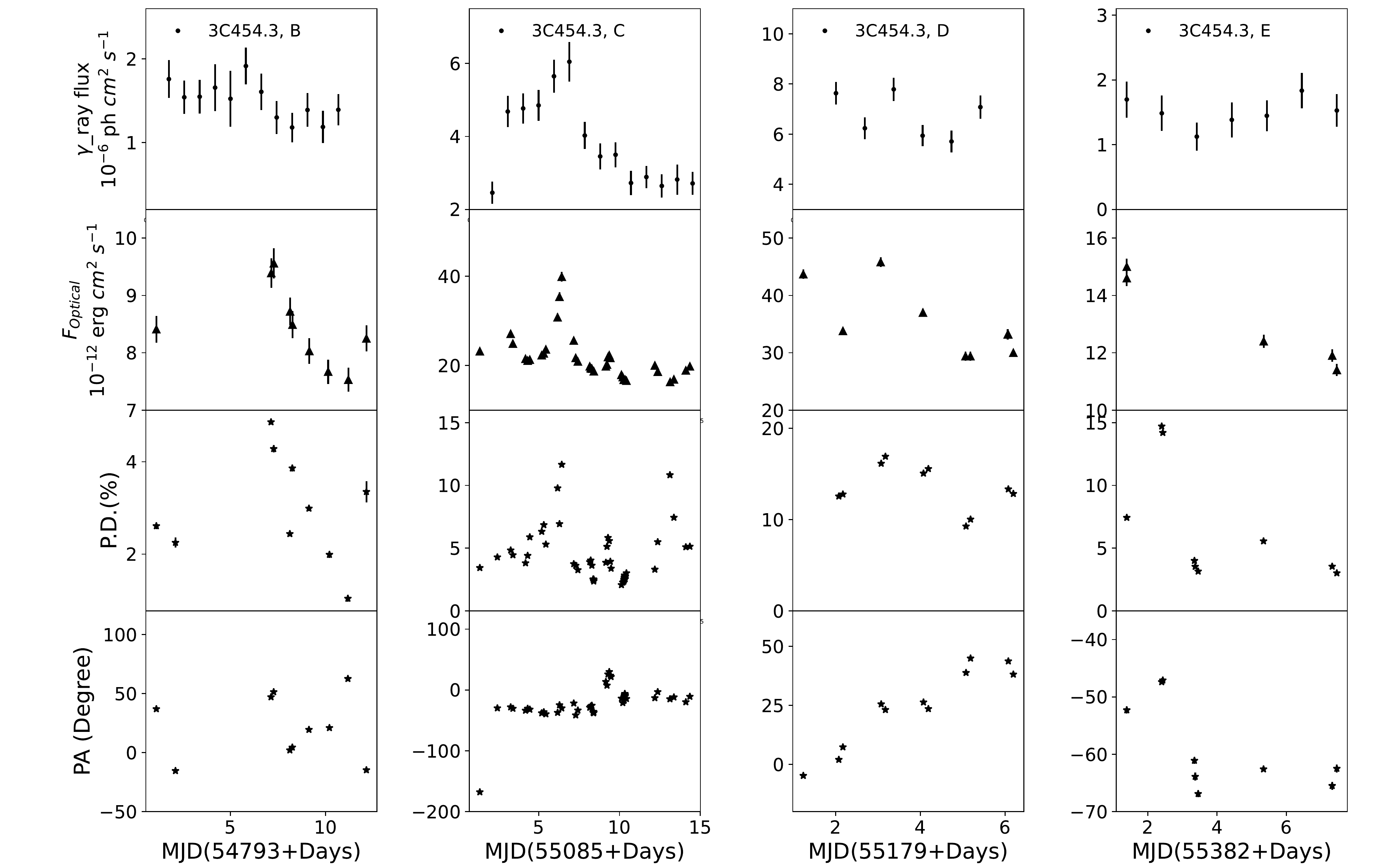}
\caption{\label{fig:epoch_lc11}Multi-wavelength light curves along with polarization measurements. The panels have the same meaning as in Fig. \ref{fig:epoch_lc1}.}
\end{figure*}
\begin{figure*}
\includegraphics[width=16cm, height=8cm]{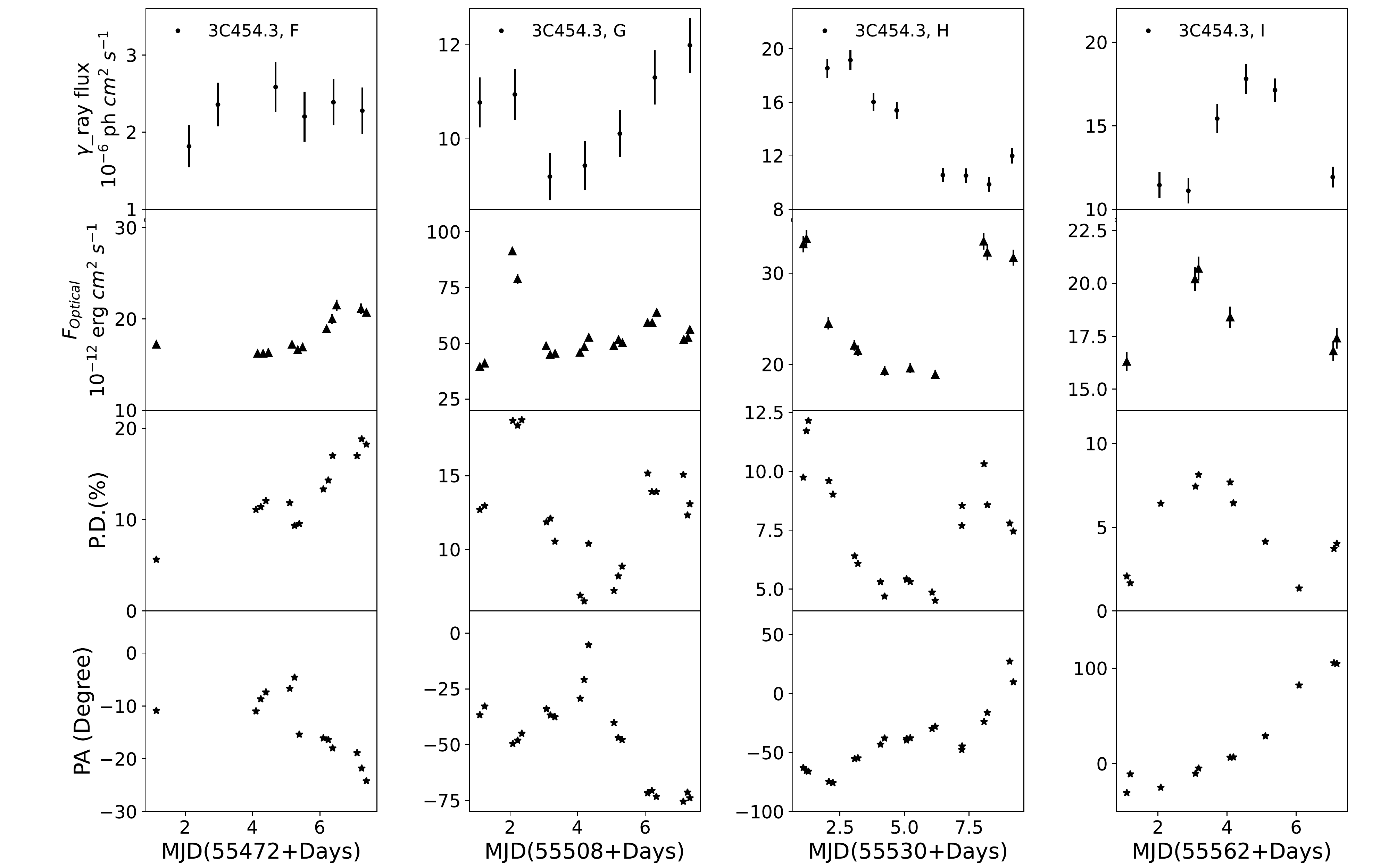}
\caption{\label{fig:epoch_lc12}Multi-wavelength light curves along with polarization measurements. The panels have the same meaning as in Fig. \ref{fig:epoch_lc1}.}
\end{figure*}
\begin{figure*}
\includegraphics[width=16cm, height=8cm]{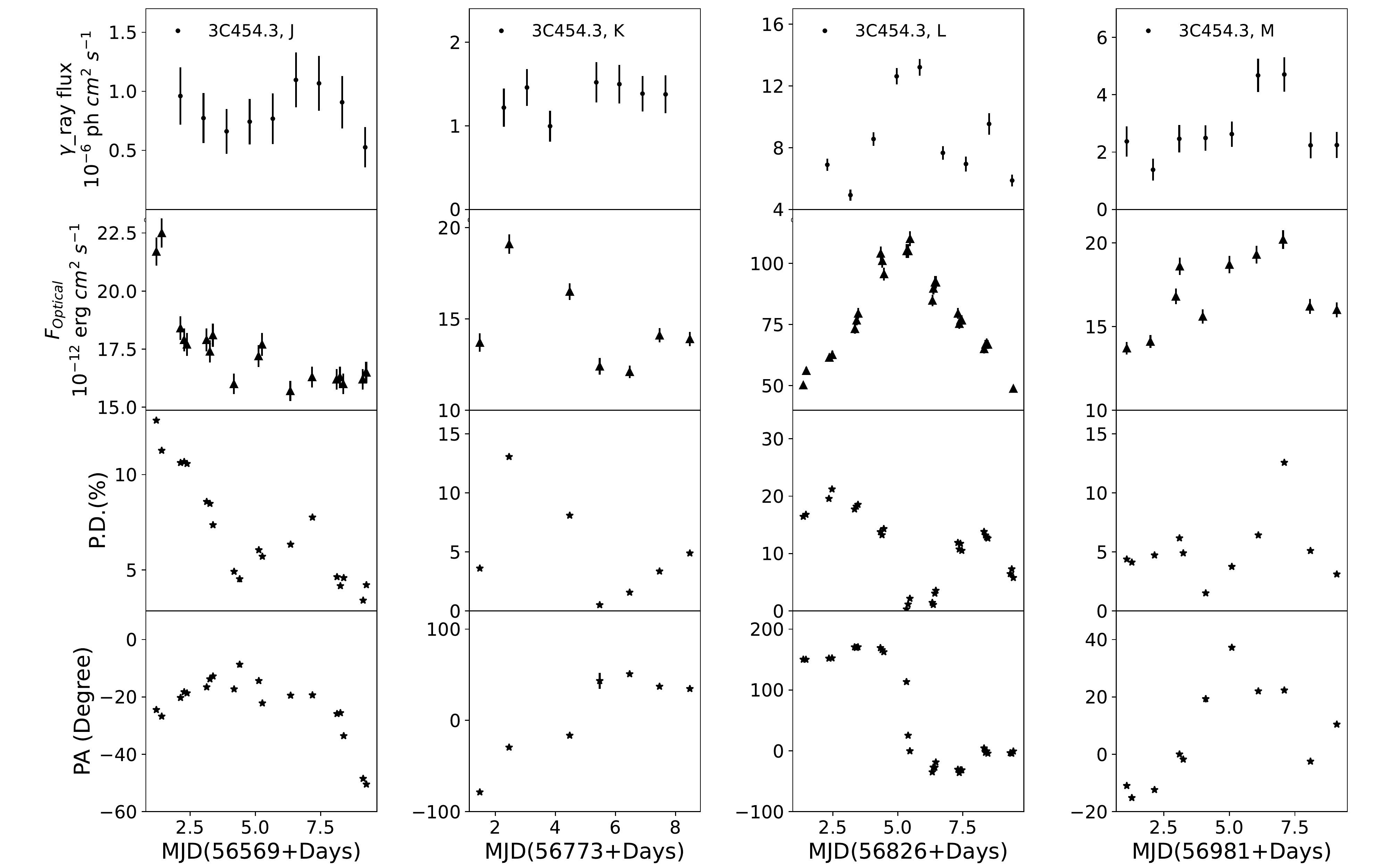}
\caption{\label{fig:epoch_lc13}Multi-wavelength light curves along with polarization measurements. The panels have the same meaning as in Fig. \ref{fig:epoch_lc1}.}
\end{figure*}
\begin{figure*}
\includegraphics[width=14cm, height=8cm]{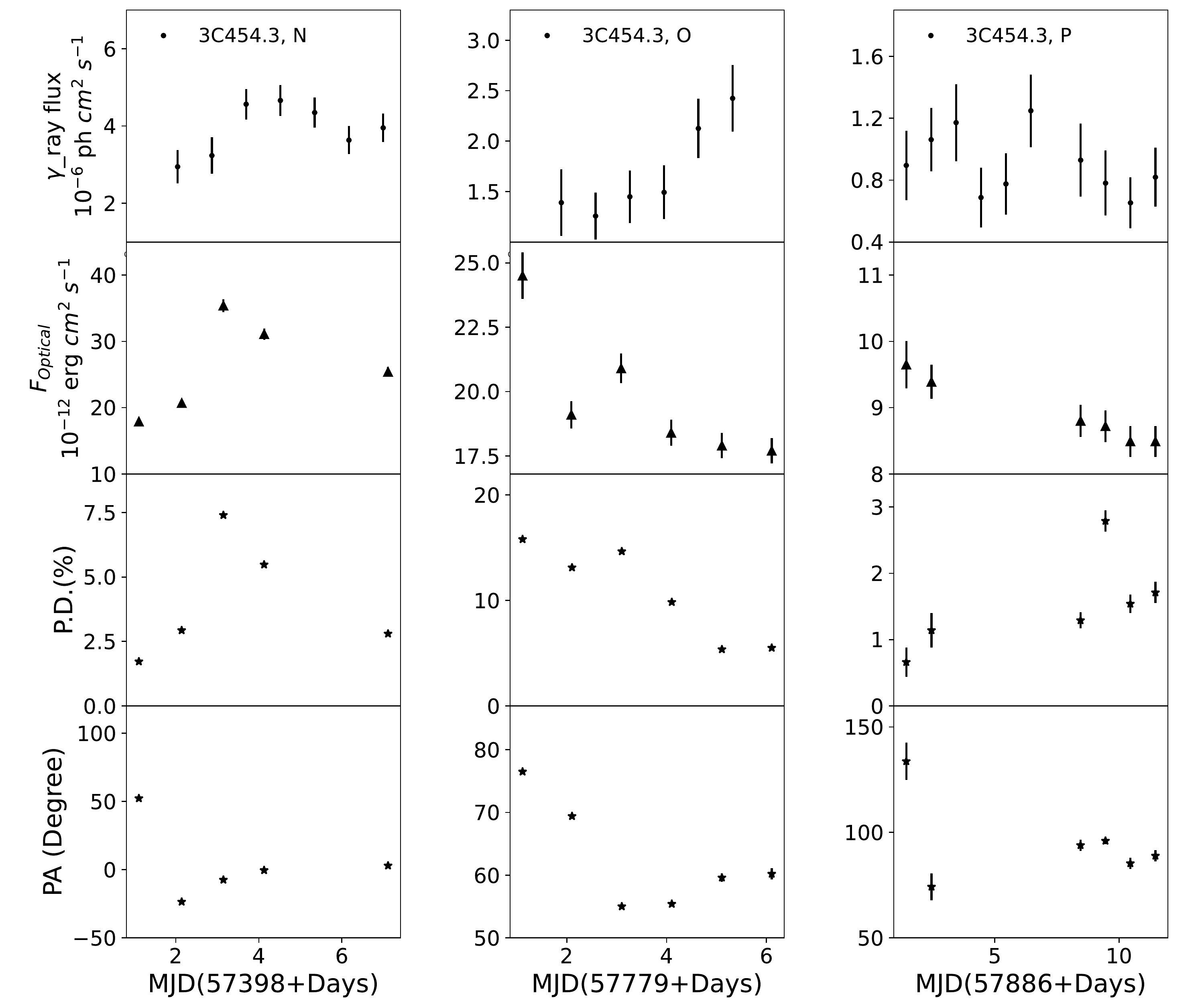}
\caption{\label{fig:epoch_lc14}Multi-wavelength light curves along with polarization measurements. The panels have the same meaning as in Fig. \ref{fig:epoch_lc1}.}
\end{figure*}

\end{subfigures}

\begin{subfigures}\label{fig:epoch_cor}
\begin{figure*}
\includegraphics[width=15cm, height=13cm]{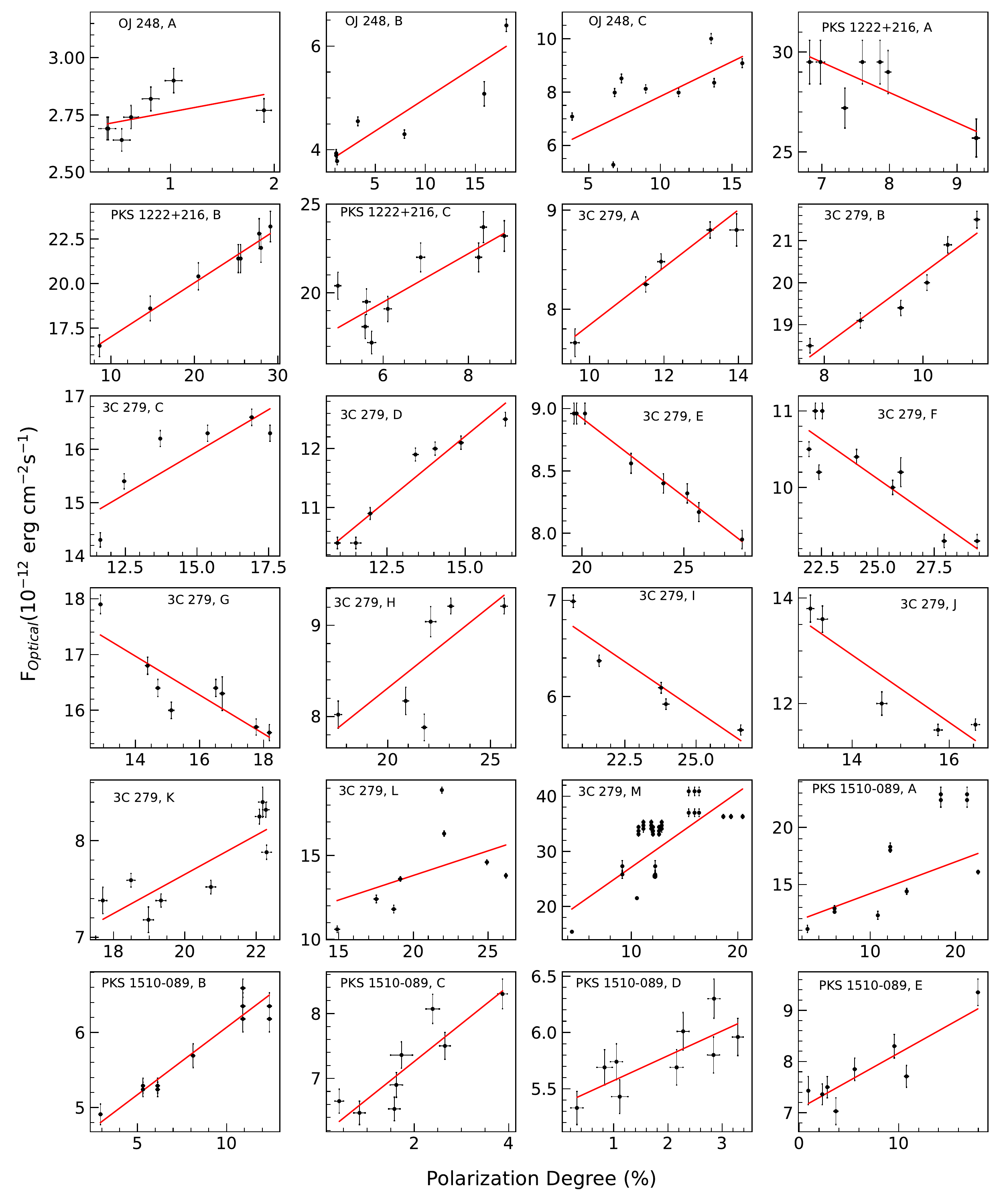}
\caption{\label{fig:epoch_cor1}Degree of polarization vs optical V-band flux plots on short-term timescales for OJ 248, PKS 1222$+$216, 3C 279 and PKS 1510$-$089. The name of the source and the epoch are given in each panel. The solid red line represents the straight line fit to the data.}
\end{figure*}
\begin{figure*}
\includegraphics[width=15cm, height=13cm]{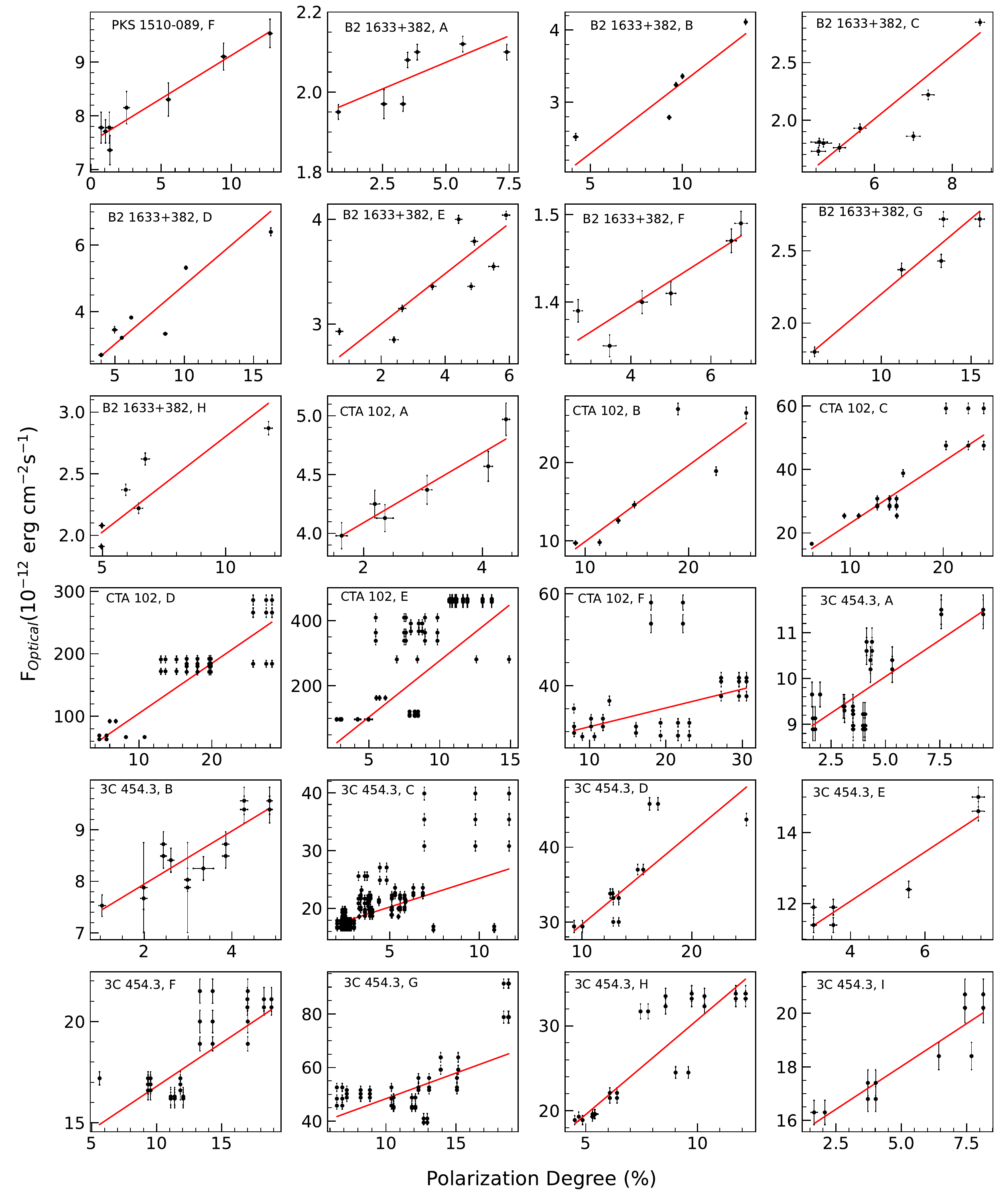}
\caption{\label{fig:epoch_cor2}Same as in Figure \ref{fig:epoch_cor1} but for PKS 1510$-$089, B2 1633$+$383, CTA 102, and 3C 454.3.}
\end{figure*}
\begin{figure*}
\includegraphics[width=15cm, height=5cm]{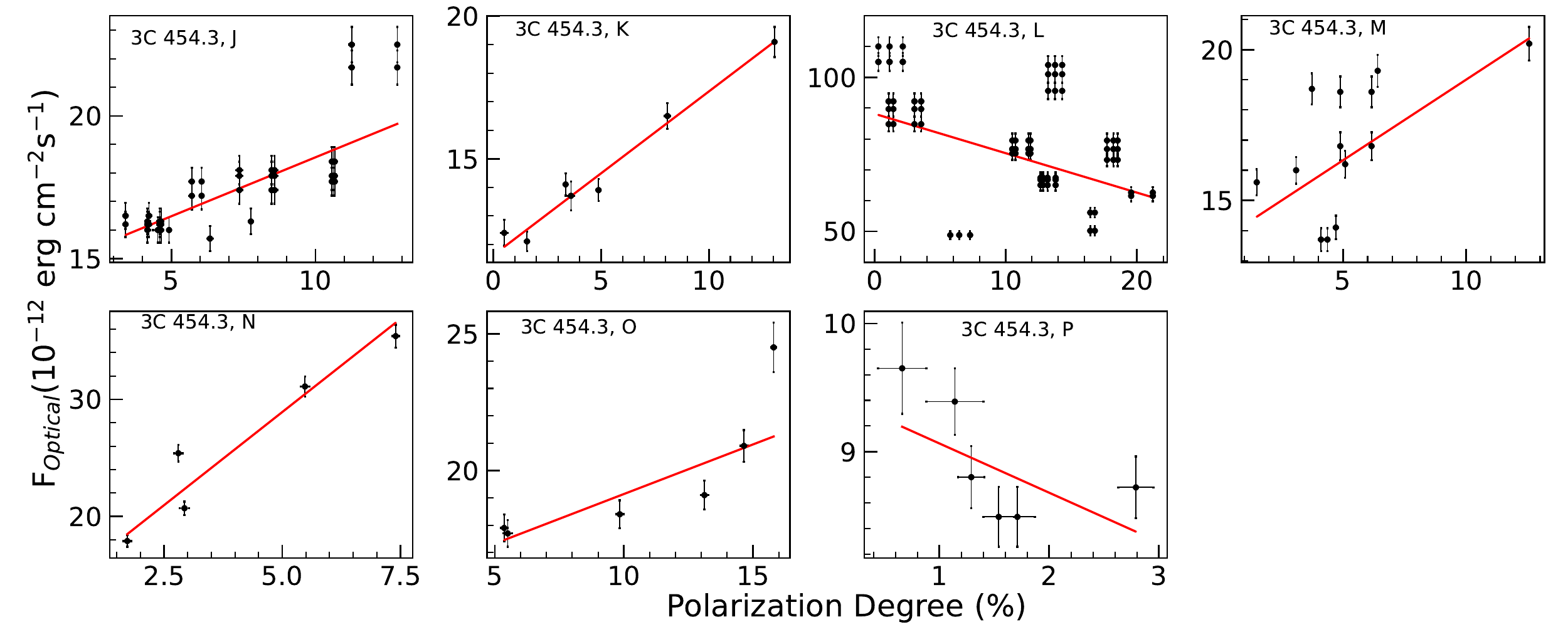}
\caption{\label{fig:epoch_cor3}Same as in Figure \ref{fig:epoch_cor1} but for 3C 454.3.}
\end{figure*}
\end{subfigures}

\section{Analysis} \label{sec:analysis}
\subsection{Correlation between optical flux and polarization}
\subsubsection{Correlation on long-term timescales}
The entire 10 year Steward Observatory data was observed in 10 different observing cycles. After the continuous monitoring of the sources for several months in an observing cycle, there is a gap of 4-6 months before the start of the next observing cycle. Long-term timescale in this work pertains to one observing cycle, which could contain data for about six months. The time period of the cycle, the minimum, maximum and average values of optical flux, and PD during each observing cycle of all the sources are given in Table \ref{tab:LTV_stats}. 

For each FSRQ, we first searched for a correlation between optical flux and PD during each of these 10 observing cycles by using the Spearman rank correlation method. The polarimetric and photometric observations from the Steward Observatory are quasi-simultaneous, therefore easily matched for a Spearman rank test. The values of Spearman rank correlation coefficient (R) and the probability of no correlation (P) are given in Table \ref{tab:LTV_stats}. We conclude on the presence of positive correlation (PC) or negative correlation (NC) between optical flux and PD at the 95\% confidence level in a cycle, only when $R > 0.5$ or $R < -0.5$, respectively, with $P < 0.05$. For those observing cycles, where a close correlation between optical flux and PD is noticed, we show in Figures \ref{fig:pol_flx_cor1} and \ref{fig:pol_flx_cor2}, the linear least-square fit between optical flux and PD. 

\subsubsection{Correlation on short-term timescales}
To explore the correlation between optical flux and PD on short-term timescales, we carefully inspected the light curves of each of the 10 observing cycles for all the sources. We found that there are almost continuous monitoring of the sources for about 10 days every month during each observing cycle \citep{2018MNRAS.479.2037P}. So, for each source, we divided the light curve of each observing cycle into segments such that each segment will have at least five measurements of optical flux, PD, and PA. Each segment can have observations that span about 10 days. Therefore, the short-term timescales considered in this work include observing periods of about 10 days. 

We then calculated the optical flux variability amplitude for each of these segments using the fractional root mean square variability ($F_{var}$) given by \cite{2003MNRAS.345.1271V} and defined as
\begin{equation}
F_{var} = \sqrt{\frac{S^2 - \overline{\sigma_{err}^2}}{{\bar{x}^2}}}.
\end{equation}
The error in $F_{var}$ is estimated as: 
\begin{equation}
err(F_{var}) =  \sqrt{\left( \sqrt{\frac{1}{2N}}\frac{\overline{\sigma_{err}^2}}{\bar{x}^2 F_{var}} \right)  ^ 2+ \left(  \sqrt{\frac{\overline{\sigma_{err}^2}}{N}} \frac{1}{\bar{x}}\right) ^2 },
\end{equation}
where $S^2$ is the sample variance of the light curve and $\overline{\sigma_{err}^2}$ is mean square error defined as $\overline{\sigma_{err}^2} = \frac{1}{N}\sum_{i=1}^{N}{\sigma^{2}_{err,i}}$. We considered a source to be significantly variable in a segment only if F$_{var}$ > 3 $\times$ err(F$_{var}$). We also searched for a correlation between optical $V-$band flux and PD in each of these segments by using the Spearman rank correlation. We considered that the optical flux and PD are well correlated if (i) the Spearman rank correlation coefficient, $R$, is either $>$ 0.5 (positive correlation) or $<$ $-$0.5 (negative correlation), and (ii) the null hypothesis probability, $P$, is $<$ 0.05 (that corresponds to 95\% confidence level). For our detailed analysis, we considered only those segments in which (a) source shows significant optical variations and (b) there is a good correlation, either positive or negative, between optical flux and polarization. Our selection criteria yielded 3 epochs for OJ 248, 3 epochs for PKS 1222$+$216, 13 epochs for 3C 279, 6 epochs for PKS 1510$-$089, 8 epochs for B2 1633$+$382, 6 epochs for CTA 102, and 16 epochs for 3C 454.3. The time duration of the epochs together with the statistics of the flux and polarization variations for each of the sources are given in Table \ref{tab:STV_stats}. 

For those epochs, we also searched for correlation between optical and $\gamma-$ray fluxes as well as between optical flux and PD. The results of the correlation analysis are given in Table \ref{tab:epoch_cor}. The $\gamma$-ray and optical light curves along with PD and PA for these epochs are plotted in Figures \ref{fig:epoch_lc1} to \ref{fig:epoch_lc14}. The degree of polarization vs optical flux plots for all the sources are plotted in Figures \ref{fig:epoch_cor1} to \ref{fig:epoch_cor3}. The values of fractional variability amplitude, F$_{var}$, and its error for the light curve of each of the epochs are given in Table \ref{tab:epoch_res}. 

\begin{table*}
\centering
\caption{\label{tab:epoch_res}Results of optical variability analysis.}
\begin{tabular} {lcccccccccc}\hline
Blazar		& Epoch & $F_{var}(\%)$ & $\tau_{var}$   & \multicolumn{3}{c}{$\alpha$} & $\eta^o$ & $\delta^o$  & R  & $B$  \\ 
        	&	&               &   (in days)             &   Min   & Max    &    Average             &            &    &    (in cm)     &   (in G)           \\ \hline 
OJ 248		& A & 2.65$\pm$0.78   &   21.16$\pm$11.46 & 0.32 & 0.47 & 0.40 & - & - & -   & - \\ 
		& B & 20.20$\pm$0.98   &   2.47$\pm$0.16 & 0.66 & 0.99 & 0.81 & 0.17 & 1.29 & 4.43 $\times$ 10$^{15}$   & 0.46 \\ 
		& C & 16.26$\pm$0.62   &   2.28$\pm$0.14 & 0.94 & 1.16 & 1.04 & 0.18 & 1.29 & 4.11 $\times$ 10$^{15}$   & 0.48 \\ 
		
PKS 1222$+$216  & A & 4.89$\pm$1.48   &   21.57$\pm$8.16 & 0.23 & 0.85 & 0.40 & - & - & -   & - \\ 
		& B & 10.17$\pm$1.35   &   10.67$\pm$6.03 & 0.23 & 0.85 & 0.37 & - & - & -   & - \\ 
		& C & 10.44$\pm$1.27   &   7.91$\pm$3.21 & 0.20 & 0.36 & 0.28 & - & - & -   & - \\ 
		  
3C 279		& A & 5.47$\pm$0.61   &   13.55$\pm$3.76 & 1.63 & 1.79 & 1.69 & 0.05 & 0.98 & 1.86$ \times$ 10$^{16}$   & 0.16 \\ 
		& B & 5.62$\pm$0.38   &   22.95$\pm$6.78 & 0.27 & 1.56 & 1.31 & 0.03 & 0.89 & 2.85$ \times$ 10$^{16}$   & 0.12 \\ 
		& C & 5.34$\pm$0.38   &   26.46$\pm$4.66 & 1.36 & 1.43 & 1.40 & 0.02 & 0.87 & 3.20$ \times$ 10$^{16}$   & 0.11 \\ 
		& D & 7.53$\pm$0.35   &   22.02$\pm$6.09 & 1.35 & 1.44 & 1.39 & 0.03 & 0.90 & 2.76$ \times$ 10$^{16}$   & 0.12 \\ 
		& E & 4.41$\pm$0.35   &   21.39$\pm$5.97 & 1.29 & 1.35 & 1.34 & 0.03 & 0.90 & 2.69$ \times$ 10$^{16}$   & 0.12 \\ 
		& F & 5.97$\pm$0.36   &   10.91$\pm$1.54 & 0.96 & 1.33 & 1.24 & 0.06 & 1.02 & 1.56$ \times$ 10$^{16}$   & 0.18 \\ 
		& G & 4.31$\pm$0.39   &   15.58$\pm$3.19 & 0.96 & 1.37 & 1.28 & 0.04 & 0.96 & 2.08$ \times$ 10$^{16}$   & 0.15 \\ 
		& H & 7.15$\pm$0.64   &   8.38$\pm$1.44  & 1.41 & 1.47 & 1.45 & 0.07 & 1.07 & 1.25$ \times$ 10$^{16}$   & 0.22 \\ 
		& I & 8.19$\pm$0.41   &   10.81$\pm$1.52 & 1.40 & 1.52 & 1.46 & 0.06 & 1.03 & 1.54$ \times$ 10$^{16}$   & 0.19 \\ 
		& J & 8.76$\pm$0.72   &   14.72$\pm$2.75 & 1.39 & 1.51 & 1.46 & 0.04 & 0.97 & 1.99$ \times$ 10$^{16}$   & 0.15 \\ 
		& K & 5.77$\pm$0.44   &   21.19$\pm$5.90 & 0.96 & 1.36 & 1.27 & 0.03 & 0.90 & 2.67$ \times$ 10$^{16}$   & 0.12 \\ 
		& L & 18.82$\pm$0.44   &   5.49$\pm$0.39 & 1.34 & 1.51 & 1.43 & 0.11 & 1.16 & 8.89$ \times$ 10$^{15}$   & 0.28 \\ 
		& M & 21.79$\pm$0.40   &   0.89$\pm$0.23 & 1.28 & 1.43 & 1.35 & 0.70 & 1.64 & 2.03$ \times$ 10$^{15}$   & 0.84 \\ 
		
PKS 1510$-$089	& A & 25.13$\pm$0.66   &   2.66$\pm$0.22 & 1.00 & 1.25 & 1.14 & 0.04 & 0.93 & 3.47$ \times$ 10$^{15}$   & 0.49 \\ 
		& B & 10.80$\pm$0.93   &   9.11$\pm$3.25 & 0.94 & 1.18 & 1.04 & - & - & -   & - \\ 
		& C & 9.32$\pm$1.00   &   10.18$\pm$4.31 & 0.83 & 1.00 & 0.93 & - & - & -   & - \\ 
		& D & 4.31$\pm$1.01   &   9.80$\pm$3.77 & 0.71 & 0.88 & 0.77 & - & - & -   & - \\ 
		& E & 8.77$\pm$1.09   &   5.12$\pm$1.04 & 0.89 & 1.00 & 0.95 & 0.02 & 0.82 & 5.83$ \times$ 10$^{15}$   & 0.33 \\ 
		& F & 8.44$\pm$1.22   &   6.49$\pm$1.91 & 0.72 & 0.83 & 0.76 & 0.01 & 0.77 & 6.97$ \times$ 10$^{15}$   & 0.29 \\ 
		
B2 1633$+$382 	& A & 3.47$\pm$0.42   &   15.44$\pm$3.12 & 0.94 & 1.06 & 1.01 & 0.12 & 1.19 & 2.57$\times$ 10$^{16}$   & 0.14 \\ 
		& B & 18.99$\pm$0.49   &   4.20$\pm$0.23 & 1.02 & 1.35 & 1.19 & 0.45 & 1.52 & 8.93$\times$ 10$^{15}$   & 0.30 \\ 
		& C & 18.89$\pm$0.60   &   3.93$\pm$0.32 & 0.89 & 1.13 & 0.98 & 0.48 & 1.57 & 8.61$\times$ 10$^{15}$   & 0.32 \\ 
		& D & 32.96$\pm$0.62   &   3.06$\pm$0.10 & 1.18 & 1.47 & 1.32 & 0.61 & 1.60 & 6.82$\times$ 10$^{15}$   & 0.37 \\ 
		& E & 12.58$\pm$0.31   &   7.81$\pm$0.79 & 1.23 & 1.40 & 1.33 & 0.24 & 1.34 & 1.46$\times$ 10$^{16}$   & 0.21 \\ 
		& F & 3.57$\pm$0.38   &   26.19$\pm$9.37 & 0.91 & 1.30 & 1.05 & - & - & -   & - \\ 
		& G & 15.51$\pm$0.83   &   7.27$\pm$1.37 & 1.03 & 1.25 & 1.16 & 0.26 & 1.37 & 1.39$\times$ 10$^{16}$   & 0.22 \\ 
		& H & 15.01$\pm$0.69   &   3.90$\pm$0.40 & 0.94 & 1.07 & 1.04 & 0.48 & 1.57 & 8.52$\times$ 10$^{15}$   & 0.32 \\ 
		
CTA 102	& A & 7.58$\pm$1.17   &   7.78$\pm$2.36 & 1.38 & 1.49 & 1.45 & 0.10 & 1.15 & 1.24$\times$ 10$^{16}$   & 0.22 \\ 
		& B & 42.65$\pm$1.14   &   2.80$\pm$0.31 & 1.87 & 2.00 & 1.92 & 0.29 & 1.35 & 5.26$\times$ 10$^{15}$   & 0.42 \\ 
		& C & 39.31$\pm$1.07   &   0.55$\pm$0.10 & 1.88 & 2.05 & 2.00 & 1.49 & 1.76 & 1.34$\times$ 10$^{15}$   & 1.13 \\ 
		& D & 43.95$\pm$0.83   &   0.26$\pm$0.02 & 1.76 & 3.30 & 2.12 & 3.13 & 1.97 & 7.16$\times$ 10$^{14}$   & 1.79 \\ 
		& E & 45.12$\pm$0.72   &   0.13$\pm$0.03 & 0.69 & 2.19 & 1.65 & 6.11 & 2.34 & 4.36$\times$ 10$^{14}$   & 2.63 \\ 
		& F & 24.14$\pm$0.81   &   1.00$\pm$0.50 & 2.10 & 2.52 & 2.21 & - & - & -   & - \\ 
 
3C 454.3	& A & 8.40$\pm$0.71   &   8.22$\pm$2.68 & 0.93 & 1.31 & 1.11 & 0.19 & 1.30 & 1.49$\times$ 10$^{16}$   & 0.21 \\ 
		& B & 6.07$\pm$1.80   &   8.24$\pm$2.71 & 0.70 & 1.10 & 0.97 & 0.19 & 1.31 & 1.50$\times$ 10$^{16}$   & 0.20 \\ 
		& C & 23.20$\pm$0.47   &   0.73$\pm$0.17 & 0.91 & 1.78 & 1.42 & 2.13 & 2.00 & 2.03$\times$ 10$^{15}$   & 0.90 \\ 
		& D & 17.10$\pm$0.72   &   1.18$\pm$0.30 & 1.20 & 1.62 & 1.52 & 1.31 & 1.81 & 2.98$\times$ 10$^{15}$   & 0.67 \\ 
		& E & 12.37$\pm$0.83   &   3.09$\pm$1.88 & 1.05 & 1.15 & 1.10 & - & - & -   & - \\ 
		& F & 11.09$\pm$0.71   &   1.76$\pm$0.95 & 1.19 & 1.46 & 1.31 & - & - & -   & - \\ 
		& G & 23.17$\pm$0.64   &   0.90$\pm$0.55 & 1.44 & 1.73 & 1.60 & - & - & -   & - \\ 
		& H & 24.19$\pm$0.86   &   2.65$\pm$0.32 & 1.44 & 1.66 & 1.57 & 0.58 & 1.55 & 5.75$\times$ 10$^{15}$   & 0.41 \\ 
		& I & 9.51$\pm$1.16   &   7.78$\pm$2.58 & 1.32 & 1.46 & 1.40 & 0.20 & 1.29 & 1.40$\times$ 10$^{16}$   & 0.21 \\ 
		& J & 10.09$\pm$0.65   &   3.58$\pm$0.70 & 1.27 & 1.50 & 1.39 & 0.43 & 1.49 & 7.45$\times$ 10$^{15}$   & 0.34 \\ 
		& K & 16.68$\pm$1.16   &   2.94$\pm$0.41 & 1.04 & 2.03 & 1.28 & 0.53 & 1.56 & 6.41$\times$ 10$^{15}$   & 0.38 \\ 
		& L & 21.98$\pm$0.56   &   0.68$\pm$0.47 & 1.29 & 1.44 & 1.38 & - & - & -  & - \\ 
		& M & 12.75$\pm$0.89   &   1.42$\pm$0.54 & 1.19 & 1.38 & 1.29 & -& - & -   & - \\ 
		& N & 27.51$\pm$1.28   &   1.87$\pm$0.14 & 1.20 & 1.33 & 1.25 & 0.83 & 1.71 & 4.45$\times$ 10$^{15}$   & 0.50 \\ 
		& O & 12.80$\pm$1.26   &   3.92$\pm$0.72 & 1.34 & 1.44 & 1.38 & 0.39 & 1.47 & 8.03$\times$ 10$^{15}$   & 0.32 \\ 
		& P & 4.56$\pm$1.33   &   70.32$\pm$25.29 & 0.59 & 0.87 & 0.75 & - & - & -   & - \\ 
 \hline
\end{tabular}
\end{table*}

\subsection{Optical variability timescale}
For each epoch, we also calculated the variability timescale as follows \citep{1974ApJ...193...43B}:
\begin{equation}
    \tau_{ij} = \frac{dt}{ln(F_i/F_j)}
\end{equation}
where $dt$ is the difference in time between any two flux measurements $F_i$ and $F_j$ such that $|F_i - F_j| > \sigma_{F_i} + \sigma_{F_j}$. The minimum from the ensemble of $\tau_{ij}$ values was taken as the minimum variability timescale $\tau_{var}$. The errors in $\tau_{ij}$ were obtained by error propagation \citep{1992drea.book.....B}. The $\tau_{var}$ values and their corresponding uncertainties are given in Table \ref{tab:epoch_res} for all the epochs of eight FSRQs. 

\subsection{Constraining physical parameters using optical variability timescale}
Due to relativistic beaming, the observed variability timescale ($\tau_{var}^o$) is shortened from the intrinsic timescale ($\tau_{var}^i$) and is given by 
\begin{equation}
\tau_{var}^o = \tau_{var}^i (1 + z)/ \delta
\end{equation}
where $\delta$ is the Doppler factor, and the superscripts $o$ and $i$ denote the observed and intrinsic values, respectively. 

Considering relativistic beaming, we estimated the values of the observed Doppler factor $\delta^o$ for each epoch of all the sources. For that we first calculated the value of the spectral index for each epoch as 
\begin{equation}\label{eq:alpha}
\alpha  = - \frac{ln\left(\frac{S_V}{S_R}\right)}{ln\left(\frac{\nu_V}{\nu_R}\right)},
\end{equation}
where S$_V$ and S$_R$ are the flux densities (S$_{\nu}$ $\propto$ $\nu^{-\alpha}$), and $\nu_V$ and $\nu_R$ are the effective frequencies of $V$ and $R$ bands, respectively. The minimum, maximum and average values of spectral indices for all the epochs are given in Table \ref{tab:epoch_res}. 

Using the fact that any observed variation with energy $L$ must occur on timescales larger than $\tau_{min} = \tau_{var}^o/(1+z)$, we calculated the inferred efficiency of accretion as \citep{2002apa..book.....F}
\begin{equation}\label{eq:eta}
\eta^o \ge 5 \times 10^{-43} L / \tau_{min}
\end{equation} 
where $L$ is the bolometric luminosity and was taken from \cite{2016ApJS..226...20F} for each source. The lower limit of accretion efficiency for all the epochs are given in Table \ref{tab:epoch_res}. In the table, the value of $\eta^o >$ 1  indicates that relativistic beaming is responsible for the observed high amplitude variation in the optical flux during that epoch. 

Relativistic beaming amplifies the observed flux relative to the intrinsic flux as $L^o = \delta^{3 + \alpha} L^i$ and shortens the observed timescale as $\tau_{min}^o = \delta^{-1} \tau_{min}^i$. Considering Equation \ref{eq:eta} as an equality to get the lower limit, we found $\eta^o = 5 \times 10^{-43} L^o/\tau_{min}^o$ and $\eta^i = 5 \times 10^{-43} L^i/\tau_{min}^i$, and thus the observed Doppler factor is given by
\begin{equation}\label{eq:delta_obs}
\delta^o \ge \left(\frac{\eta^o}{\eta^i}\right)^{1/(4+\alpha)}
\end{equation}
Since the value of $\eta^i$ can range from 0.007 (nuclear fusion) to 0.32 (accretion), we adopted a geometric mean value of $\eta^i = 0.05$ \citep[e.g.][]{1980A&A....88...23P,2002apa..book.....F} and using Equation \ref{eq:delta_obs} estimated the value of observed Doppler factor for each epoch. The derived lower limits of $\delta^o$ for each epoch of all the sources are given in Table \ref{tab:epoch_res}. 

Using the values of observed Doppler factor $\delta^o$ and the minimum variability timescale $\tau_{var}^o$, we set an upper limit on the size of the emission region as 
\begin{equation}
R \leq c \tau_{var}^o \frac{\delta^o}{1+z}
\end{equation}
The upper bound on the size of the emitting region for each epoch is given in Table \ref{tab:epoch_res}.

In the observer's frame, the lifetime of the synchrotron electrons within the jet is given by \citep{2008ApJ...672...40H}
\begin{equation}\label{eq:sync}
t_{syn} \approx 4.75 \times 10^2 \left(\frac{1+z}{\delta^o \nu_{GHz} B^3}\right)^{1/2} days,
\end{equation}
Considering that the observed variations in the optical flux of these eight FSRQs are due to synchrotron cooling, we estimated the value of the magnetic field using Equation \ref{eq:sync} and requiring that the observed minimum variability timescale has to be larger than or equal to the lifetime of the synchrotron electrons, $t_{syn}$. The calculated values of the magnetic field for the eight FSRQs are given in Table \ref{tab:epoch_res}.

\begin{subfigures}\label{fig:q_u}
\begin{figure*}
\centering
\includegraphics[width=18cm, height=14cm]{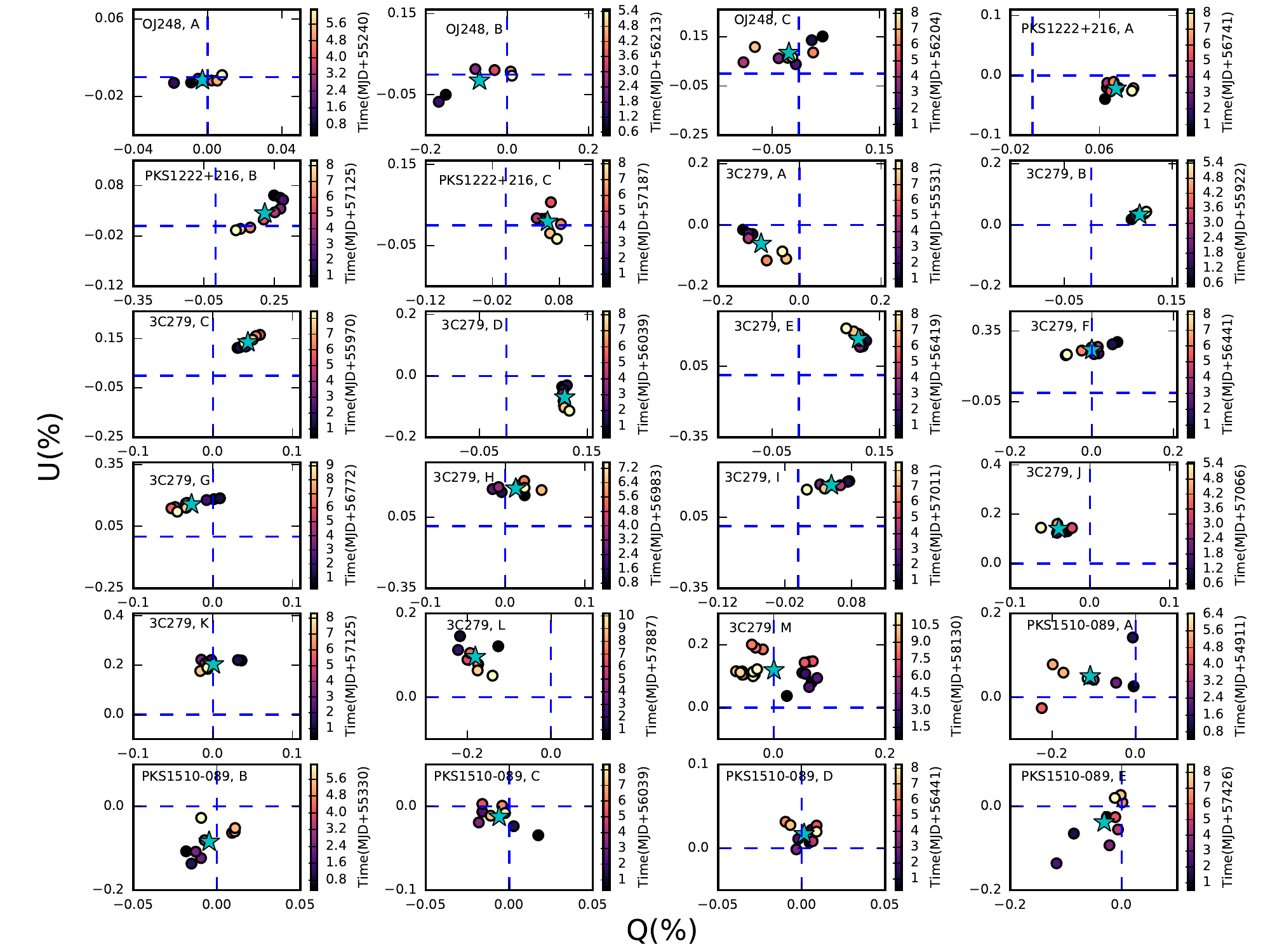}
\caption{\label{fig:q_u1}Stokes Q vs Stokes U plots for OJ 248, PKS 1222$+$216, 3C 279 and PKS 1510$-$089. The Q=0 and U=0 are shown as blue dashed lines and the centroid is marked with a cyan star in each plot.}
\end{figure*}
\begin{figure*}
\centering
\includegraphics[width=18cm, height=14cm]{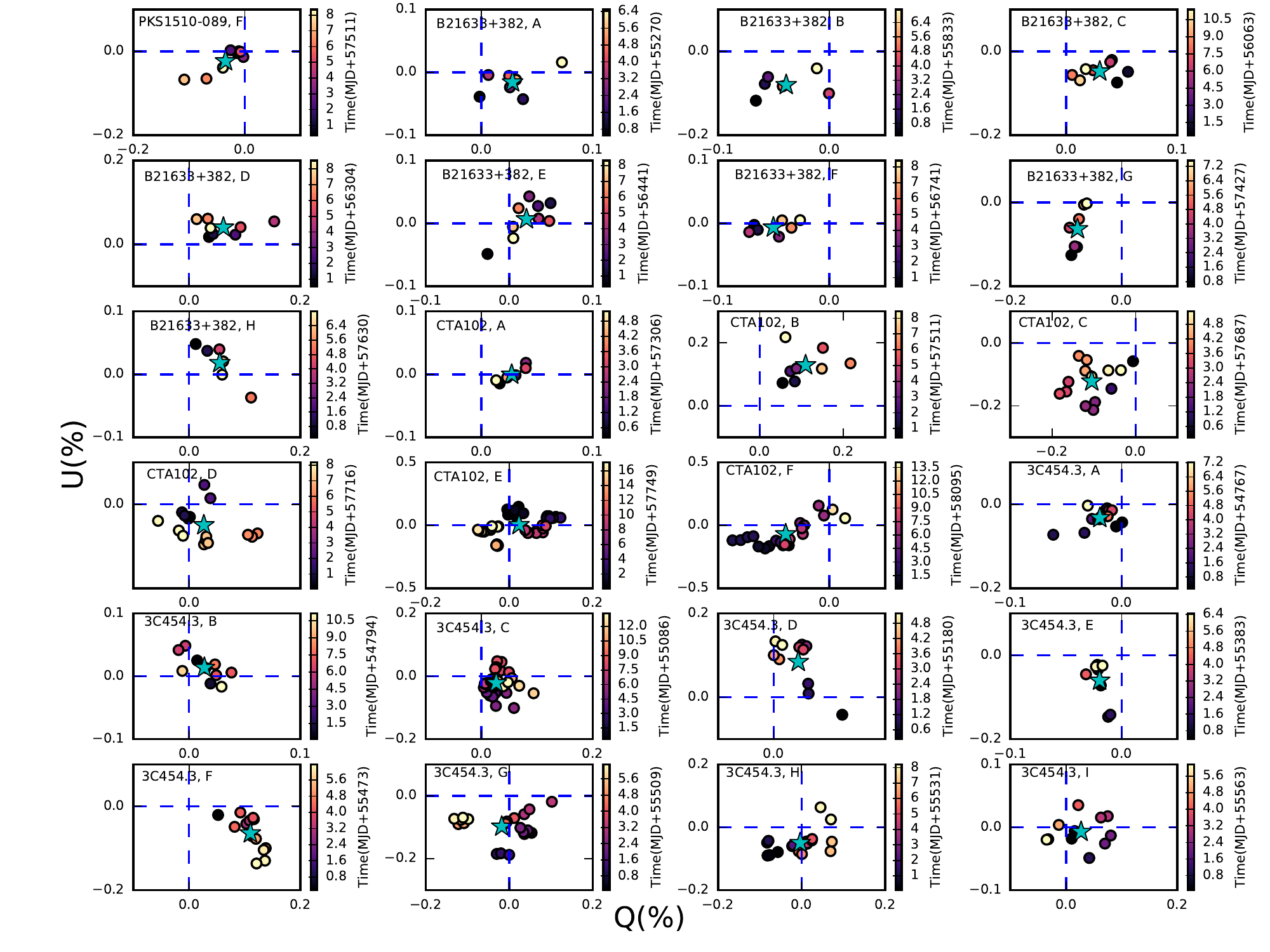}
\caption{\label{fig:q_u2}Same as in Figure \ref{fig:q_u1} but for PKS 1510$-$089, B2 1633$+$383, CTA 102, and 3C 454.3.}
\end{figure*}
\begin{figure*}
\centering
\includegraphics[width=18cm, height=6cm]{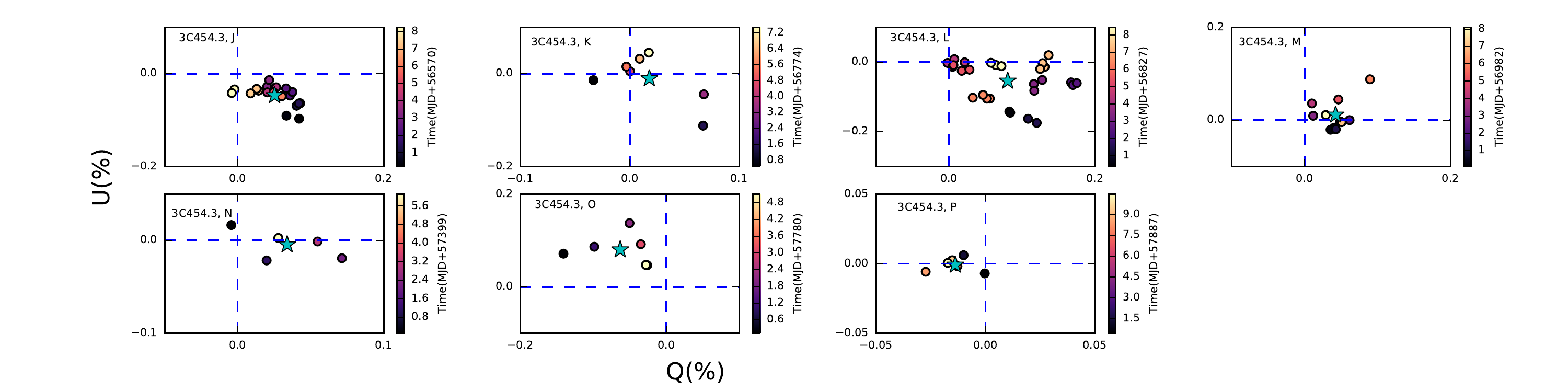}
\caption{\label{fig:q_u3}Same as in Figure \ref{fig:q_u1} but for 3C 454.3.}
\end{figure*}
\end{subfigures}

\section{Results} \label{sec:res}
\subsection{OJ 248} 
The FSRQ OJ 248, at z = 0.941 \citep{2010MNRAS.405.2302H}, was first identified by \cite{1968ApJ...154..413H}. Later, it was observed at high-energy $\gamma$-ray by the Energetic Gamma Ray Experiment Telescope (EGRET) instrument onboard the Compton Gamma-ray Observatory \citep{1999ApJS..123...79H}. OJ 248 is found to be significantly variable in the optical band \citep{1997A&AS..121..119V,1998A&AS..127..445R}. During its 2012-2013 outburst, a significant change in its polarization degree was observed with an erratic correlation between its polarization degree and optical brightness \citep{2015MNRAS.450.2677C}. 

During the entire 10 year monitoring by Steward Observatory, the optical V-band flux of OJ 248 varied from (2.08$-$10.04) $\times$ 10$^{-12}$ erg cm$^{-2}$ s$^{-1}$, while the degree of polarization ranged from 0.03\% to 18.09\%. On short-term timescales, it also showed flux variations with minimum and maximum F$_{var}$ values of 2.65\% and  20.20 \%, respectively. We found positive correlations between optical flux and PD in three out of ten observing cycles, while a negative correlation between the two was observed in one observing cycle. During the remaining six cycles no significant correlation was detected between optical flux and polarization variations. 
The source was mostly undetected in $\gamma-$rays, except during cycle C5. In cycle C5, few moderate flares are seen in $\gamma$-rays, while in the optical band, such short term flares are seen over long term flux changes. On shorter timescales, optical flux was found to be positively correlated with PD during all three epochs. At epoch A, though variations were noticed in the optical band, the source was not detected in $\gamma$-rays. While in epoch B, the source showed flux variations in the $\gamma$-rays as well as in the optical frequencies, in epoch C, the $\gamma$-ray brightness of the source is constant within errors, but, variations were seen in the optical flux and polarization. Thus on both short and long term timescales, varied correlations were found between optical and $\gamma$-ray flux variations. 
We also noticed an abrupt rotation of the PA in epoch A, during which the PA changes by 335 degrees within a day. 

Using optical variability timescales, we derived the values of $\eta\sim$(0.17$-$0.18), $\delta^o\sim$1.29, $R\sim($0.41$-$0.43)$\times$ 10$^{16}$ cm, and $B\sim$(0.46-0.48) G.
   
\subsection{PKS 1222+216}
PKS 1222+216, at a redshift z = 0.432 \citep{1966ApJ...145..654B}, was detected in $\gamma$-rays by EGRET \citep{1999ApJS..123...79H} and at VHE by the Major Atmospheric Gamma-ray Imaging Cherenkov Telescope \citep[MAGIC;][]{2011ApJ...730L...8A}. It has been studied over a wide range of frequencies by \cite{2014ApJ...786..157A}. The weak correlation observed between its optical and $\gamma$-ray flux exposes the signatures of polarized optical synchrotron emission from the jet and blue thermal emission from the accretion disk \citep{2011arXiv1110.6040S}. However, a time delay was also noticed between optical and $\gamma-$ray flux variations \citep[e.g.][]{2014ApJ...797..137C,10.1093/mnras/stv2653}.

Steward Observatory observed PKS 1222+216 from MJD 54948 to MJD 58306. During this period, it showed optical flux variations with V-band flux ranging from (8.52$-$49.92) $\times$ 10$^{-12}$ erg cm$^{-2}$ s$^{-1}$, while the minimum and maximum values of PD recorded were 0.25\% and 29.1\%, respectively. We also found flux variations on shorter timescales, the values of F$_{var}$ ranged from 4.89$-$10.44\%.

On long-term timescales, we found a significant positive correlation between flux and PD in five observing cycles (C2, C3, C5, C9 and C10), while in one cycle (C6) these two were anti-correlated. During cycle C2, there is a gradual change in both the optical and $\gamma$-ray brightness, with short term flares superimposed on the long term changes. In C3, while the brightness in the $\gamma$-ray is nearly constant, the optical brightness is changed by a factor of 2. In C5, while the brightness of the source in $\gamma$-rays is nearly constant, variations were noticed in the optical. In cycle C6, though moderate variation is seen in $\gamma$-ray, optical flux variations are more than a factor of two. In C10, while the source is undetected in $\gamma$-rays during most of the cycle, modest variation is seen in the optical. On short-term timescales, in two out of three epochs flux and PD were positively correlated while in one epoch anti-correlation between the two is observed. We didn't find any significant correlation between $\gamma$-ray and optical flux on short-term timescales on any epoch. There can be two possibilities; either there was no correlation between the optical and $\gamma-$rays \citep{2011arXiv1110.6040S} or the correlation could not be detected due to the time delay between the two fluxes \citep{10.1093/mnras/stv2653}. However, during epoch A, after the initial four days, these two quantities show indications of anticorrelation. 
No abrupt rotation of the PA was noticed in any epoch.

\subsection{3C 273}
3C 273, the first quasar discovered by \cite{1963Natur.197.1040S}, is located at a redshift $z$ = 0.158. It is also the first quasar detected in the $\gamma$-ray (50 to 500 MeV) energy band \citep{1978Natur.275..298S}. It has been studied for flux variations in the optical band by several authors \citep[e.g.][]{2014ApJS..213...26F,2017ApJS..229...21X,2019ApJ...880..155L}. Between 2008 and 2015, a low degree of optical polarization was observed and its optical emission was found to be dominated by the accretion disk during this period \citep{2020MNRAS.497.2066F}.

During the 10 year period of investigation, the optical flux from 3C 273 varied between (110.60$-$181.90) $\times$ 10$^{-12}$ erg cm$^{-2}$ s$^{-1}$. During the same period, its PD varied from 0.04\% to 1.58\%. On long-term timescales, we found a significant positive correlation between optical flux and PD variations only in two observing cycles (C1 and C2), while in the remaining eight cycles no significant correlation was observed. 

\subsection{3C 279}
3C 279, at a redshift of $z$ = 0.538 \citep{1965ApJ...142.1673B}, was among the sources discovered in the $\gamma$-rays (30 MeV to over 5 GeV) by EGRET \citep{1992ApJ...385L...1H}. It has been regularly detected in the 100 MeV to 300 GeV range by {\it Fermi}. It has been studied extensively across all the wavelengths \citep[e.g.][and references therein]{2018ApJ...858...80R}. It showed a significant change in the position angle, which coincided with the $\gamma$-ray flare from 18 February 2009 to 10 March 2009 \citep{2010Natur.463..919A}. During 2008$-$2012, 3C 279 revealed different polarization behaviours at low optical flux and flaring states. \cite{2016A&A...590A..10K} proposed that the polarization behaviour of the source at two optical flux states could be due to two different processes, a stochastic process during low optical flux state and a deterministic process during flaring. From its six years of multi-wavelength photometric and spectropolarimetric observations \cite{2018MNRAS.479.2037P} concluded that either the $\gamma$-ray emission mechanism or the $\gamma$-ray emission site may change with time. \cite{2018ApJ...858...80R} found a negative correlation between optical flux and PD in 3C 279 during the period 2013$-$2014. Recently, \cite{2020MNRAS.492.3829L} found different types of behaviour between optical flux and PD with no preferred correlations by analysing different activity states of the source.

During the 10 years of the monitoring period (MJD 54794$-$58310), the optical V-band flux of 3C 279 ranged from (1.37$-$104.80) $\times$ 10$^{-12}$ erg cm$^{-2}$ s$^{-1}$, while the degree of polarization varied from 0.63\% to 34.5\%. It also showed variations on shorter timescales with minimum and maximum flux variability amplitudes of 4.31\% and 21.79\%, respectively.

On long-term timescales, we noticed no significant correlation between flux and PD variations except in two observing cycles (C4 and C6). The variations between flux and PD are random for most of the observing cycles. In C4, though variations were seen in the optical band, while no variations were seen in the $\gamma$-ray band. Optical PD was found to change between 3-18\%, which followed the optical flux changes. In C6, small amplitude flares (lasting several days) were seen in $\gamma$-rays, and optical flux changes seem to follow the $\gamma$-ray variations. During this cycle, optical flux and PD were anti-correlated. On short-term timescales, in eight out of a total of thirteen epochs, we found a positive correlation between PD and flux, while in the remaining five occasions an anti-correlation was observed between PD and flux. We found a significant positive correlation between $\gamma$-ray and optical flux in epochs L and M. For the same epochs, \cite{2019ApJ...880...32L} and \cite{2020ApJ...890..164P} too have noticed a correlation between optical and $\gamma-$ray light curves with zero time lag. 
No rotation of the PA was noticed in any epoch. 

We estimated the values of $\eta\sim$(0.02$-$0.70), $\delta^o\sim$(0.87$-$1.64), $R\sim($0.20$-$3.20)$\times$ 10$^{16}$ cm, and $B\sim$(0.11$-$0.84) G.

\subsection{PKS 1510$-$089}
PKS 1510$-$089, at a redshift of z = 0.361 \citep{1996AJ....112...62T}, was discovered optically as a quasar by \cite{1966AuJPh..19..559B}. It is an extremely powerful $\gamma$-ray source detected by {\it Fermi} and a highly polarized source \citep{1984ApJ...279..485S,1993ApJS...87..451H}. It is a well-observed source across all frequencies, from radio to $\gamma$-rays \citep[e.g.][]{1986ApJ...300..216M,1997MNRAS.288..920L,2007A&A...464..175B,2008A&A...488..867N,2010ApJ...716..835A}. A significant rotation of its optical polarization angle was observed in 2009 during the optical polarimetric monitoring between 2009 and 2013, which was due to the ejection of new polarized component \citep{2018Galax...6...18B}. The optical and $\gamma-$ray fluxes of PKS 1510$-$089 was found to be correlated with zero time lag \citep[e.g.][]{10.1093/mnras/stv2653,2017A&A...601A..30C}.

PKS 1510$-$089 was observed by Steward Observatory between MJD 54829 and MJD 58306. During this period, its V-band flux varied from (3.94$-$50.47) $\times$ 10$^{-12}$ erg cm$^{-2}$ s$^{-1}$, while the PD ranged from 0.2\% to 25.82\%. It also showed variations on shorter timescales with the minimum and maximum F$_{var}$ values of 4.31\% and 25.13\%, respectively.

Optical flux is found to be positively correlated with PD in five out of ten observing cycles, while in the remaining cycles no significant correlation between flux and PD was observed. During each of these five cycles (C1, C2, C6, C7, and C9), flux variations were seen in both the optical and $\gamma$-ray bands. Close correlations between these two energies could not be ascertained due to scarcity of optical observations, except during cycle C7, where variations in the optical band have a resemblance to variations in the $\gamma$-ray band. On short-term timescales, optical flux is positively correlated with PD in all six epochs. 
We noticed a significant positive correlation between $\gamma$-ray and optical fluxes during epoch A, while in the remaining epochs B, C, D, E and F no significant correlation between these two quantities was found. 
We did not find any abrupt rotation of the PA during any epoch. 

We derived values of $\eta\sim$(0.01$-$0.04), $\delta^o\sim$(0.77$-$0.93), $R\sim($0.35$-$0.70)$\times$ 10$^{16}$ cm, and $B\sim$(0.29$-$0.49) G.

\subsection{B2 1633+382}
B2 1633+382, located at a redshift of z = 1.814 \citep{2017A&A...597A..79P}, was identified as a quasar by \cite{1973A&A....27..475P}. It was discovered to be highly variable in the optical B-band by \cite{1977A&A....59..419B} and, further corroborated by \cite{1997A&AS..121..119V} by observing the intraday variability in the optical B-band. An evidence of quasi-periodicity of $\sim$ 1.9 yr was found in its optical total and polarized flux by \cite{2020MNRAS.492.5524O}. The variations in optical and $\gamma-$ray light curves of B2 1633+382 were found to be correlated with a time delay of zero within uncertainties \citep[e.g.][]{2019ARep...63..378H,2021MNRAS.504.2509W}.

Steward Observatory monitored the source from MJD 54745 to MJD 58306. During this 10 year observing period, it showed optical flux variations with V-band flux ranging from (1.26$-$11.65) $\times$ 10$^{-12}$ erg cm$^{-2}$ s$^{-1}$. During the same period, the source also showed variations in optical polarization with minimum and maximum polarization degrees of 0.36\% and 27.26\%, respectively. On shorter timescales, we also noticed flux variations with minimum and maximum F$_{var}$ values of 3.47\% and 32.96\%. 

On long-term timescales, in six out of ten observing cycles, optical flux was found to be positively correlated with PD, while in the rest four cycles no significant correlation was found between the two. Of the six cycles when optical flux and PD are correlated, during C3, C6 and C10, variations were found in the $\gamma$-ray band, while during the other three cycles, namely C1, C4 and C8, the source was not found to show variations in the $\gamma$-ray band. In cycle C6, we observed a gradual decline in brightness in both the optical and $\gamma$-ray bands. We found a positive correlation between optical flux and PD in all eight epochs on short-term timescales. A significant positive correlation between $\gamma$-ray and optical fluxes was observed only in epoch E, while, in the other epochs no significant correlation was seen between the two quantities.
No swing of the PA was noticed in any of these epochs.

The derived values of $\eta$, $\delta^o$, $R$, and $B$ are (0.12$-$0.61), (1.19$-$1.60), (0.68$-$2.57)$\times$ 10$^{16}$ cm, and (0.14$-$0.37) G, respectively.

\subsection{CTA 102}
CTA 102 is a highly polarized FSRQ \citep{1981ApJ...243...60M} located at a redshift of z = 1.037 \citep{1965ApJ...141.1295S}. It was first identified as a radio quasar by the Caltech Radio Survey \citep{1960PASP...72..237H}. It has been studied across the entire EM spectrum for flux variations \citep[e.g.][]{2018A&A...617A..59K,2018ApJ...866...16P,2020ApJ...891...68C}. It has shown minute scale variability in the optical band \citep{2009AJ....138.1902O}. The intraday variability was also observed in its optical polarized emission during the 2012 $\gamma$-ray outburst \citep{2015ApJ...813...51C}. A positive correlation was observed between its optical and $\gamma-$ray fluxes which was consistent with a zero time lag within error bars \citep[e.g.][]{2019MNRAS.490.5300D,2020ApJ...891...68C}.

CTA 102 was observed by Steward Observatory during all its observing cycles except in cycle C1. During its monitoring period (MJD 55124$-$58306), its optical flux varied from  (3.05$-$483.3) $\times$ 10$^{-12}$ erg cm$^{-2}$ s$^{-1}$, while the degree of polarization ranged from 0.08\% to 36.8\%. The minimum and maximum values of F$_{var}$ on short-term timescales are 7.58\% and 45.12\%, respectively.

We found a positive correlation between optical flux and PD in five out of nine observing cycles, while no significant correlation was found in the remaining four cycles. On short-term timescales, we noticed positive correlation between flux and PD in all six epochs. In three (D, E, and F) out of six epochs, we found a significant positive correlation between $\gamma$-ray and optical flux variations. Moreover, in epoch B, an indication of correlation between optical and $\gamma-$ray flux variations was seen but it was not significant. While in epoch A, there was an optical flare, the source did not show variations in the $\gamma$-ray band.
We noticed an abrupt change in the values of the PA in epochs A, D, and E. During epoch A, a change of 345 degrees was observed in the values of PA from MJD 57307.19 to MJD 57308.19 ($\sim$ 1 day), which is accompanied by an increase in the optical flux and PD. However, the $\gamma$-ray flux was constant within error bars during this period. In epoch D, an abrupt change of 277 degrees in $\sim$ 20 hrs was found in the PA, while a PA rotation of 150 degrees was observed in epoch E. 

We obtained values of $\eta\sim$(0.10$-$6.11), $\delta^o\sim$(1.15$-$2.34), $R\sim($0.04$-$1.24)$\times$ 10$^{16}$ cm, and $B\sim$(0.22$-$2.63) G.

\subsection{3C 454.3}
3C 454.3, at a redshift z = 0.859, was first observed as a variable and bright $\gamma$-ray source by EGRET \citep{1993ApJ...407L..41H}. It is a highly polarized quasar and has been studied over a wide range of energies \citep[e.g.][]{2009ApJ...697L..81B,2010ApJ...721.1383A,2017MNRAS.464.2046K,2017MNRAS.470.3283S}. In 2005, \cite{2006A&A...453..817V} detected an extraordinary optical flare. By analysing its various long-term epochs, \cite{2019MNRAS.486.1781R} found both positive and negative correlation between its optical flux and PD. The optical and $\gamma-$ray flux variations of 3C 454.3 were found to be well correlated with zero time delay \citep[e.g.][]{10.1093/mnras/stv2653,2021ApJ...906....5A}.

During the 10 years (MJD 54743$-$58306) of monitoring by Steward Observatory, the optical V-band flux of 3C 454.3 varied between (5.40$-$109.80) $\times$ 10$^{-12}$ erg cm$^{-2}$ s$^{-1}$. During the same period, its optical polarization varied from 0.13\% to 24.97\%. We also searched for flux variations on shorter timescales and found the values of F$_{var}$ in the range of $\sim$ 4.56\%$-$27.51\%. 

On long-term timescales, we observed a positive correlation between optical flux and PD in seven out of ten observing cycles, while in the other three cycles no significant correlation was detected between the two. Among the seven cycles where correlation between optical flux and PD were noticed, in six cycles, optical and $\gamma$-ray flux variations were correlated, while in cycle C4, while variations were seen in the optical band the source was not detected in the $\gamma$-rays. 
On short-term timescales, in fourteen out of sixteen epochs we found a positive correlation between flux and PD, while in the other two epochs optical flux and PD were anti-correlated. Among the 16 epochs, during the seven epochs, namely, A, C, D, G, L, M and N, the optical and $\gamma$-ray flux variations were found to be positively correlated, while during epoch O, we found anti-correlation between optical and $\gamma$-ray flux variations. 
In epoch L, the source showed a flare at both optical and $\gamma$-ray bands. At the peak of the optical flare abrupt rotation of $\sim 206$ degrees was observed that happened in about three days (MJD 56829-56832). Similar rotation of the PA was also reported by \cite{2020ApJ...902...61L}. In epoch C, an abrupt change of 138 degrees within a day was also observed in the PA. 

We derived values of $\eta\sim$(0.19$-$2.13), $\delta^o\sim$(1.30$-$2.00), $R\sim($0.20$-$1.50)$\times$ 10$^{16}$ cm, and $B\sim$(0.20$-$0.90) G.

\section{Discussion}\label{sec:dis}
We investigated the correlation between optical flux and polarization variations on long-term as well as on short-term timescales in eight $\gamma-$ ray bright FSRQs using $\sim$ 10 years of data from the Steward Observatory. We found a positive correlation between optical flux and PD in 34 observing cycles, negative correlation in 3 observing cycles and no correlation in 42 observing cycles. The optical flux positively correlated with PD in 47 out of 55 short-term epochs, while an anti-correlation was detected between the two in the remaining 8 epochs. We also searched for correlations between $\gamma-$ ray and optical flux on short-term timescales and found that these two are positively correlated in 14 epochs and negatively correlated in one epoch, while in the remaining 40 epochs, no significant correlation between these two fluxes was found. The results of our analysis for individual sources are summarized in Table \ref{tab:res_sum}. In the table, the results for short-term analysis are not given for 3C 273 as our selection criteria yielded no epochs for short-term analysis. 

\begin{table*}
\centering
\caption{\label{tab:res_sum}Summary of our analysis.}
\begin{tabular} {lccccccccccc}\hline
Blazar		& \multicolumn{4}{c}{Long-term timescale} & \multicolumn{3}{c}{Short-term timescale} & $\eta^o$ & $\delta^o$  & R  & $B$  \\ \cmidrule[0.03cm](r){2-5}\cmidrule[0.03cm](r){6-8}
        	& No. of observing cycles &  PC   &  NC &  No Corr. & No. of epochs & PC & NC             &	        &             &    ($\times$ 10$^{16}$ cm)     &   (in G)  \\ \hline 
OJ 248		& 10 & 3 & 1 & 6 &  3	& 3 & 0 &   0.17$-$0.18 & 1.29          &   0.41$-$0.43         & 0.46$-$0.48          \\
PKS 1222$+$216  & 10 & 5 & 1 & 4 &  3	& 2 & 1 &      -        &	 -      &     -                 &      -    \\
3C 273		& 10 & 2 & 0 & 8 &  0	& 0 & 0 &      -        &       -       &     -                 &   -      \\  
3C 279		& 10 & 1 & 1 & 8 &  13	& 8 & 5 & 0.02$-$0.70   & 0.87$-$1.64   &  0.20$-$3.20          &   0.11$-$0.84   \\
PKS 1510$-$089	& 10 & 5 & 0 & 5 &  6	& 6 & 0 & 0.01$-$0.04   & 0.77$-$0.93   &  0.35$-$0.70          &   0.26$-$0.49       \\
B2 1633$+$382 	& 10 & 6 & 0 & 4 &  8	& 8 & 0 & 0.12$-$0.61   & 1.19$-$1.60   &  0.68$-$2.57          &   0.14$-$0.37        \\
CTA 102	& 9  & 5 & 0 & 4 &  6	& 6 & 0 & 0.10$-$6.11   & 1.15$-$2.34   &  0.04$-$1.24          &   0.22$-$2.63        \\
3C 454.3	& 10 & 7 & 0 & 3 &  16	& 14 & 2 & 0.19$-$2.13  & 1.29$-$2.00   &  0.20$-$1.50          &   0.20$-$0.90        \\
 \hline
\end{tabular}
\end{table*}

Flux variations in blazars are explained by models based on extrinsic as well as intrinsic mechanisms. In the extrinsic scenario, the observed flux variations are caused by geometrical effects like swing in the jet viewing angle or the change in the Doppler factor due to bending of the relativistic jets \citep[e.g.][]{1992A&A...259..109G,1999A&A...347...30V}. However, in intrinsic models the flux variations are associated with factors intrinsic to the source like magnetic instabilities \citep{2014ApJ...780...87M}, motion of shocks within the jets (shock-in-jet models; \citealt{1985ApJ...298..114M}). 

In the shock-in-jet scenario, a shock propagating along the relativistic jet accelerates the particles to high energies which are then cooled down by emission mechanisms resulting in a change in flux. At the same time, it also aligns the magnetic field, thereby creating an ordered magnetic field. The change in the ordering of the magnetic field produces change in the degree of polarization. So, the observed correlation between optical flux and PD can reasonably be explained by the shock-in-jet model.

\begin{table*}
\centering
\caption{\label{tab:op_vlbi}Comparison between the optical and the 15 GHz VLBI polarization observations.}
\resizebox{0.7\textwidth}{!}{
\begin{tabular}{lccccccc} \hline
Blazar      	&           \multicolumn{4}{c}{Optical}  &  \multicolumn{3}{c}{VLBI}  \\
		&  Epoch  & Period        & PD (\%)         & PA (degree)           & Epoch    & PD (\%) & PA (degree) \\ \hline
3C 279       & D & 56039-56047 & 13.07$\pm$0.03 & 164.23$\pm$0.08 &  56046.0  &  5.8 &  140.0 \\ 
PK1510-089   & A & 54911-54917 & 13.53$\pm$0.04 & 72.47$\pm$0.11  &  54915.0  &  3.8 &  67.0 \\
   & C & 56039-56047 & 1.89$\pm$0.05  & 113.96$\pm$1.16 &  56046.0  &  1.4 &  178.0 \\
   & D & 56441-56449 & 1.86$\pm$0.04  & 46.28$\pm$1.35  &  56445.0  &  3.9 &  13.0 \\
3C 454.3     & N & 57399-57405 & 4.07$\pm$0.04  & 112.70$\pm$0.41 &  57403.0  &  1.3 &   109.0   \\
\hline
\end{tabular}}
\end{table*}
Variations in the degree of polarization can also be explained by changes in the spectral index of the emitting electrons in the relativistic jet \citep[e.g.][]{2018ApJ...858...80R}. The maximum degree of polarization for a power law ($dN/dE \propto  E^{-p}$, where $p = 2\alpha +1$) distribution of relativistic electrons is given by
\begin{equation}\label{eq:delta_pd}
PD = \frac{(\alpha +1)}{\alpha + 5/3}
\end{equation}
where $\alpha$ is the spectral index. 
For each epoch of all the FSRQs, we estimated the values of the spectral index using Equation \ref{eq:alpha}. As can be seen from Table \ref{tab:epoch_res}, the maximum change in the spectral index ($\Delta\alpha$) during all the epochs is 0.33 for OJ 248, 0.62 for PKS 1222$+$216, 1.29 for 3C 279, 0.27 for PKS 1510$-$089, 0.39 for B2 1633$+$382, 1.51 for CTA 102, and 0.99 for 3C 454.3 and the corresponding expected maximum change in the PD, calculated using Equation \ref{eq:delta_pd}, are 4\%, 9\%, 14\%, 2\%, 3\%, 6\%, and 7\%, respectively. But the maximum variation observed in PD in this work is 16.98\% for OJ 248, 20.47\% for PKS 1222$+$216, 11.3\% for 3C 279, 19.98\% for PKS 1510$-$089, 12.25\% for B2 1633$+$382, 28.84\% for CTA 102, and 20.92\% for 3C 454.3
We noticed that the change expected in PD due to the change in the spectral index is much lower than the observed variation in PD for all the sources except 3C 279. 
It indicates that the change in the power-law spectral index of the emitting electrons within the jet cannot solely explain the observed variations in PD in this work.

The observed anti-correlation between optical flux and PD can also be explained by the Turbulent, Extreme Multi-zone (TEMZ, \citealt{2014ApJ...780...87M}) model. According to this model, there can be more turbulent cells (small scale emission regions) during outbursts compared to other times.  These multiple cells can have different polarization angles leading to a lower degree of polarization with increasing flux. To test the hypothesis of the presence of more than one emission component, for example, a stable polarized component plus other multiple components, we plot the observed Q and U Stokes parameters for each epoch of all the sources in Figures \ref{fig:q_u1} to \ref{fig:q_u3}. In each plot, the Q, U centroid significantly deviates from (0,0) implying the presence of a stable polarized component with the PA aligned along the jet direction \citep{2009A&A...503..103B}. On the other hand, if the enhanced flux variations coincide with the emergence of a new VLBI knot (a new emission region), the observed variation in flux can be due to this new emission region. If the magnetic field in this new blob of emission is aligned with the large scale magnetic field a correlation between flux and PD is expected. Alternatively, if the direction of the magnetic field of the new emitting blob is either chaotic or if it is misaligned with the large scale magnetic field, an anti-correlation between flux and PD can be expected. We found overlapping VLBI observations at 15 GHz \citep{2018ApJS..234...12L} on 5 of the 55 epochs studied in this work. A comparison between optical and 15 GHz VLBI polarization measurements for these five epochs is shown in Table \ref{tab:op_vlbi}. The values of PA found from VLBI observations, that could provide parsec scale resolution, are similar to that found from optical observations. On these five epochs, we found a positive correlation between optical flux and PD. Therefore, the varied correlations between optical flux and PD (both correlation and anti-correlation) observed on short-term timescales in these FSRQs could be due to the presence of multiple emission components in their jets. We note that the values of optical PA and PD do not exactly match with that of the measurements from VLBI. This is because the VLBI measurements are at a particular epoch while the optical measurements are the average over a period of about 10 days that encompass the epoch of VLBI observation. Also, the optical and radio polarization signatures can be different, as the radio emission generally comes from a larger region in the jet.

Fast flares in the $\gamma$-ray band are also explainable by magnetic reconnection in the emission region \citep{2018ApJ...862L..25Z, 2019MNRAS.484.1378G, 2020ApJ...900L..23H}. The light curves that are expected from magnetic reconnection seem to match with observations \citep{2019MNRAS.482...65C}. It has been found that optical polarization angle swings are associated with multiwavelength flares \citep{2016MNRAS.463.3365A, 2018MNRAS.474.1296B}. Also, from observations, it is found that PA swings are accompanied by a drop in PD \citep{2016MNRAS.457.2252B}. Thus it is likely that PA swings affect the magnetic field in the emission region of blazars. Recently, \citep{2020ApJ...901..149Z} has investigated the effect of magnetic reconnection on the observed polarization signatures. Among the observable signatures of magnetic reconnection are (a) swings in PA during flares and (b) a decrease in PD during PA swings. From our analysis of flux and polarization behaviour over times scales of the order of days for a total of 55 epochs in eight FSRQs, we found abrupt polarization swings in a total of 5 epochs of 3 FSRQs, namely, OJ 248, CTA 102, and 3C 454.3. 
In one epoch of the source 3C 454.3, a PA swing of $\sim$ 206 degrees in $\sim$ 3 days was observed. This PA swing is simultaneous to the peak of the optical and $\gamma-$ray flares, and during the PA swing, the PD was also found to be the minimum (See Figure \ref{fig:epoch_lc13}). \cite{2020ApJ...902...61L} suggested that this drastic drop in PD and the simultaneous PA swing during the flare could be either due to shock moving in a helical path within jet or due to change in emission dominance between standing shock and moving shock polarized components with the orthogonal magnetic field orientations. Such PA swings that are observed along with $\gamma$-ray flares \citep{2008Natur.452..966M, 2010ApJ...710L.126M} suggest that they are physically connected to gamma-ray activity \citep{2015MNRAS.453.1669B, 2018MNRAS.474.1296B}. 

Our knowledge of the flux and polarization variability characteristics of these eight FSRQs is drawn from the analysis of different modes of multi-wavelength data that are scanty and with poor time resolution. Near simultaneous observations over several wavelengths and covering different timescales (with good time sampling) on a large sample of blazars are needed to pinpoint the exact causes (and thereby constrain available models) for different correlations between flux and polarization in blazars as well as the connection between optical polarization properties with the observed high energy $\gamma$-ray emission.

\section{Summary}\label{sec:summary}
Using about 10 years of optical photometric and polarimetric data from Steward Observatory, and $\gamma-$ ray data from {\it Fermi}-LAT, for the eight $\gamma-$ray bright FSRQs, we probed for (i) the correlation between optical flux and polarization on both long-term and the short-term timescales, (ii) correlation between $\gamma$-ray flux and optical flux on short-term timescales. We summarize the findings of this work below.
\begin{enumerate}
\item On long-term timescales, all the sources showed variations in the $\gamma$-ray flux, optical flux and polarization degree. In 34 out of the total 79 observing cycles of all the sources, we found a positive correlation between optical flux and degree of polarization, while in three cycles, we noticed anti-correlation between optical flux and polarization degree. In the remaining 42 observing cycles, the variations in optical flux and polarization degree were found to be random with no significant correlation.
\item On short-term timescales, the optical flux significantly correlated with PD in 47 out of 55 epochs, while in the other 8 epochs we found anti-correlations between the optical flux and PD. 
\item On short-term timescales, out of total 55 epochs, in 14 epochs the $\gamma$-ray flux is found to be correlated with the optical flux. In only one epoch, optical flux is anti-correlated with $\gamma$-ray flux variations, while on the remaining 40 epochs, no correlation is observed between optical and $\gamma$-ray flux variations.   
\item On short-term timescales, the values of optical flux variability amplitude of these eight FSRQs ranged from 2.65\% to 45.12\% and the derived values of the magnetic fields varied from 0.11 to 2.63 G.
\end{enumerate}

\section*{Acknowledgements}
We thank the anonymous referee for his/her comments that helped to improve the manuscript. This work has made use of Fermi data collected from the Fermi Science Support Center (FSSC) supported by the NASA Goddard Space Flight Center. Data from the Steward Observatory spectropolarimetric monitoring project were used. This program is supported by Fermi Guest Investigator grants NNX08AW56G, NNX09AU10G, NNX12AO93G, and NNX15AU81G.

\section*{Data Availability}
The data underlying this article are publicly available from the Fermi-LAT\footnote{\url{https://fermi.gsfc.nasa.gov/ssc/data/access/}} and Steward Observatory\footnote{\url{http://james.as.arizona.edu/~psmith/Fermi}} data archives. 

\bibliographystyle{mnras}
\bibliography{ref} 

\bsp	
\label{lastpage}
\end{document}